\documentclass[journal, transmag]{IEEEtran}

\usepackage{mathptmx} 
\usepackage{times} 
\usepackage{amsmath} 
\usepackage{amsbsy} 
\usepackage{amssymb}
\usepackage{mathrsfs}
\usepackage{comment}
\usepackage{xfrac}
\usepackage{multirow}
\usepackage[table]{xcolor}
\usepackage[export]{adjustbox}
\usepackage{tikz}
\usetikzlibrary{external,positioning,decorations.pathreplacing,shapes,arrows,patterns}
\usepackage[outline]{contour}

\usepackage{boxedminipage}
\usepackage{pgfplots}
\pgfplotsset{compat=1.14}
\usepackage{graphicx}
\usepackage{pstool}
\usepackage[latin1]{inputenc}
\usepackage{xifthen}
\usepackage{epic}
\usepackage{subcaption}
\usepackage[outdir=./]{epstopdf}

\usepackage{float}
\floatstyle{boxed}
\newfloat{textbox}{thp}{lop}
\usepackage{capt-of}

\newcommand{\mmse}{\mathsf{mmse}}

\newcommand{\nyq}{\mathsf{Nyq}}
\newcommand{\smp}{\mathsf{smp}}
\newcommand{\qnt}{\mathsf{qnt}}
\newcommand{\PCM}{\mathsf{PCM}}
\newcommand{\Yv}{\mathbf{Y}}
\newcommand{\SI}{\mathsf{SI}}

\newcommand{\intfstofs}{\int_{-\frac{f_s}{2}}^\frac{f_s}{2} }
\newcommand{\Ltwo}{\mathrm{L}_2}

\tikzstyle{int}=[draw, fill=blue!10, minimum height = 1cm, minimum width=1.5cm,thick ]
\tikzstyle{sint}=[draw, fill=blue!10, minimum height = 0.5cm, minimum width=0.8cm,thick ]
\tikzstyle{sum}=[circle, fill=blue!10, draw=black,line width=1pt,minimum size = 0.5cm, thick ]
\tikzstyle{ssum}=[circle, fill=blue!10,draw=black,line width=1pt,minimum size = 0.1cm]
\tikzstyle{int1}=[draw, fill=blue!10, minimum height = 0.5cm, minimum width=1cm,thick ]
\tikzstyle{enc}=[draw, fill=blue!10, minimum height = 2.7cm, minimum width=1cm,thick ]
\tikzstyle{int}=[draw, fill=blue!10, minimum height = 1cm, minimum width=1.5cm,thick ]

\title{\LARGE \bf
Analog-to-Digital Compression: A New \\
Paradigm for Converting Signals to Bits}

\author{ 
\IEEEauthorblockN{
Alon Kipnis, Yonina C. Eldar and  Andrea J. Goldsmith}

\thanks{A. Kipnis and A. J. Goldsmith are with the Department of Electrical Engineering, Stanford University, Stanford, CA 94305 USA. 

Y. C. Eldar is with the Department of Electrical Engineering, Technion - Israel Institute of Technology Haifa 32000, Israel.

This work was supported in part by the NSF under grant CCF-1320628, under the NSF Center for Science of Information (CSoI) grant CCF-0939370, and under BSF Transformative Science Grant 2010505.}
}

\begin{document}
\graphicspath{{../Figures/}}
\maketitle

%%%%%%%%%%%%%%%%%%%%%%%%%%%%%%%%%%%%%%%%%%%%%%%%%%%%%%%%%%%%%%%%%%%%%%%%%%%%%%%%

%%%%%%%%%%%%%%%%%%%%%%%%%%%%%%%%%%%%%%%%%%%%%%%%%%%%%%%%%%%%%%%%%%%%%%%%%%%%%%%%

\section{INTRODUCTION}
\label{sec:Intro}

\begin{figure}
\begin{center}
\includegraphics[scale = 0.4]{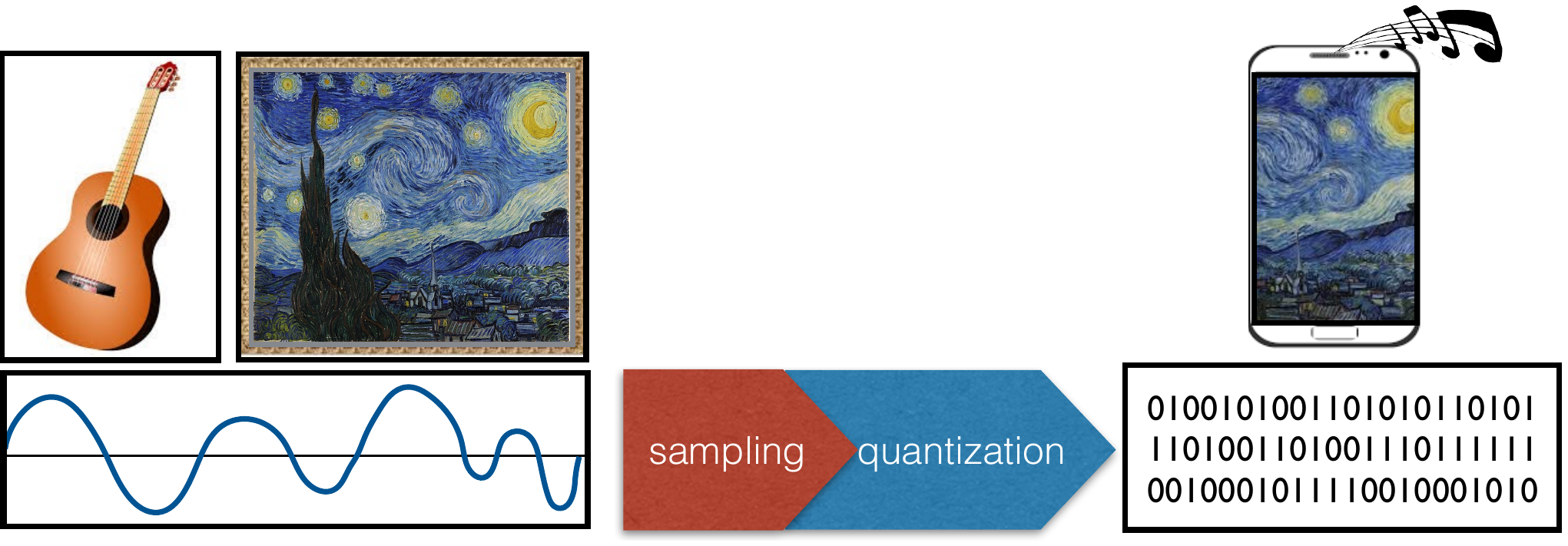}
\end{center}
\caption{Analog-to-digital conversion is achieved by combining sampling and quantization. 
\label{fig:motivation}}
\end{figure}

%\IEEEPARstart{P}
Processing, storing and communicating information that originates as an analog signal involves conversion of this information to bits. This conversion can be described by the combined effect of sampling and quantization, as illustrated in Fig.~\ref{fig:motivation}. The digital representation is achieved by first sampling the analog signal so as to represent it by a set of discrete-time samples and then quantizing these samples to a finite number of bits. Traditionally, these two operations are considered separately. The sampler is designed to minimize information loss due to sampling based on characteristics of the continuous-time input. The quantizer is designed to represent the samples as accurately as possible, subject to a constraint on the number of bits that can be used in the representation. The goal of this article is to revisit this paradigm by illuminating the dependency between these two operations. In particular, we explore the requirements on the sampling system subject to constraints on the available number of bits for storing, communicating or processing the analog information. 
%Consequently, the information loss is a result of possibly sub-optimal sampling and quantization. 
%In particular, the joint effect of these two operation leads to a critical sampling rate that attains the optimal performance for sampling signals under a constraint on the bit-resolution in representing the signal. For example, in uniform sampling bandlimited signals, this new critical rate is below the Nyquist rate and converges to the latter as the bit-precision increases.
% This dependency arise from considering the joint effect of these two operations, rather than each one separately. 
\par
As motivation for jointly optimizing sampling and quantization, consider the minimal sampling rate that arises in classical sampling theory due to Whittaker, Kotelnikov, Shannon and Landau \cite{wiiittaker1915functions, MR0028549, 1447892}. These works establish the Nyquist rate or the spectral occupancy of the signal as the critical sampling rate above which the signal can be perfectly reconstructed from its samples. This statement, however, focuses only on the critical sampling rate required to perfectly reconstruct a bandlimited signal from its discrete samples; it does not incorporate the quantization precision of the samples and does not apply to signals that are not bandlimited. It is in fact impossible to obtain an exact representation of any continuous-amplitude sequence of samples by a digital sequence of numbers due to finite quantization precision, and therefore any digital representation of an analog signal is prone to error. That is, no continuous amplitude signal can be reconstructed from its quantized samples with zero distortion regardless of the sampling rate, even when the signal is bandlimited. This limitation raises the following question:
In converting a signal to bits via sampling and quantization at a given bit precision, can the signal be reconstructed from these samples with minimal distortion based on sub-Nyquist sampling? 
In this article we discuss this question by extending classical sampling theory to account for quantization and for non-bandlimited inputs. Namely, for an arbitrary stochastic input and given a total budget of quantization bits, we consider the lowest sampling rate required to sample the signal such that reconstruction of the signal from its quantized samples results in minimal distortion. As we shall see, without assuming any particular structure on the input analog signal, this sampling rate is often below the signal's Nyquist rate.
\\

\begin{figure}
\begin{center}
\begin{tikzpicture}[node distance=2cm,auto,>=latex]
\draw[dotted, line width = 1] (2,0.8) -- (2,-3.5)  node[left,align = center, xshift = 0cm] {analog~ \\ (continuous-time)} node[right,align = center, xshift = 0.5cm] {analog~ \\ (discrete-time)};    
\draw[dotted, line width = 1] (5,0.8)  -- (5,-3.5)  node[right] {~digital};

\node at (0,0) (source) {$X(t)$} ;
\node [int,right of = source, node distance =2 cm, align = center] (sampler) {sampler};
 \node [int] (enc) [right of = sampler, node distance = 3cm]{$\mathrm{encoder}$};
\node [right of = enc, node distance = 2cm] (right_edge) {};
\node [below of = right_edge, node distance = 2cm] (right_b_edge) {};

\node [right] (dest) [below of=source, node distance = 2cm]{$\hat{X}(t)$};
\node [int, minimum width=4.5cm] (dec) [xshift = -1.5cm,below of=enc, node distance = 2cm] {$\mathrm{decoder}$};

\draw[-,line width=2pt] (sampler) -- node[above] {$f_s \left[ \frac{\mathrm{smp}}{\mathrm{sec}}\right]$ } (enc);
    \draw[-,line width=2pt] (enc) -- node[above] {$R [\frac{\mathrm{bits}}{\mathrm{sec}}]$} (right_edge);
    
  \draw[-,line width = 2]  (right_edge.west) -| (right_b_edge.east);
    \draw[->,line width = 2]  (right_b_edge.east) -- (dec.east);
%   \draw[->,line width=2pt] (enc) -- node[above] {$R [\frac{\mathrm{bits}}{\mathrm{sec}}]$} (dec);
   \draw[->,line width=2pt] (dec) -- (dest);
   %{$Y[n]=X(n/f_s)$} (enc);
    \draw[->,line width=2pt] (source) -- (sampler);
    \draw[<->,dashed,line width=1pt] (source) -- node {distortion} (dest);
\end{tikzpicture}
\end{center}
\caption{\label{fig:combined_sampling_source_coding} Analog-to-digital compression (ADX) and reconstruction setting. Our goal is to derive the minimal distortion between the signal and its reconstruction from any encoding at bitrate $R$ of the samples of the signal taken at sampling rate $f_s$.}
\end{figure}
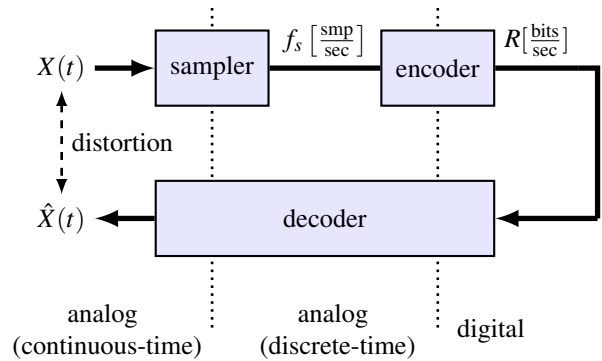

The minimal distortion achievable in the presence of quantization depends on the particular way the signal is quantized or, more generally, encoded, into a sequence of bits. Since we are interested in the fundamental distortion limit in recovering an analog signal from its digital representation, we consider all possible encoding and reconstruction (decoding) techniques. As an example, in Fig.~\ref{fig:motivation} the smartphone display may be viewed as a reconstruction of the real world painting The Starry Night from its digital representation. 
%More precisely, the image on the smartphone is a recovery of the analog image associated with the light reflected back from the painting at a particular location and lighting condition. 
No matter how fine the smartphone screen, this recovery is not perfect since the digital representation of the analog image is not accurate, so that loss of information occurs during the transformation from analog to digital. Our goal is to analyze this loss as a function of hardware limitations on the sampling mechanism and the number of bits used in the encoding. It is convenient to normalize this number of bits by the signal's free dimensions, that is, the dimensions along which new information is generated. For example, the free dimensions of a visual signal are usually the horizontal and vertical axes of the frame, and the free dimension of an audio wave is time. For simplicity, we consider analog signals with a single free dimension, the dimension of \emph{time}. %Most of the conclusions of our discussion are naturally extended to signals with multiple free dimension. 
Therefore, our restriction on the digital representation is given in terms of its \emph{bitrate} -- the number of bits per unit time. 
\par 
For an arbitrary continuous-time random signal with known statistics, the fundamental distortion limit due to the encoding of the signal using a limited bitrate is given by Shannon's distortion-rate function (DRF) \cite{Shannon1948,shannon1959coding,berger1971rate}. This function provides the optimal tradeoff between the bitrate of the signal's digital representation and the distortion in recovering the original signal from this representation. Shannon's DRF is described only in terms of the distortion criterion, the probability distribution on the continuous-time signal, and the maximal bitrate allowed in the digital representation. Consequently, the optimal encoding scheme that attains Shannon's DRF is a general mapping from continuous-time signal space to bits that does not consider practical constraints in implementing it. In practice, the encoding of an analog signal into bits entails first sampling the signal and then representing the samples using a limited number of bits. Therefore, in practice, the minimal distortion in recovering analog signals from their bit representation considers the digital encoding of the signal \emph{samples}, with a constraint on both the \emph{sampling rate} and the \emph{bitrate} of the system. Here the sampling rate $f_s$ is defined as the number of samples per unit time of the continuous-time source signal and the bitrate $R$ is the number of bits per unit time used in the representation of these samples. The resulting system describing our problem is illustrated in Fig.~\ref{fig:combined_sampling_source_coding}, and is referred to as the \emph{analog-to-digital compression} (ADX) setting. \par
The digital representation in this setting is obtained by transforming a continuous-time continuous-amplitude random source signal $X(t)$ through a concatenated operation of a \emph{sampler} and an \emph{encoder}, resulting in a bit sequence. For instance, when the input signal $X(t)$ is observed over a time interval of length $T$, then the sampler produces $\lfloor f_s T \rfloor$ samples and the encoder maps these samples to $\lfloor TR \rfloor$ bits. 
%; the sampler is any transformation of a continuous-time signal to a discrete-time signal, and the encoder is any mapping of the discrete-time signal to a sequence of bits.
The \emph{decoder} estimates the original analog signal from this bit sequence. The \emph{distortion} is defined to be the mean squared error (MSE) between the input signal $X(t)$ and its reconstruction $\widehat{X}(t)$. Since we are interested in the fundamental distortion limit subject to a sampling constraint, we allow optimization over the encoder, decoder and the time horizon $T$. In addition, we also explore the optimal sampling mechanism, but limit ourselves to the class of linear and continuous deterministic samplers \cite{eldar2015sampling}. Namely, each sampler in this class is a linear continuous mapping of signals over time lag $T$ to $\mathbb R^{\lfloor f_s T \rfloor}$.
%The conclusions we obtain based on these expression extend to other fidelity criterions for measuring performance of systems. %Some of these different criterions are discussed in Section~\ref{sec:applications}.
\par
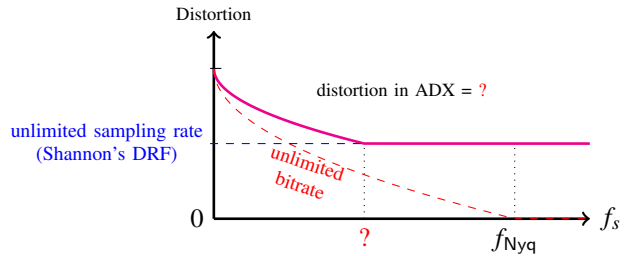
\begin{figure}[h]
\begin{center}
\begin{tikzpicture}[scale=1]

\draw[->,line width=1pt]  (0,0) node[left] {$0$}--(5,0) node[right] {$f_s$};
\draw[->,line width=1pt]  (0,0)--(0,2.5) node[above] {\scriptsize Distortion};

\draw[magenta, line width = 1pt] plot[domain=0:2, samples=100] (\x*\x/2,2-\x/2) -- plot[domain=2.5:5, samples=100] (\x, 1) node[above, xshift=-2.5cm, yshift = 0.5cm] {\color{black} \scriptsize distortion in ADX = $\color{red}?$};
 
\draw[red, dashed] plot[domain=0:4, samples=100] (\x*\x/4,2-\x/2) --  plot[domain=4:5, samples=100] (\x, 0);	

\draw[dotted] (2,0) node [below,color = red] {$?$}-- (2,1); 

\draw[dotted] (4,0) node [below] {$f_{\nyq}$}-- (4,1); 

\draw[blue,dashed] (-0.05,1) node[left, align = center, scale = 0.75,blue] {unlimited sampling rate \\ (Shannon's DRF) } -- (2,1);
\draw (-0.05,2) -- (0.1,2);
\node [rotate=-18.5, red, align = center, scale = 0.75,yshift = -0.15] at  (1.2,0.6) (Sx_or) {unlimited \\
bitrate};
\end{tikzpicture}
\caption{\label{fig:contribution} 
The minimal sampling rate for attaining the minimal distortion achievable in the presence of quantization is usually below the Nyquist rate, whereas sampling at the Nyquist is necessary to attain zero distortion without quantization constraints. }
\end{center}
\end{figure}
The minimal distortion in ADX is bounded from below by two extreme cases of the sampling rate and the bitrate, as illustrated in Fig.~\ref{fig:contribution}: (1) when the bitrate $R$ is unlimited, the minimal ADX distortion reduces to the MSE in interpolating a signal from its samples at rate $f_s$. (2) When the sampling rate $f_s$ is unlimited or above the Nyquist rate of the signal, the ADX distortion reduces to Shannon's DRF of the signal. Indeed, in this situation the optimal encoder can recover the original continuous-time signal without distortion, and then encode this recovery in an optimal manner according to the scheme that attains Shannon's DRF. Our goal is therefore to characterize the MSE due to the joint effect of a finite bitrate constraint and sampling at a sub-Nyquist sampling rate or for signals that are not bandlimited. In particular, we are interested in the minimal sampling rate for which Shannon's DRF, describing the minimal distortion subject to a bitrate constraint, is attained. As illustrated in Fig.~\ref{fig:contribution}, and as will be explained in more detail below, this sampling rate is usually below the Nyquist rate of the signal. We denote this minimal sampling rate as the \emph{critical sampling rate} subject to a bitrate constraint, since it describes the minimal sampling rate required to attain the optimal performance in systems operating under quantization or bitrate restrictions. Therefore, the critical sampling rate extends the minimal-distortion sampling rate considered by Shannon, Nyquist and Landau. It is only as the bitrate goes to infinity that sampling at the Nyquist rate is necessary to attain minimal (namely zero) distortion for general input distributions.
%new paradigm that we denote as the ADX paradigm
\par
Figure~\ref{fig:combined_sampling_source_coding} represents a general block diagram for systems that process information through sampling and are limited in the number of bits they can transmit per unit time, the amount of memory they use, or the number of states they can assume. Therefore, the critical sampling rate that arises in this setting describes the fundamental limit of sampling in systems like audio and video recorders, radio receivers, and digital cameras. Moreover, this model also includes signal processing techniques that use sampling and operate under bitrate constraints, such as artificial neural networks \cite{lecun2015deep}, financial markets analyzers \cite{karatzas1998methods}, and techniques to accelerate operations over large datasets by sampling \cite{MAL-035}. 
In the box {\bf System Constraints on Bitrate}, we list a few scenarios where sampling and bitrate restrictions arise in practice. Other applications of the ADX paradigm are discussed in Section~\ref{sec:applications}. \\

% TEXTBOX BEGINS
\begin{textbox}
\begin{center}
{\bf System Constraints on Bitrate} \\
\end{center}
The ADX setting of Fig.~\ref{fig:combined_sampling_source_coding} is relevant to any system that processes information by sampling and is subject to a bitrate constraint. Following is a list of three possible restrictions on a system's bitrate that arise in practice:
\begin{itemize}
\item {\bf Memory} -- Digital systems often operate under a constraint on the amount of memory or the states they can assume. Under such a restriction, the bitrate is the normalized amount of memory used over time (or the dimension of the source signal). For example, consider a system of $K$ states that analyzes information obtained by observing an analog signal for $T$ seconds. The maximal bitrate of the system is $R=\log_2(K)/T$. %Therefore, if this system takes a decision based on the continuous-time waveform, the uncertainty in this decision is greater than any decision that is obtained based on the original signal perturbed by Shannon's DRF of the signal at bitrate $R$. %The minimal sampling rate required in order to reliably transform the analog signal to the appropriate state is $f_R$. That is, by sampling at this rate and according to the optimal sampling scheme, no additional distortion is added to the system on top of Shannon's DRF. 
\item {\bf Power} -- Emerging sensor network technologies, such as those developed for biomedical applications and ``smart cities'', use many low-cost sensors to collect data and transmit it to remote locations \cite{YICK20082292}. These sensors must operate under severe power restrictions, hence they are limited by the number of comparisons in their analog-to-digital (ADC) operation. These comparisons are typically the most energy consuming part of the ADC unit, so that the total power consumption in an ADC is proportional to the number of comparisons \cite[Sec 2.1]{ginsburg2007energy}. In general, the number of comparisons is proportional to the bitrate, since any output of bitrate $R$ is generated by at least $R$ comparisons (although the exact number depends on the particular implementation of the ADC and may even grow exponentially in the bitrate \cite{761034}). Therefore, power restrictions lead to a bitrate constraint and to a MSE distortion floor given by Shannon's DRF of the analog input signal. \par
~~~An important scenario of power-restricted ADCs arises in wireless communication using millimeter waves \cite{rappaport2014millimeter}. Severe path-loss of electromagnetic waves in these frequencies are compensated by using a large number of receiver antennas. Each antenna is associated with an RF chain that includes an ADC unit. Due to the resulting large number of ADCs, power consumption is one of the major engineering challenges in millimeter wave communication. %Indeed, current implementations are limited to quantization with very low resolution (1-3 bits). 
\item {\bf Communication} -- Low power sensors may also be limited by the rates of communication available to send their digital sensed information to a remote location. For example, consider a low-energy device collecting medical signals and transmitting its measurements wirelessly to a central processor (e.g. a smartphone). The communication rate from the sensor to the central processor depends on the capacity of the channel between them, which is a function of the available transmit power for communication. When the transmit power is limited, so is the capacity. As a result, the data rate associated with the digital representation of the sensed information cannot exceed this capacity limit since, without additional processing, there is no point in collecting more information than what can be communicated. 
\end{itemize}
\end{textbox}
% TEXTBOX ENDS

To derive the critical sampling rate we rely on the following two steps:
\begin{enumerate}
\item[(i)] Given the output of the sampler, derive the optimal way to encode these samples subject to the bitrate $R$, so as to minimize the MSE distortion in reconstructing the original continuous-time signal. 
\item[(ii)] Derive the optimal sampling scheme that minimizes the MSE in (i) subject to the sampling rate constraint. 
\end{enumerate}
When the analog signal can be perfectly recovered from the output of the sampler, the fundamental distortion limit in step (i) depends only on the bitrate constraint, and leads to Shannon's DRF. We explore this function as well as the optimal encoding to attain it in Section~\ref{sec:theoretical_ADC}. In Section~\ref{sec:sampling_rate_distortion} we characterize the distortion in step (i) and the optimal sampling structure in step (ii). Applications of the ADX framework and the critical sampling rate that attains the minimal distortion are discussed in Section~\ref{sec:applications}.
\par
Before exploring the minimal distortion limit in the ADX setting, it is instructive to consider the distortion in a particular system implementing a simple version of a sampler, an encoder and a decoder. This analysis is carried out in Section~\ref{sec:PCM} for pulse code modulation (PCM). Although this system does not implement the optimal sampling and encoding scheme, it illustrates an instance where, as a result of the bitrate constraint, sampling below the Nyquist rate is optimal. In addition, this analysis provides a simple way to introduce the notions of sampling, quantization and bitrate, and serves as a basis for the generalization of the sampling and encoding operations to the optimal ones that are discussed in Sections \ref{sec:theoretical_ADC} and \ref{sec:sampling_rate_distortion}.

\section{Analog-to-Digital Compression via Pulse-Code Modulation \label{sec:PCM}}
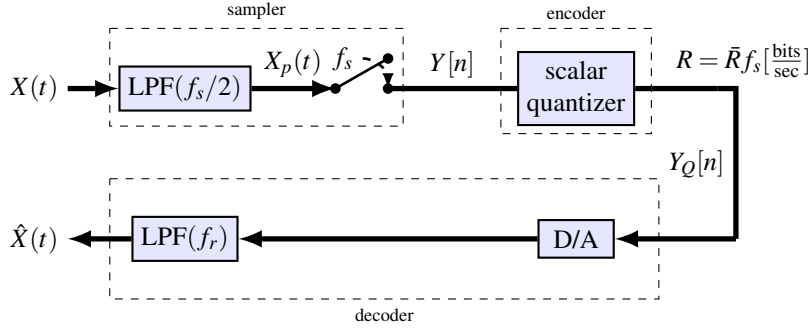
\begin{figure*}
\begin{center}
\begin{tikzpicture}[node distance=2cm,auto,>=latex]
  \node at (0,0) (source) {$X(t)$} ;
  \node[int1, right of = source, node distance =2cm] (anti) {$\mathrm{LPF}(f_s/2)$};   
 \node [coordinate,right of = anti, node distance = 4cm] at (0,0) (smp_in) {};
  \node [coordinate, right of = smp_in,node distance = 0.7cm] (smp_out){};
	\node [coordinate,above of = smp_out,node distance = 0.4cm] (tip) {};
\fill  (smp_out) circle [radius=2pt];
\fill  (smp_in) circle [radius=2pt];
\fill  (tip) circle [radius=2pt];
\node[left,left of = tip, node distance = 0.6 cm] (ltop) {$f_s$};
\draw[->,dashed,line width = 1pt] (ltop) to [out=0,in=90] (smp_out.north);
\draw[line width=1pt]  (smp_in) -- (tip);
\draw[dashed] (smp_in)+(-3,0.8) -- node[above] {\scriptsize sampler} +(0.9,0.8) --  +(0.9,-0.5)-- +(-3,-0.5) -- +(-3,0.8);

\node[int1,right of=smp_out, node distance = 2.5cm, align = center] (enc) {scalar \\  quantizer};
%\node [above of=plus_out, node distance=1.3cm] (eta) {$\eta[n]$};

\draw[dashed] (enc)+(-1,0.8) -- node[above, align = center] {\scriptsize encoder} +(1,0.8) -- +(1,-0.6)-- +(-1,-0.6) -- +(-1,0.8);

\node [right of = enc, node distance = 2cm] (right_edge) {};
\node [below of = right_edge, node distance = 2cm] (right_b_edge) {};
\node [right] (dest) [below of=source, node distance = 2cm]{$\hat{X}(t)$};

\node [int1] (LPF) [below of=anti, node distance = 2cm, align = center] {$\mathrm{LPF}(f_r)$};

\node[int1,below of = enc, node distance = 2cm] (dec) {D/A};

\draw[-,line width=2pt] (smp_out) -- node[above] {$ Y[n]$} (enc);
\draw[-,line width=2pt] (enc) -- node[above, xshift = 0.9cm]{$R=\bar{R} f_s [\frac{\mathrm{bits}}{\mathrm{sec}}]$}(right_edge);

\draw[dashed] (LPF)+(-1,0.8) -- +(6.3,0.8)-- +(6.3,-0.8)-- node[below, align = center] {\scriptsize decoder} +(-1,-0.8) -- +(-1,0.8);

%\draw[dotted, line width = 1] (0.9,0.3) -- (0.9,-3)  node[left] {analog~} ; 
%\draw[dotted, line width = 1] (7.2,0.3)  -- (7.2,-3)  node[right] {~digital} ;

\draw[-,line width = 2]  (right_edge.west) -|  (right_b_edge.east) node[left, yshift = 1cm] {$Y_Q[n]$};
\draw[->,line width = 2]  (right_b_edge.east) -- (dec);
\draw[->,line width = 2]  (dec) -- (LPF);
\draw[->,line width=2pt] (LPF) -- (dest);
\draw[->,line width=2pt] (source) -- (anti);
\draw[->,line width=2pt] (anti) -- node[above] {$\scriptsize X_p(t)$} (smp_in);
\end{tikzpicture}
\end{center}
\caption{\label{fig:PCM_system} Pulse-code modulation  and reconstruction system.}
\end{figure*}

A particular example for a system incorporating a sampler, an encoder and a decoder, is given in Fig.~\ref{fig:PCM_system}. This system converts the analog signal $X(t)$ to a digital representation $Y_Q[n]$ by a uniform sampler followed by a scalar quantizer. This conversion technique is known as PCM \cite{black1947pulse,1697556}; and we refer to 
\cite[Sec I.A]{gray1998quantization} for a historical overview. The bitrate in this system is defined as the average number of bits per unit time required to represent the process $Y_Q[n]$. The goal of our analysis is to derive the MSE distortion in recovering the analog input signal $X(t)$ under a constraint $R$ on this bitrate, assuming a particular sampling rate $f_s$ of the sampler. We denote this distortion by $D_{\PCM}(f_s,R)$. Since the system in Fig.~\ref{fig:PCM_system} is a special case of Fig.~\ref{fig:combined_sampling_source_coding}, the function $D_{\PCM}(f_s,R)$ is lower bounded by the minimal distortion in the ADX, obtained by optimizing over all encoders and decoders,
subject only to a sampling rate constraint $f_s$ and a bitrate constraint $R$. \par
%Although the sampling and encoding scheme used by the PCM are not optimal in general, the characterization of the minimal distortion in this relatively familiar setting provides a good introduction for our more general setting. \\
We analyze the system of Fig.~\ref{fig:PCM_system} assuming a stochastic continuous-time continuous-amplitude source signal $X(t)$ at its input. This signal is first filtered using a \emph{pre-sampling} low-pass filter (LPF) to yield $X_{p}(t)$. The filtered signal is then sampled uniformly at rate $f_s$ samples per second. Each sample $Y[n]$ is mapped using a \emph{scalar quantizer} to $Y_Q[n]$, which is the nearest value to $Y[n]$ among a prescribed set of $K$ \emph{quantization levels}. The box {\bf Scalar Quantization} provides more details on the operation of the scalar quantizer. Since each of the quantization levels can be assigned a finite digital number, we say that the process $Y_Q[n]$ is a digital representation of $Y[n]$. As explained in {\bf Scalar Quantization}, the selection of the quantization levels and the length of the digital number assigned to each of them may also be subject to optimization. Henceforth, we assume that $\bar{R}$ is the expected number of bits per sample assigned to represent the quantization levels (the expectation is with respect to the distribution of the source signal). Using this notation, the \emph{bitrate} of the digital representation, namely, the number of bits per unit time required to represent the process $Y_{Q}[n]$, is defined as $R = \bar{R}f_s$.

The process of recovering the analog source signal $X(t)$ from the digital sequence $Y_{Q}[n]$ is described at the bottom of Fig.~\ref{fig:PCM_system}: the digital discrete-time sequence of quantized values $Y_Q[n]$ is first converted to a continuous-time impulse-train using a digital-to-analog (D/A) unit, and then filtered using an ideal LPF with cutoff frequency $f_r$. In the time domain, this LPF is equivalent to an ideal sinc interpolation between the analog sample values to create a continuous-time signal bandlimited to $(-f_r,f_r)$. The result of this interpolation is denoted by $\hat{X}(t)$. We measure the \emph{distortion} of the system by the MSE between $X(t)$ and $\hat{X}(t)$ averaged over time, namely
\begin{equation} \label{eq:distortion_pcm}
D_{\PCM}(f_s,R) \triangleq \lim_{T\rightarrow \infty} \frac{1}{T}\int_{-T/2}^{T/2} \mathbb E \left(X(t)-\hat{X}(t) \right)^2 dt.
\end{equation}
Note that letting the time grow symmetrically in both directions simplifies some of the expressions, however our results remain valid even if time grows in one direction. It is in general possible to use a different decoding scheme that would lead to a lower MSE under the same sampling and bitrate constraint. Indeed, the expression in \eqref{eq:distortion_pcm} is minimized by using the conditional expectation of $X(t)$ given $Y_Q[n]$ as the reconstruction signal, rather than using $\hat{X}(t)$. However,
the non-linearity introduced by the scalar quantizer makes the exact analysis of the distortion under the conditional expectation a difficult task \cite{gray1998quantization} and therefore, for simplicity, we focus here on interpolation by lowpass filtering. \\ %Some sense on how much this reconstruction scheme is sub-optimal compared to reconstruction using the conditional expectation in the oversampling regime can be learned from \cite{485717}. This work shows that the optimal reconstruction of oversampled signals vanishes at a rate not faster then $1/f_s^2$, whereas the reconstruction scheme we use here decreases as $1/f_s$.

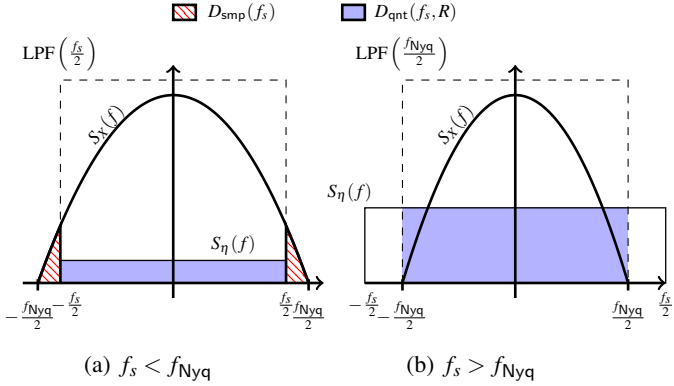
\begin{figure}
\begin{center}
\begin{subfigure}[h]{0.23\textwidth}
\begin{tikzpicture}[scale=1]
\node at (0,0) (origin) {};

\fill[fill=blue!30] (-1.5,0) --(-1.5,0.3) --(1.5,0.3)--(1.5,0) -- cycle;

\draw[line width=0.5pt] (-1.5,0) node[below,xshift = 0.1cm] {\scriptsize $-\frac{f_s}{2}$}-- node[above,yshift=0.1cm,xshift=2.3cm] {\scriptsize $S_{\eta}(f)$}(-1.5,0.3)-- (1.5,0.3)--(1.5,0) node[below]{\scriptsize $\frac{f_s}{2}$}--cycle;

\draw[dashed] (-1.5,0) --(-1.5,2.7) node[above] {\scriptsize $\mathrm{LPF}\left(\frac{f_s}{2} \right)$} --(1.5,2.7)--(1.5,0) -- cycle;

\draw[line width=1pt] plot[domain=-1.8:1.8, samples=100]  (\x, {-(\x)*(\x)*2.5/1.8/1.8+2.5 }) ;

\draw [line width=1pt, pattern=north west lines, pattern color=red] (0,3.5) rectangle  (0.3,3.7) node[left, xshift = 1.2cm, yshift = -0.1cm] {\scriptsize $D_{\smp}(f_s)$};

\draw [line width=1pt, pattern=north west lines, pattern color=red]  plot[domain=-1.8:-1.5, samples=100]  (\x, {-(\x)*(\x)*2.5/1.8/1.8+2.5 }) -- (-1.5,0) ;

\draw [line width=1pt, pattern=north west lines, pattern color=red] (1.5,0) -- plot[domain=1.5:1.8, samples=100]  (\x, {-(\x)*(\x)*2.5/1.8/1.8+2.5 }) ;

\draw[line width=1pt] (-1.8,-0.1) node[below,xshift=-0.1cm] {\scriptsize $-\frac{f_{\nyq}}{2}$}-- (-1.8,0.05);
\draw[line width=1pt] (1.8,-0.1) node[below,xshift=0cm] {\scriptsize $\frac{f_{\nyq}}{2}$}-- (1.8,0.05);

\node[above,yshift=0.45cm,xshift=0.2cm, rotate=50] at (-0.9,1.5) {\scriptsize $S_{X}(f)$};
%\node[right,yshift=0.25cm] at (0,1) {\scriptsize $\sigma_\eta^2$};
%\draw[line width=1pt] (-0.1,0.9) -- (0.1,1.1);

\draw[->,line width=1pt] (origin)+(0,-0.2) -- +(0,2.9) node[above] {};
\draw[->,line width=1pt] (origin)+(-2,0) -- +(2,0);
\end{tikzpicture}
\caption{ $f_s <  f_{\nyq}$ \label{fig:sub_Nyquist}}
\end{subfigure}
\begin{subfigure}[h]{0.23\textwidth}
\begin{tikzpicture}[scale=1]
\node at (0,0) (origin) {};

\fill[fill=blue!30] (-1.5,0) --(-1.5,1) --(1.5,1)--(1.5,0) -- cycle;

\draw[line width=0.5pt] (-2,0) node[below,xshift = 0cm] {\scriptsize $-\frac{f_s}{2}$}-- node[above,yshift=0.45cm,xshift=-0.2cm] {\scriptsize $S_{\eta}(f)$}(-2,1)-- (2,1)--(2,0) node[below]{\scriptsize $\frac{f_s}{2}$}--cycle;

\draw[dashed] (-1.5,0) --(-1.5,2.7) node[above] {\scriptsize $\mathrm{LPF}\left( \frac{f_{\nyq}}{2} \right)$} --(1.5,2.7)--(1.5,0) -- cycle;

\draw[line width=1pt] plot[domain=-1.5:1.5, samples=100]  (\x, {-(\x)*(\x)*2.5/2.25+2.5 }) ;

\draw [fill=blue!30, line width=1pt] (-2.3,3.5) rectangle  (-2,3.7) node[right, xshift = 0cm, yshift = -0.1cm, align = center] {\scriptsize $D_{\qnt}(f_s,R)$};
 	
\draw[line width=1pt] (-1.5,-0.1) node[below,xshift=0.cm] {\scriptsize $-\frac{f_{\nyq}}{2}$}-- (-1.5,0.05);
\draw[line width=1pt] (1.5,-0.1) node[below,xshift=0cm] {\scriptsize $\frac{f_{\nyq}}{2}$}-- (1.5,0.05);

\node[above,yshift=0.45cm,xshift=0.2cm, rotate=50] at (-0.8,1.5) {\scriptsize $S_{X}(f)$};
%\node[right,yshift=0.25cm] at (0,1) {\scriptsize $\sigma_\eta^2$};
%\draw[line width=1pt] (-0.1,0.9) -- (0.1,1.1);

\draw[->,line width=1pt] (origin)+(0,-0.2) -- +(0,2.9) node[above] {};
\draw[->,line width=1pt] (origin)+(-2,0) -- +(2,0) ;
\end{tikzpicture}
\caption{ $f_s > f_{\nyq}$   \label{fig:super_Nyquist}}
\end{subfigure}
\caption{ \label{fig:PCM_spectral} Spectral representation of the distortion in PCM \eqref{eq:distortion_pcm}: (a) sampling below the Nyquist rate introduces sampling distortion $D_{\smp}(f_s,R)$. (b) Sampling distortion vanishes when sampling above the Nyquist rate, but the contribution of the in-band quantization noise $D_{qnt}(f_s,R)$ increases due to lower bit-precision of each sample.}
\end{center}
\end{figure}

We now turn to analyze the distortion in \eqref{eq:distortion_pcm} as a function of the sampling rate $f_s$ and the bitrate $R$. We assume that $X(t)$ is a stationary stochastic process with a symmetric power spectral density (PSD) $S_X(f)$, and denote its bandwidth by $f_{\nyq}/2$. If $X(t)$ is not bandlimited then we use the notation $f_{\nyq}=\infty$. In either case we assume that $X(t)$ is bounded in energy, and denote
\[
\sigma^2 = \mathrm{var}\,X(t)= \int_{\mathbb R} S_X(f)df. 
\]
We further assume that the PSD $S_X(f)$ is unimodal, in the sense that its energy distribution is decreasing as one moves away from the origin, as given, for example, in  Fig.~\ref{fig:PCM_spectral}. Under this assumption, the pre-sampling filter that minimizes the distortion, among all linear time-invariant filters, is a LPF with cutoff frequency $f_s/2$ \cite{KipnisBitrate}. Henceforth we assume that this filter is used. Finally, we pick the cutoff frequency $f_r$ of the reconstruction filter to match the bandwidth of the low-pass filtered signal. This cutoff frequency is therefore the minimum between $f_s/2$ and the bandwidth of $X(t)$ which equals $f_{\nyq}/2$. \par
As a result of these assumptions, the only distortion introduced in the sampling process is due to the pre-sampling filter, and only in the case where $f_s$ is smaller than the Nyquist rate of $X(t)$. In fact, this distortion is exactly the energy in the part of the spectrum of $X(t)$ blocked by the pre-sampling filter. We therefore write
\[
D_{\smp}(f_s) \triangleq \sigma^2 - \int_{-\frac{f_s}{2}}^{\frac{f_s}{2}} S_X(f) df. 
\]
Note that $D_{\smp}(f_s)$ equals zero when $f_s$ is above the Nyquist rate of $X(t)$. 
\par

In order to analyze the distortion due to quantization, we represent the output of the quantizer as 
\begin{equation} \label{eq:quantization _noise_model}
Y_{Q}[n] =  Y[n] + \eta[n], \quad n=0,1,\ldots,
\end{equation}
where $\eta[n] = Y_{Q}[n] - Y[n]$ is the \emph{quantization noise}. Since there is no aliasing in the sampling operation, the reconstruction filter applied to $Y[n]$ leads to the signal $X_{p}(t)$ at the output of the first LPF. Since the quantizer is a deterministic function of $Y[n]$, the process $\eta[n]$ is stationary and we denote its PSD by $S_\eta(f)$ ($S_\eta(f)$ is periodic with period $f_s$). Nevertheless, an exact description of the statistics of $\eta[n]$ turns out to be a surprisingly difficult task. As a result, many approximations to its statistics have been developed \cite{59924,gray1998quantization}. Most of these approximations provide conditions under which the spectrum of $\eta[n]$ is white (i.e. different elements of $\eta[n]$ are uncorrelated) \cite{930935}. One of the widely used approximations was provided by Bennet \cite{BLTJ1340}, who showed that when the distribution of the input to the quantizer $Y[n]$ is continuous and the quantization levels are uniformly distributed, the spectrum of the quantization noise $S_\eta(f)$ converges to a constant as the quantizer resolution $\bar{R}$ increases. Another way to achieve uniform spectral distribution of $\eta[n]$ is by dithering the signal at the input to the quantizer, i.e., by adding a psuedo-random noise signal \cite{schuchman1964dither}. For simplicity, our analysis below assumes that $S_\eta(f)$ is a constant, although deviation from this rule would not affect our general conclusions. Regardless of this assumption and as explained in the box {\bf Scalar Quantization}, the variance of this noise $\eta[n]$ is proportional to the variance of the process $Y[n]$ at the input to the quantizer and is exponentially decreasing with the number of quantization bits $\bar{R}$, namely,
\begin{equation} \label{eq:quantization_noise_variance}
\mathrm {var}(\eta[n]) = c_Q \mathrm {var}(Y[n]) 2^{-2\bar{R}}. 
\end{equation} 
The proportionality constant $c_Q$ depends on the actual digital label assigned to each quantization level. At high quantization precision $\bar{R} = R/f_s$ and using a uniform quantizer, the value of the constant corresponding to optimal encoding converges to $c_Q = \frac{\pi e}{6}$. This value of $c_Q$ is used in our figures. \par

Under the assumption that the PSD of $\eta[n]$ is constant over the entire discrete-time frequency range with variance \eqref{eq:quantization_noise_variance}, and using the fact that the variance of $Y[n]$ equals the variance of the low-pass filtered version of $X(t)$, the contribution of the quantization to the distortion in \eqref{eq:distortion_pcm} is given by 
\begin{align} 
D_{\qnt}(f_s,R) &\triangleq \int_{-f_r}^{f_r} S_\eta(f) df \label{eq:distortion_quantizer} \\
& =  c_Q  \left(\frac{\min\left\{f_s , f_{\nyq}  \right\}}{f_s} \int_{-\frac{f_s}{2}}^\frac{f_s}{2} S_X(f) df   \right) 2^{-2 R / f_s } , \nonumber
\end{align}
where the term in the brackets represents the variance of $Y[n]$ or the energy of the signal at the output of the reconstruction LPF (the $\min$ is because the LPF at the sampler is in use only if the sampling rate is lower than Nyquist). The overall distortion of PCM is therefore
\begin{align}
D_{\PCM}(f_s,R) & = D_{\smp}(f_s) + D_{\qnt}(f_s,R)  \label{eq:dist_sum_pcm}.
\end{align}
The important observation from this expression is that under a fixed bitrate $R$, the distortion due to quantization increases as the sampling rate $f_s$ increases. This increase in $f_s$ means less quantization bits are available to represent each sample, and therefore the distortion due to quantization is larger. On the other hand, the distortion due to sampling decreases as $f_s$ increases and in fact vanishes as $f_s$ exceeds the Nyquist rate. A spectral interpretation of the function $D_{\PCM}(f_s,R)$ is illustrated in 
%These two trends can also be seen by considering the spectrum of the sampled source signal and the approximate spectrum of the quantization noise, for two representative cases of the sampling frequency, spectral interpretation of which is illustrated in
Fig.~\ref{fig:PCM_spectral}. This figure shows the spectrum of the sampled source signal and the spectrum of the quantization noise under the high resolution approximation for two representative cases of the sampling frequency:
\begin{itemize}
\item[(a)] {\bf Sub-Nyquist sampling:} the distortion due to sampling $D_{\smp}(f_s)$ is the part of $S_X(f)$ not included in the \emph{sampling interval} $\left( -f_s/2, f_s/2 \right)$. The distortion due to quantization is relatively low since the small value of $f_s$ allows the quantization of each sample with the relatively high resolution of $\bar{R} = R/f_s$ bits. 
\item[(b)] {\bf Super-Nyquist sampling:} the distortion due to sampling $D_{\smp}(f_s)$ is zero, but the distortion due to quantization $D_{\qnt}(f_s)$ is affected by the reduction in the bit-resolution that decreases linearly in $f_s$, since $\bar{R} = R/f_s$. 
\end{itemize}
\begin{figure}
\begin{center}
\begin{tikzpicture} 
\node at (3,2) {\includegraphics[trim=0cm 0cm 0cm 0cm,  ,clip=true,scale=0.4]{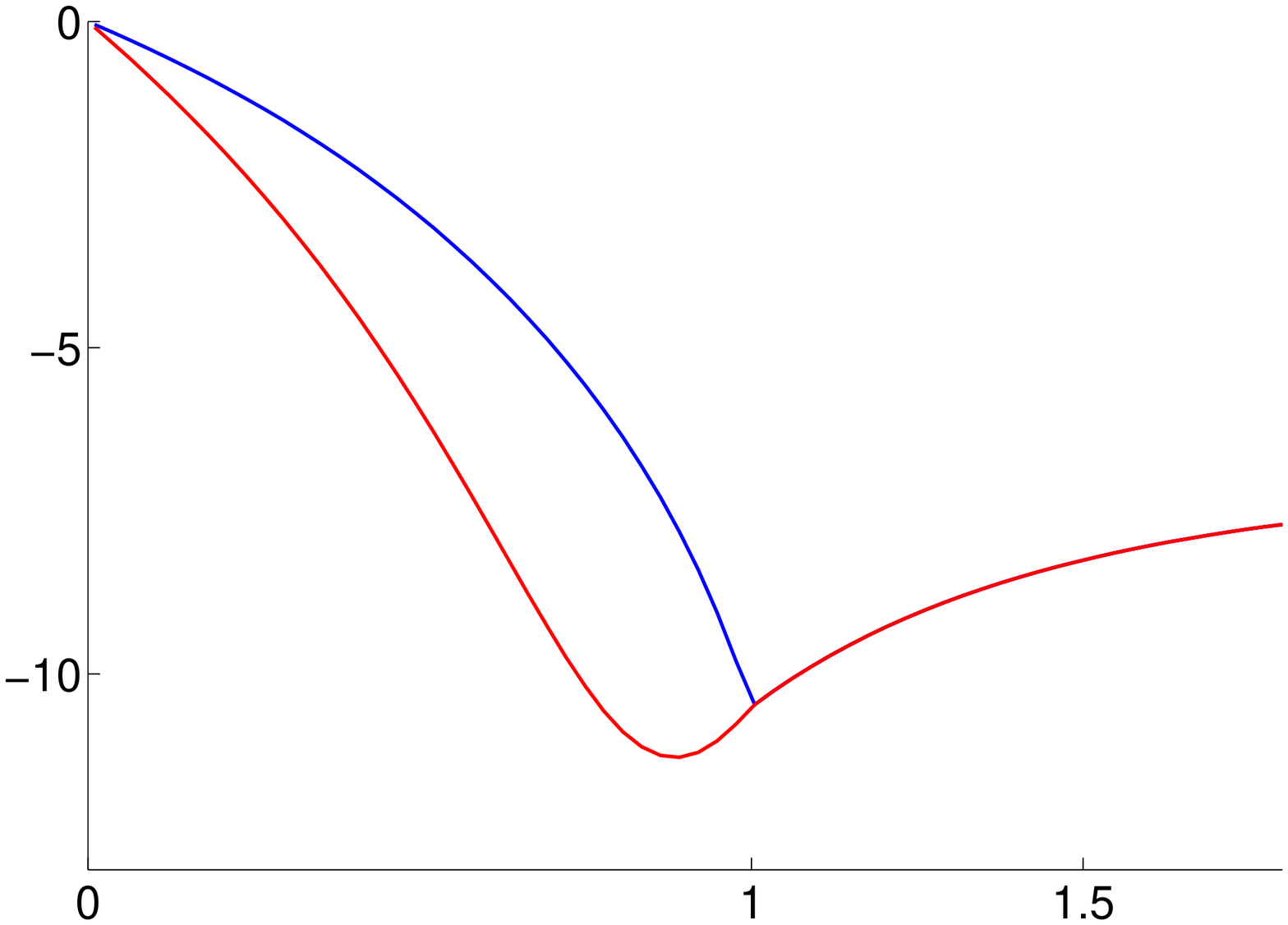}};

%\draw[->] (0,-0.9) -- node[right] {$\frac{f_s}{f_{\nyq}}$} (7.5,-0.9); 
\node[right] at (2.5,-1.2) {$f_s/f_{\nyq}$}; 
\node[above, xshift = -0.2cm, yshift = -0.7cm,rotate = 90] at (-0.7,3) {\small $D_{\PCM}(f_s,R)$ [dB] };

%\node[rotate=-40] at (-3,2.4) {\small { \color{blue} $S_{\Pi}(f)$ }} ;
%\node[rotate=-65] at (-3,-0.4) {\small {\color{red} $S_{\Lambda}(f)$ }} ;
%\node[rotate=-45] at (-3,1.4) {\small { $S_\Omega(f)$ }};

\node at (4,4) (rect_psd)
 {\includegraphics[width=1.8cm,height=1.6cm]{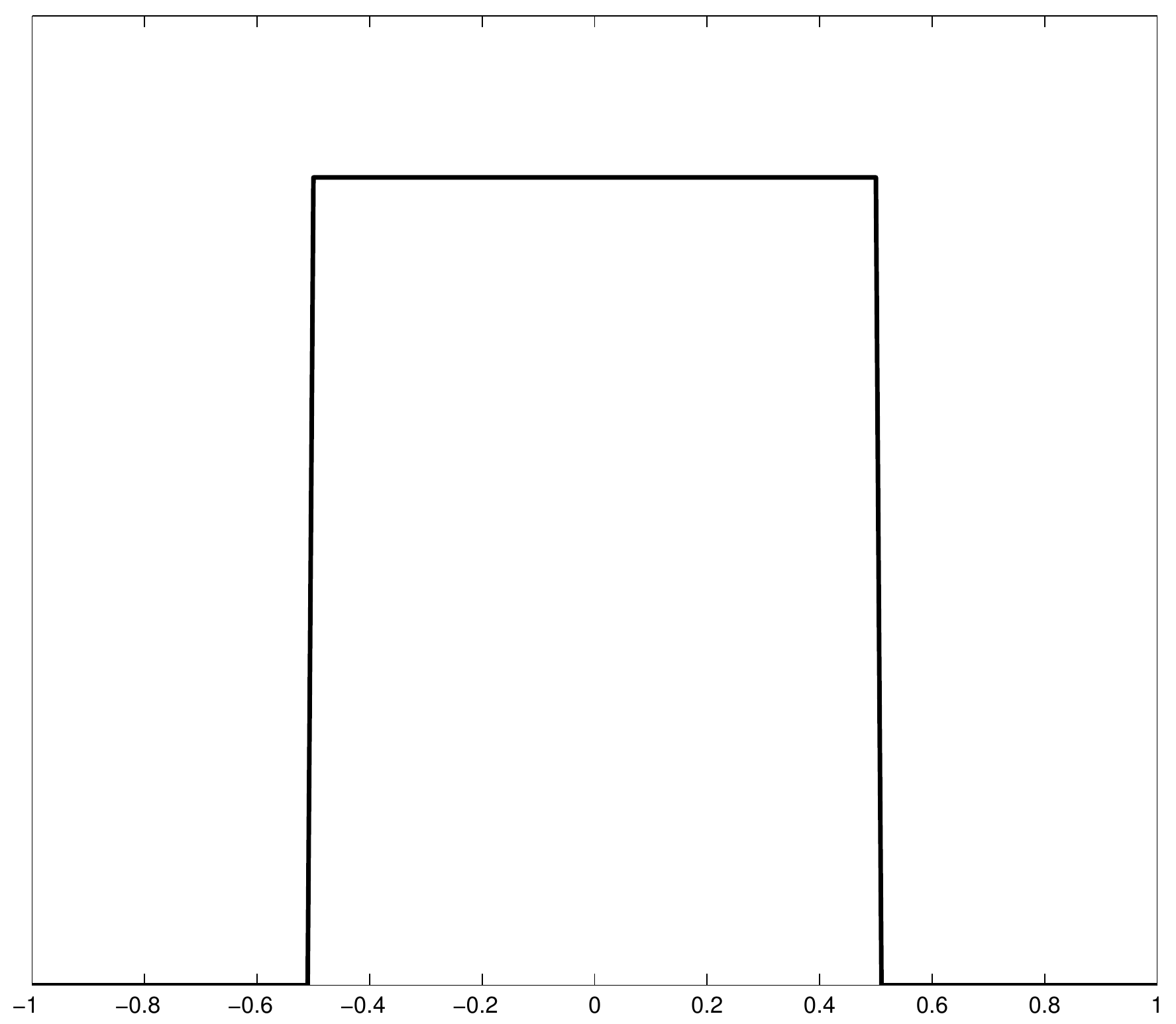} };
\node at (6,4) (triangle_psd) {\includegraphics[scale=0.1]{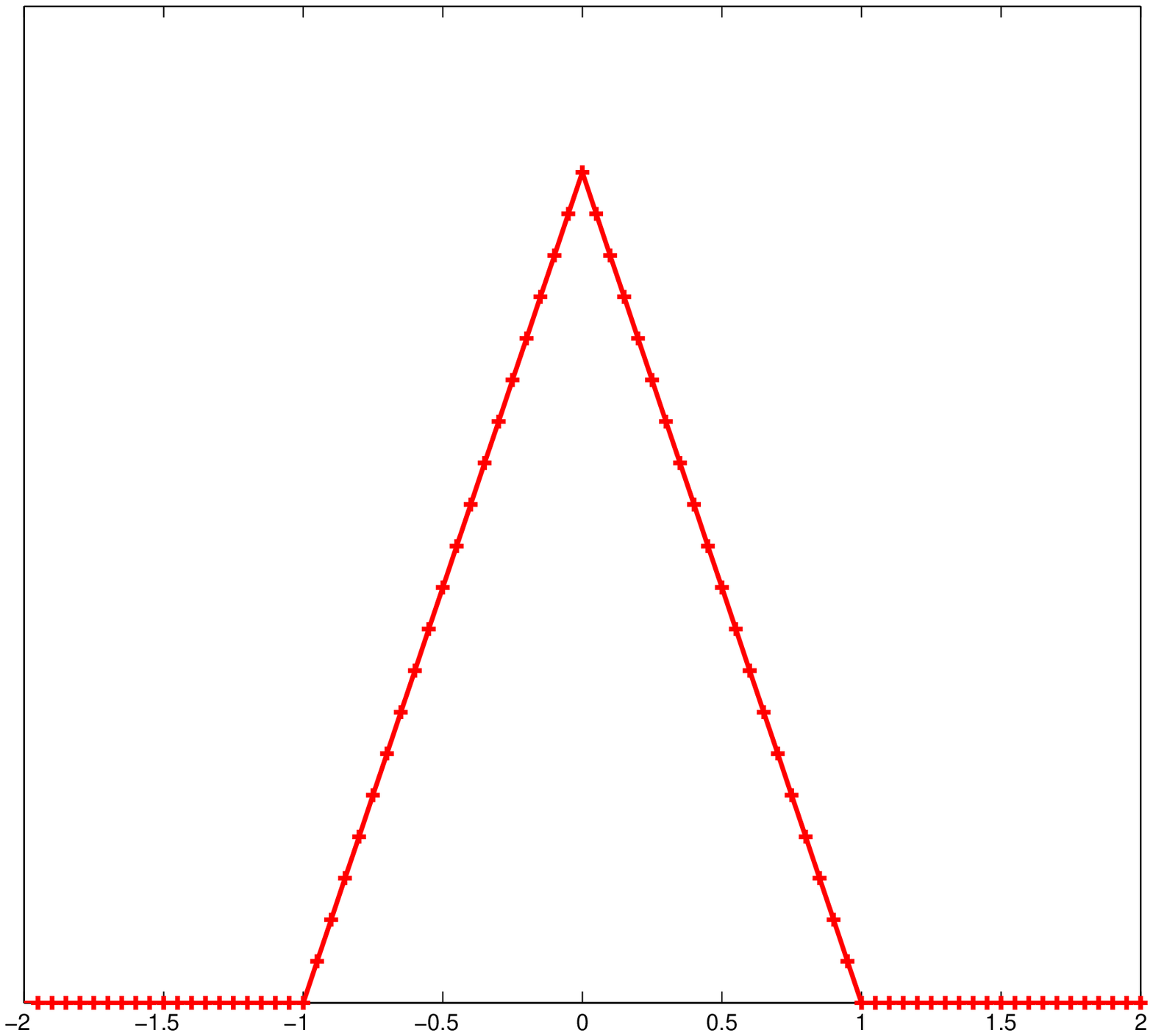} };
%\node at (6,5) (GaussMarkov_psd) {\includegraphics[width=1.8cm,height=1.6cm]{GaussMarkov_psd} };
%\node[yshift = +1cm] at (GaussMarkov_psd) {\scriptsize { $S_\Omega(f)$ }};
%\contourlength{1.5pt}
%\node at (GaussMarkov_psd) {\contour{white}{\scriptsize { $f_0=f_B$ }}};
\node[yshift = +1cm] at (triangle_psd) {\scriptsize { $S_{\Lambda}(f)$ }};
\node[yshift = +1cm] at (rect_psd){\scriptsize { $S_{\Pi}(f)$ }};

%\node at (-1.46,-1.5) {\color{red} \large $\star$};
%\node at (-1.05,-1.85) {\color{blue} \large $\star$};
%\node at (-0.9,-0.36) { \large $\star$};

%\node at (-1.25,-2.45) {\color{red}  $\diamond$};
%\node at (-1.05,-1.86) {\color{blue}  $\diamond$};
%\node at (-0.23,-1.16) { \large $\diamond$};

\end{tikzpicture}
\caption{\label{fig:various_PSD} Distortion in PCM as a function of sampling rate $f_s$ for a fixed bitrate $R$ and the PSDs in the small frames. With a nonuniform energy distribution, the optimal sampling rate of PCM is below the Nyquist rate. 
%The dashed curves are the corresponding DRFs $D(f_s,R)$ describing the minimal distortion under any encoding-decoding scheme. The optimal sampling rates $f_s^\star$ and $f_{R}$ corresponds to the $\star$s and $\diamond$s, respectively.
}
\end{center}
\end{figure}
It follows from the description above that there exists a sampling rate that balances the two error contributions from quantization and sampling to minimize the total distortion in \eqref{eq:dist_sum_pcm}. This sampling rate can be seen in Fig.~\ref{fig:various_PSD}, where the distortion $D_{\PCM}(f_s,R)$ is illustrated versus the relative sampling rate $f_s/f_{\nyq}$ for two PSDs. For the PSD $S_\Pi(f)$ with uniform energy distribution, the sampling rate that minimizes the distortion is exactly the Nyquist rate. For the triangular PSD $S_{\Lambda}(f)$, the optimal sampling rate is below the Nyquist rate. In general, it is shown in \cite{KipnisBitrate} that under similar assumptions, the sampling rate that minimizes the distortion in PCM is always at or below the Nyquist rate. This rate is in fact strictly smaller than the Nyquist rate when the energy of the signal is not uniformly distributed over its spectral support, as in $S_\Lambda(f)$ of Fig.~\ref{fig:various_PSD}. % but decreases gradually to zero at the Nyquist rate (that is, when $S_X(f)$ is continuous at $f = f_{\nyq}$).
%Since both $D_{\smp}(f_s)$ and $D_{\qnt}(f_s,R)$ are differentiable functions of $f_s$, the optimal sampling rate corresponds to an equilibrium point where both derivatives are equal in their absolute value. Moreover, for a bandlimited $X(t)$, $D_{\smp}(f_s)$ is zero for any $f_s > f_{\nyq}$, and this equilibrium point is always attained at or below the Nyquist rate. In fact, this point is strictly smaller than the Nyquist rate when energy of the signal is not uniformly distributed over its spectral support, but decreases gradually to zero at the Nyquist rate (that is, when $S_X(f)$ is continuous at $f = f_{\nyq}$). 
Going back to our general question, PCM illustrates an instance where, as a result of a bitrate constraint, sampling below the Nyquist rate is optimal. \par
Another conclusion from our analysis is that under a fixed bitrate, the distortion in PCM increases as a result of oversampling. This phenomena is explained by the increasing correlation between consecutive time samples at a super-Nyquist sampling rate, since the covariance function of a bandlimited signal is continuous  \cite{335948, 370112}. This correlation is not exploited by the quantizer, which maps two similar samples to the same digital value,  leading to a redundant digital representation of the analog signal. Since the overall bitrate is limited, this redundancy in representation is translated to a higher distortion compared to the distortion in a less redundant representation obtained at a lower sampling rate. In fact, it is well-known that the sampling rate that minimizes the distortion in PCM also maximizes the entropy rate of the process post-quantization, i.e. of $Y_Q[n]$ \cite{gray1998quantization}. Therefore, we conclude that the most efficient representation of the analog signal in PCM under a bitrate constraint is attained by sampling at or below the Nyquist rate. \par
%We note that a similar conclusion can be made if the scalar quantizer is replaced by a sigma-delta modulator \cite{}. \\
The above conclusions imply that we can readily improve the performance of  PCM by providing a more compact representation of the signal in terms of bitrate under the same distortion level, in one of the following ways: (1) Reduce the correlation between consecutive quantizer outputs by using a whitening transformation as in 
\emph{transform coding} \cite{gray1998quantization} or by a delta feedback loop as in \emph{sigma-delta modulation} \cite{1092194, kipnis2015optimal}. (2) Compress the digital process $Y_Q[n]$ using a universal lossless compressor, such as \emph{Lempel-Ziv} \cite{1055714, Ziv78compressionof} or \emph{context-tree weighting} \cite{willems1995context}. (3) Aggregate a large block of, say, $N$ samples of $Y[n]$ and represent these samples using a single index out of $2^{\bar{R}N}$ possible values. \\

This last technique, known in general as \emph{vector quantization} \cite{gray1998quantization}, does not assume any restrictions on the mapping from the samples to the digital representation except the size of the block. It therefore covers a wide range of quantization techniques operating at bitrate $R$, and includes (1) and (2) as special cases. 
As we shall see in the next section, this technique leads to the most general way to encode any discrete-time process to a digital representation subject only to a bitrate constraint. Moreover, combined with an optimal mechanism to represent the analog signal as a bit sequence, this encoding technique attains the minimal distortion in encoding $X(t)$, described by Shannon's DRF $D(R)$. 
%In the next section we go back to the analog-to-digital compression setting of Fig.~\ref{fig:combined_sampling_source_coding} and consider the minimal distortion subject only to a bitrate constraint. 

\begin{textbox}
\begin{center}
 {\bf Scalar Quantization} \\
\end{center}
Consider the problem of representing a random number $X$ drawn from a continuous distribution using another number taken from a finite alphabet of $K$ elements $X_Q \in \left\{x_1,\ldots,x_K \right\}$. Since an exact representation of $X$ cannot be attained due to cardinality limitations, the goal is to minimize 
\begin{equation} \label{eq:scalar_quantization_error}
\mathbb E \left(X - X_Q \right)^2. 
\end{equation}
The mapping of $X$ to $X_Q$ is called \emph{quantization}. When the representation of a sequence of random numbers is considered, we use the term \emph{scalar} quantization to denote the fact that the same quantization mapping is applied to each element of the sequence, independently of the previous elements. \par
Assuming that the quantizer inputs are independent, the estimation of each input sample from the output of the quantizer is based only on one of these $K$ states $\left\{x_1,\ldots,x_K \right\}$. Evidently, minimal estimation error is attained by mapping $X$ to the reconstruction value $x_i$ that minimizes \eqref{eq:scalar_quantization_error}. As a result, the procedure of optimizing a scalar quantizer of $K$ states can be described by selecting the optimal $K$ reconstruction values. Given the distribution of the input, this optimal set may be attained by an iterative procedure known as the Lloyd or, more commonly, the \emph{K-Means}, algorithm \cite{1056489, 1057548}. \par
The number of bits or \emph{bit-resolution} of the quantizer is the number of binary digits that represent $X$ at its output by assigning a different label to each state. Clearly, the output of a $K$-state quantizer can be encoded with $\lceil \log_2 K \rceil$ binary digits. However, this number may be reduced \emph{on average} if the labels of the states consist of binary numbers of different length. For example, by using uniform quantization levels to quantize a non-uniformly distributed input, we may label those states that are more likely with binary numbers shorter than those numbers assigned to less likely states. These numbers must satisfy the condition that no member is a prefix of another member, so that the sequence of states can be uniquely decoded. This procedure is denoted \emph{variable length} scalar quantization, compared to \emph{fixed length} quantization, in which the labels are all binary numbers of the same length. \par
Interestingly, the average MSE over an i.i.d. sequence using a variable length scalar quantizer may be strictly smaller than with a fixed length scalar quantizer for the same average number of bits, even if the levels in the latter were optimized for the input distribution using the Lloyd algorithm. For example, with input taken from a standard normal distribution, the average MSE attained by a variable length scalar quantizer with equally spaced reconstruction levels and an optimal labeling of these levels converges to $(\pi e/6) 2^{-2\bar{R}} \approx 1.42 \times 2^{-2\bar{R}}$ as $\bar{R}$ becomes large \cite{gish1968asymptotically}. In fact, it is also shown in \cite{gish1968asymptotically} that a uniform quantizer with optimal labeling converges to the optimal variable length quantizer as $\bar{R}$ increases. On the other hand, the distortion attained by a fixed length quantizer under an optimal selection of the $K = 2^{\bar{R}}$ reconstruction levels converges to $(\sqrt{3} \pi/2) 2^{-2\bar{R}} \approx 2.72 \times 2^{-2\bar{R}}$ \cite{1056489}. \par
As explained in the box {\bf The Source Coding Problem and the Distortion-Rate Function}, a lower MSE for the same average number of bits per source sample $\bar{R}$ can be attained by using a \emph{vector quantizer}, i.e., by considering the joint encoding of multiple samples from a sequence of samples, rather than one sample at a time. 
\end{textbox}

\section{Minimal Distortion subject to a Bitrate Constraint\label{sec:theoretical_ADC} }
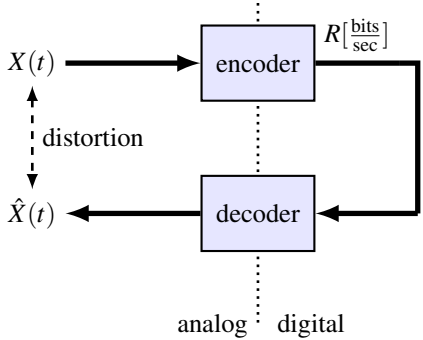
\begin{figure}
\begin{center}
\begin{tikzpicture}[node distance=2cm,auto,>=latex]
\draw[dotted, line width = 1] (3,0.8)  -- (3,-3.5)  node[left] {analog~} node[right] {~digital};

\node at (0,0) (source) {$X(t)$} ;
 \node [int] (enc) [right of = source, node distance = 3cm]{$\mathrm{encoder}$};
\node [right of = enc, node distance = 2cm] (right_edge) {};
\node [below of = right_edge, node distance = 2cm] (right_b_edge) {};

\node [right] (dest) [below of=source, node distance = 2cm]{$\hat{X}(t)$};
\node [int] (dec) [below of=enc, node distance = 2cm] {$\mathrm{decoder}$};

    \draw[-,line width=2pt] (enc) -- node[above] {$R [\frac{\mathrm{bits}}{\mathrm{sec}}]$} (right_edge);
    
  \draw[-,line width = 2]  (right_edge.west) -| (right_b_edge.east);
    \draw[->,line width = 2]  (right_b_edge.east) -- (dec.east);
%   \draw[->,line width=2pt] (enc) -- node[above] {$R [\frac{\mathrm{bits}}{\mathrm{sec}}]$} (dec);
   \draw[->,line width=2pt] (dec) -- (dest);
   %{$Y[n]=X(n/f_s)$} (enc);
    \draw[->,line width=2pt] (source) -- (enc);
    \draw[<->,dashed,line width=1pt] (source) -- node {distortion} (dest);
\end{tikzpicture}
\end{center}
\caption{\label{fig:source_coding_system} Encoding with full continuous-time source signal information.}
\end{figure}

We now go back to the ADX setting of Fig.~\ref{fig:combined_sampling_source_coding}. In this section we consider the minimal distortion that can be attained when the only restriction is the bitrate $R$ of the resulting digital representation. In other words, we consider the minimal distortion assuming that the encoder operates directly on the continuous-time process $X(t)$, as illustrated in Fig.~\ref{fig:source_coding_system}. \par

\begin{figure*}
\begin{center}
\includegraphics[scale = 0.3]{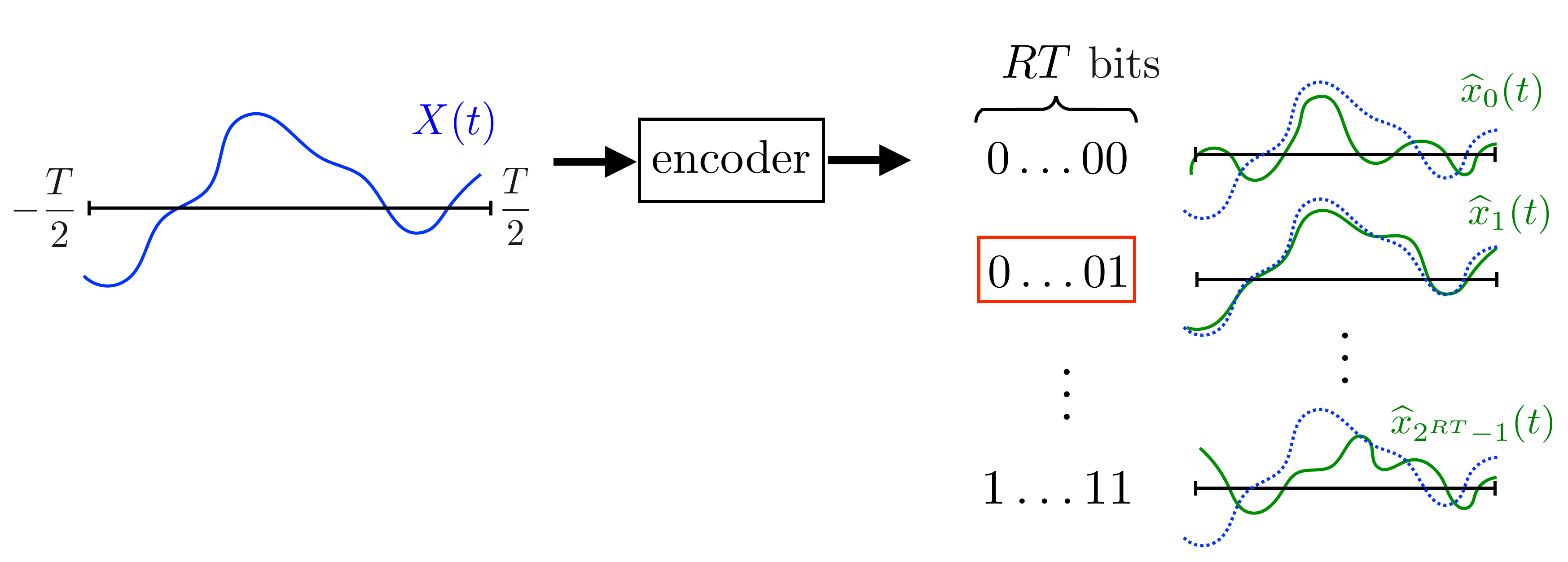}
\caption{
\label{fig:shannons_DRF_encoding} 
The optimal encoding with $TR$ bits is obtained by mapping the source signal realization to the index of the pre-determined reconstruction waveform closest to this realization. The optimal set of reconstruction waveforms and the resulting average distortion is given by Shannon's source coding theorem.
}
\end{center}
\end{figure*}
This encoder observes a realization $x(t)$ of the process $X(t)$ over some finite time horizon $T$, and then represents its observation using $\lfloor TR \rfloor$ bits. The number of possible states this encoding can take is therefore $2^{\lfloor TR\rfloor}$. As illustrated in Fig.~\ref{fig:shannons_DRF_encoding}, 
without losing generality we can assume that each reconstruction waveform produced by the decoder is only a function of one of these states, so there are at most $2^{\lfloor TR \rfloor }$ possible reconstruction waveforms. Moreover, any encoder that strives to attain the minimal MSE in this system would map the input signal to the state $i$ associated with the reconstruction waveform $\widehat{x}_i(t)$ that is closest to the input in the distance defined by the $\Ltwo$ norm over the interval $[-T/2,T/2]$, as derived from our distortion criterion.
%\begin{equation} \label{eq:distortion}
%\left\|x-\widehat{x}_i \right\|_T^2 = \frac{1}{T} \int_{-%\frac{T}{2}}^\frac{T}{2} \left(x(t) - \widehat{x}_i(t) %\right)^2 dt
%\end{equation}
Therefore, the only freedom in designing the optimal encoding scheme is in deciding on the set of reconstruction waveforms $\left\{ \widehat{x}_i(t),\,t\in [-T/2,T/2],\,i=1,\ldots,2^{\lfloor TR \rfloor} \right\}$, which we denote as \emph{codewords}.\par
The procedure for selecting these codewords and the resulting minimal MSE distortion are given by Shannon's classical source coding theorem \cite{Shannon1948,shannon1959coding} and its extensions to continuous alphabets \cite{perez1964extensions,Berger1968254}. According to this theorem, a near optimal set of codewords is obtained by $2^{\lfloor TR \rfloor}$ independent random draws from a distribution on the set of functions over $[-T/2,T/2]$ with a finite $\Ltwo$ norm, such that the mutual information of the joint distribution of the input and the reconstruction waveforms is limited to $\lfloor TR \rfloor$ bits. Moreover, Shannon's theorem also provides the asymptotic minimal MSE obtained by using this set of codewords, denoted as \emph{Shannon's} or the \emph{information} \emph{distortion-rate function} (DRF) of the source signal $X(t)$ at bitrate $R$. \\

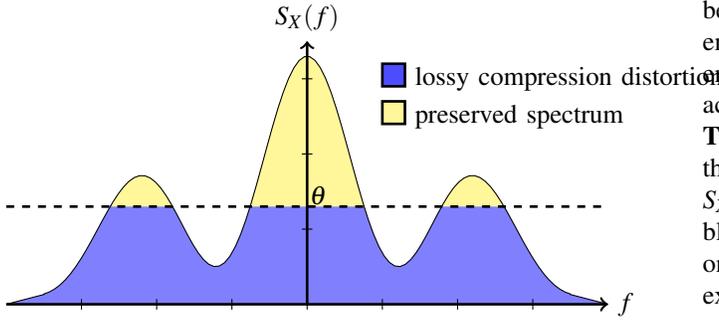
\begin{figure}
\begin{center}
\begin{tikzpicture}[scale=1]

 \fill[fill=blue!50] (-4,0) -- (-3.5,0.13)--plot[domain=-3.5:3.5, samples=100] (\x, {(0.6+2.7*cos(24*3.14*\x)^2)*exp(-\x*\x/8)}) -- (3.5,0.13)--(4,0);	

 \fill[fill=yellow!50] (-2.6,1.3) -- plot[domain=-2.6:-1.8, samples=100] (\x, {(0.6+2.7*cos(24*3.14*\x)^2)*exp(-\x*\x/8)}) --(-1.8,1.3);	

 \fill[fill=yellow!50] (1.8,1.3) -- plot[domain=1.8:2.6, samples=100] (\x, {(0.6+2.7*cos(24*3.14*\x)^2)*exp(-\x*\x/8)}) --(2.6,1.3);	

 \fill[fill=yellow!50] (-0.75,1.3) -- plot[domain=-0.75:0.75, samples=100] (\x, {(0.6+2.7*cos(24*3.14*\x)^2)*exp(-\x*\x/8)}) --(0.75,1.3);	

 \draw (-4,0) -- (-3.5,0.13)--plot[domain=-3.5:3.5, samples=100] (\x, {(0.6+2.7*cos(24*3.14*\x)^2)*exp(-\x*\x/8)}) -- (3.5,0.13)--(4,0);	

\foreach \x/\xtext in {-3,-2,-1,0,1,2,3}
 \draw[shift={(\x,0)}] (0pt,2pt) -- (0pt,-2pt) node[below] {};
  \foreach \y/\ytext in {1/,2/,3/}
    \draw[shift={(0,\y)}] (2pt,0pt) -- (-2pt,0pt) node[left] {};

\draw [fill=blue!70, line width=1pt] (1,2.9) rectangle  (1.3,3.2) node[right, xshift = 0cm, yshift = -0.2cm, align = right] {lossy compression distortion};

\draw [fill=yellow!50, line width=1pt] (1,2.4) rectangle (1.3,2.7) node[right, xshift=0cm, yshift = -0.2cm, align = right] { preserved spectrum};

\node at (0.15,1.45) {$\theta$};
\draw[dashed, line width=1pt] (-4,1.3) -- (4,1.3);
\draw[->,line width=1pt]  (-4,0)--(4,0) node[right] {$f$};
\draw[->,line width=1pt]  (0,0)--(0,3.5) node[above] {$S_X(f)$};
\end{tikzpicture}
\end{center}
\vspace{-10pt}

\caption{ \label{fig:waterfilling}
Reverse water-filling interpretation of \eqref{eq:skp}: water is poured into the area bounded by the graph of $S_{X}(f)$ up to level $\theta$. The bitrate $R$ is tied to the water level $\theta$ through the preserved part of the spectrum \eqref{eq:skp_R}. The lossy compression distortion $D$ is given by \eqref{eq:skp_D}.
}
\end{figure}

Shannon's source coding theorem with respect to a discrete-time identically distributed and independent (i.i.d.) process is explained in the box {\bf Source Coding and the Distortion-Rate Function}. In the case of a continuous-time Gaussian stationary input signal with PSD $S_X(f)$, a closed form expression for Shannon's DRF was derived by Pinsker and Kolmogorov \cite{1056823}, and is given by the following parametric form:
\begin{subequations}
\label{eq:skp}
\begin{align}
D(R_\theta) & = \int_{-\infty}^\infty \min \left\{S_X(f), \theta  \right\}df \label{eq:skp_D}\\
R_\theta & = \frac{1}{2} \int_{-\infty}^\infty \log^+ \left[S_X(f)/\theta \right]df, \label{eq:skp_R}
\end{align}
\end{subequations}
where $[x]^+$ is the maximum between $x$ and zero. The parametric form of \eqref{eq:skp} has the graphical interpretation given by Fig.~\ref{fig:waterfilling}, denoted as the \emph{water-filling} scheme: the distortion \eqref{eq:skp_D} may be seen as if water is being poured into the area bounded by the graph of $S_X(f)$ up to level $\theta$. The distortion in \eqref{eq:skp_D} is the total volume of the water. The bitrate is determined by integration over the preserved part through \eqref{eq:skp_R}. As explained in the box {\bf The Water-filling Scheme}, this approach is obtained as the solution of an optimization problem involving the allocation of the rate of the codes to describe different frequency components of the signal according to their respective energy (components with higher energy are given higher code rate). As a result, in addition to the minimal distortion subject only to the bitrate constraint, the water-filling interpretation provides the optimal coding scheme that attains this minimal distortion \cite{Berger1998}: independent spectral components of the signal are represented using independent bitstreams, where the rate of each bitstream is determined according to the water-filling principle. \par
The Pinsker-Kolmogorov expression \eqref{eq:skp} is easily adjusted to account for a distortion criteria that assigns different weights $W(f)\geq 0$ to each spectral component. This spectral weighting is useful in applications where some tones are of different importance than others, such as in psychoacoustic consideration in the digital encoding of audio signals \cite{zwicker2013psychoacoustics}. The adjustment of the expression for the minimal distortion required due to this importance weighting is achieved by evaluating the distortion equation \eqref{eq:skp_D} with respect to $W(f)S_X(f)$ rather than $S_X(f)$, in a similar way to the procedure explained in \cite{1057738}. This different weighting emphasizes the generality of the lossy compression principle: under a strict bitrate budget, part of the analog signal must be removed due to lossy compression, and this part is the least important one to our application. The Pinsker-Kolmogorov expression, with a possible spectral re-weighing, provides a mechanism to determine those parts of the signal that should be removed in an optimal encoding subject to the bitrate constraint. \\
%explain why Gaussian signal
\begin{textbox}
\begin{center}
 {\bf Source Coding and the Distortion-Rate Function} \\
\end{center}
The \emph{source coding problem} addresses the encoding of a random source sequence so as to attain the minimal distortion over all possible encoding and reconstruction schemes, under a constraint on the average bits per source symbol in this encoding. In the box {\bf Scalar Quantization} we considered the encoding of such sequences subject to the additional restriction that each source symbol is encoded independently of the other. By removing this restriction and considering the joint encoding of $n$ independent source symbols, we can attain smaller distortion using the same average number of bits. For this reason, the \emph{source coding problem} with respect to a real i.i.d. sequence $X_1, \ldots, X_n$ is defined as determining the minimal MSE attainable under all possible encoder mappings of a realization of this sequence to an index out of $2^ {\lfloor n \bar{R} \rfloor}$ possible indices, and all reconstruction decoder mappings from this set of indices back to $\mathbb R^n$. This minimal value is called the \emph{operational} DRF of the i.i.d. distribution of the sequence at code-rate $\bar{R}$, and denoted by $\delta_n(\bar{R})$. \\

In his \emph{source coding theorem}, Shannon showed that as the number of jointly described source symbols $n$ goes to infinity, the operational distortion-rate function $\delta_n(\bar{R})$ converges to the \emph{informational} distortion-rate function. The latter is defined as
\begin{equation}
\label{eq:drf_scalar}
D(\bar{R}) \triangleq \inf \, \mathbb E \left(X-\widehat{X} \right)^2,
\end{equation}
where the infimum is over all joint probability distributions $p(x,\hat{x})$ such that their marginal over the $x$ coordinate coincides with the distributions of $X_1$, and their \emph{mutual information} does not exceed $\bar{R}$. For example, when the source sequence is drawn from a standard normal distribution, the result of the optimization above leads to 
\begin{equation}
D(\bar{R}) = 2^{-2\bar{R}}. \label{eq:drf_iid_Gausian}
\end{equation}
Comparing with the distortion under scalar quantization in the box {\bf Scalar Quantization}, this value is strictly smaller than the minimal distortion in encoding the same sequence using either fixed or variable bit-length scalar quantization. This difference is explained by the fact that as $n$ goes to infinity, the law of large numbers implies that the probability mass of $n$ i.i.d. copies of a random variable of bounded variance concentrates around the edges of an $n$-dimensional sphere of radius equal to the square root of this variance. Thus, these $n$ copies can be represented in a more compact manner than with independent representations of each coordinate, as in scalar quantization \cite{csiszar1997information}.
\end{textbox}
%TEXT BOX ENDS

The DRF in \eqref{eq:skp} provides the minimal MSE (MMSE) distortion attainable in encoding a Gaussian stationary signal $X(t)$ at bitrate $R$. In fact, it provides the minimal distortion in any system that is used to recover a length $T$ realization of $X(t)$ having no more than $2^{\lfloor TR \rfloor}$ states. A special case of such a system is PCM of Section~\ref{sec:PCM}, and therefore, when $X(t)$ is a Gaussian process, the distortion in \eqref{eq:dist_sum_pcm} is bounded from below by \eqref{eq:skp}. \\

In general, the optimal encoder that attains Shannon's DRF operates in continuous-time: upon receiving a realization of $X(t)$ over $[-T/2,T/2]$, the encoder compares this realization to each of the $2^{\lfloor TR \rfloor}$ reconstruction waveforms \cite{perez1964extensions,Berger1968254}. We note, however, that Shannon's DRF is attainable even if this encoder is required to first map the analog waveform to a discrete-time sequence. Indeed this discrete-time sequence can be the random coefficients in the analog signal's expansion according to some pre-determined orthogonal basis. Consequently,
%Since this expansion provides a linear isomorphism between waveforms in $\Ltwo[-T/2,T/2]$ and discrete sequences in $\ell_2(\mathbb N)$, 
encoding and decoding may be performed with respect to this discrete sequence without changing the fundamental distortion limit described by the DRF in \eqref{eq:skp}. We emphasize that the equivalence between analog signals and coefficients in their basis expansion holds regardless of whether the original process $X(t)$ is bandlimited or not \cite{bharucha1970representation}. \par
One commonly-used example for such an orthogonal basis is the Karhunen-Lo\`{e}eve (KL) basis \cite{gelfand1959calculation}. The KL basis functions are chosen as the eigenfunctions of the bilinear kernel defined by the covariance of $X(t)$. As a result, the coefficients in this expansion are orthogonal to each other and, in fact, independent in our case of Gaussian signals. This fact implies that the KL expansion decomposes the process $X(t)$ over the interval $\left[-T/2,T/2\right]$ into a discrete Gaussian sequence of independent random variables, where the variance of each element is proportional to the eigenvalue associated with the eigenfunction. Since $X(t)$ is stationary, multiple sequences of this type obtained from different length $T$ blocks of $X(t)$ are identically distributed, and therefore can be encoded using the same block $T$ encoder that essentially encodes multiple discrete Gaussian sequences. The optimal encoding of such a sequence using $\lfloor TR\rfloor$ bits is achieved according to the water-filling principle as described in the box {\bf The Water-Filling Scheme}. Moreover, as $T$ goes to infinity, the density of the KL eigenvalues is described by the PSD $S_X(f)$ of $X(t)$, and the average distortion in encoding each block converges to \eqref{eq:skp} \cite{gallager1968information}. The above coding procedure is one way to show that Pinsker and Kolmogorov's water-filling expression \eqref{eq:skp} is attainable. \\ %An alternative decomposition of the signal that yields another coding scheme that attains \eqref{eq:skp} is given in \cite{Berger1998}. 

In order to implement any of the optimal encoding schemes of the analog signal described above, it is required to represent it first by a discrete sequence of coefficients. However, the implementation of this transformation is subject to practical limitations; in particular, realizable hardware such as filters and pointwise samplers are limited in the number of coefficient values they produce per unit time \cite{eldar2015sampling}. That is, for a time lag $T$, there exists a number $f_s$ such that any system consisting of these operations does not produce more than $Tf_s$ (analog) samples. In the next section we explore the minimal distortion that can be attained under this restriction. In particular, we are interested in the minimal sampling rate $f_s$ that is required to achieve Shannon's DRF.

\begin{textbox} \begin{center}
{\bf  The Water-filling Scheme } \\
\end{center} 
In the box {\bf Source Coding and the Distortion-Rate Function} we explored the encoding of an i.i.d. Gaussian sequence using a code of rate $\bar{R}$ bits per sample. We now extend this source coding problem to consider the joint encoding of $m$ i.i.d. sequences taken from $m$ Gaussian distributions with variances $\sigma_1^2, \ldots, \sigma_m^2$, using a total of $\bar{R}$ bits and under a sum MSE criterion. \par
From \eqref{eq:drf_iid_Gausian} we see that it is possible to describe the $i$th sequence using $\bar{R}_i$ bits per symbol such that $\sum_i R_i \leq \bar{R}$, and the overall distortion with respect to all sequences is 
\begin{equation} \label{eq:drf_vector_opt}
D(R_1,\ldots,R_m) = \sum_{i=1}^m \sigma_i^2 2^{-2R_i}.
\end{equation}
The problem we consider next is how to allocate the total bit-budget $\bar{R}$ in a way that minimizes the overall distortion. This is a convex problem whose solution can be expressed by the following parametric expression \cite[Exm. 5.2]{boyd2004convex}:
\[
R^\star_i = \frac{1}{2} \log_2^+ \left[\sigma_i^2/\theta \right]
\]
where $\theta$ is chosen to satisfy the constraint $R = \sum_{i=1}^m R^\star_i$. The resulting DRF is 
\[
D(\bar{R}) = D(R^\star_1,\ldots,R^\star_m) = \sum_{i=1}^m \min\left\{\sigma_i^2, \theta \right\}. 
\] 
This parametric expression for the DRF is referred to as a \emph{water-filling scheme}. The parameter $\theta$ may be interpreted as a water-level, such that $D(\bar{R})$ is obtained by summing the part of the variances that are below this level:
%as described in the Fig.~\ref{fig:waterfilling_eigenvalues}.  \\
\begin{center}
\begin{tikzpicture}
\draw[->,line width = 1] (0,0) -- (0,2.2) node[above] {$\sigma_i^2$};
\draw[->,line width = 1] (0,0) -- (5.8,0) node[right] {$i$} ;

\draw  [fill = yellow!50, line width = 1]  (0.1,0) rectangle (0.6,2);
\draw  [fill = yellow!50, line width = 1] (0.9,0) rectangle (1.4,1.7) ;
\draw  [fill = yellow!50, line width = 1] (1.7,0) rectangle (2.2,1.5);
\draw  [fill = blue!70, line width = 1] (0.1,0) node[below, xshift=0.2cm] {$1$} rectangle (0.6,1);
\draw  [fill = blue!70, line width = 1] (0.9,0) node[below, xshift=0.2cm] {$2$} rectangle (1.4,1);
\draw  [fill = blue!70, line width = 1] (1.7,0) node[below, xshift=0.2cm] {$\cdots$} rectangle (2.2,1);
\draw  [fill = blue!70, line width = 1] (2.5,0)  rectangle (3,0.8);
\draw  [fill = blue!70, line width = 1] (3.3,0)  rectangle (3.8,0.3);
\draw  [fill = blue!70, line width = 1] (4.1,0)  rectangle (4.6,0.1);
\draw  [fill = blue!70, line width = 1] (4.9,0)  rectangle (5.4,0.05);
\draw [dashed, line width = 1]  (-0.1,1) node[left, xshift = 0cm] {$\theta$} -- (5.5,1);

\draw [fill=blue!70, line width=1pt] (2,2.1) rectangle (2.3,2.4) node[right, xshift = 0cm, yshift = -0.2cm, align = right] {lossy compression distortion};

\draw [fill=yellow!50, line width=1pt]  (2,2.5) rectangle  (2.3,2.8) node[right, xshift=0cm, yshift = -0.2cm, align = right] { preserved spectrum};
\end{tikzpicture}
\end{center}
Intuitively, components with higher variance are described with more bits since they have a higher impact on the total distortion. An interesting property of the water-filling scheme is that when $R$ is small, the optimal coding does not allocate any bit-budget to some of the components with the lowest variance. This means that no information is sent on these low-variance components. \\

When the source is a stationary process, the DRF is described by water-filling over the PSD of the process as in \eqref{eq:skp}. In this case, different frequency sub-bands correspond to different independent signal components, and \eqref{eq:skp} is obtained by solving an optimization similar to the one above \cite{Berger1998}.
\end{textbox}

%According to the water-filling scheme, the encoding of each sub-band is performed independently of the rest of the sub-bands using an average number of bits proportional to the log ratio of the band's power to the water-level \cite{Berger1998}. Encoding a sub-band is equivalent to encoding a sequence of i.i.d Gaussian random variables, and this can be achieved using a vector quantizer (see Subsection ~\ref{subsec:quantization}). The encoding obtained this way leads to an independent stream of bits for each sub-band, where the bitrate of each stream is determined by the waterfilling scheme. In general, it is not necessary to code each frequency band separately since any optimal vector quantization applied to the random coefficients of an orthogonal decomposition of the signal (for example, as in the KL transform) will approach the DRF as $n$ goes to infinity \cite{neuhoff2013information}.  \\

\section{Analog-to-Digital Compression via Sampling \label{sec:sampling_rate_distortion}}
We have seen that the optimal tradeoff between MSE distortion and bitrate in the digital representation of an analog signal is described by Shannon's DRF of the signal. In this section we explore the minimal distortion under the additional constraint that the digital representation must be a function of the samples of the analog signal, rather than the analog signal itself.%, as in the ADX setting of Fig.~\ref{fig:combined_sampling_source_coding}

\subsection{Lossy Compression from Samples \label{subsec:combined_sampling}}
In the ADX setting of Fig.~\ref{fig:combined_sampling_source_coding}, the encoder observes samples of the source signal $X(t)$, and is required to encode these samples so that $X(t)$ can be estimated from this encoding using minimal MSE. Specifically, assuming that the sampler observes $X(t)$ for $t\in[-T/2,T/2]$, we denote by $\Yv$ the $\lfloor Tf_s \rfloor$ dimensional random vector resulting from sampling $X(t)$ at rate $f_s$. The encoder maps the vector $\Yv$ to a digital word of length $\lfloor TR \rfloor $, and delivers this sequence without errors to the decoder. The latter provides an estimate $\widehat{X}(t)$ for $X(t)$, $t\in[-T/2,T/2]$, based only on the digital sequence and the statistics of $X(t)$. The distortion between $X(t)$ and its reconstruction for a fixed sampler $S$ is defined by
\begin{equation} 
D_S(f_s,R) = \inf  \frac{1}{T}\int_{-T/2}^{T/2} \mathbb E\left(X(t) - \widehat{X}(t) \right)^2 dt \label{eq:dist_def}.
\end{equation}
The infimum in \eqref{eq:dist_def} is over encoders, decoders and time-horizons $T$. We note that under the assumption that $X(\cdot)$ and its samples are stationary, any finite time-horizon encoding strategy may be transformed into an infinite time-horizon strategy by applying it to consecutive blocks. As a result, increasing the time horizon cannot increase the distortion and the minimum over the time horizon in \eqref{eq:dist_def} can be replaced by the limit $T\rightarrow \infty$. \par
%Each sampler in this class defines a bounded linear function from the space of signals with domain $[-T/2,T/2]$ to $\mathbb R^{\lfloor T f_s \rfloor}$. Namely, denote the space of source signals as $\mathcal X_T$, then a linear continuous sampler is a member of $\mathcal X_T^{* \lfloor Tf_s \rfloor}$, where $\mathcal X_T^*$ is the dual space of $\mathcal X_T$ \cite{rudin2006functional}. \par
%
%In addition to the fundamental distortion limit under all linear continuous samplers, we are also interested in the minimal distortion that can be attained for a specific class of samplers $S$. For this reason, we define the function $D_S(f_s,R)$ as the minimal value of \eqref{eq:dist_def} over all encoders, decoders and samplers in $S$.
As an example, in PCM encoding described in Section~\ref{sec:PCM}, $S$ is a pointwise sampler at sampling rate $f_s$ preceded by a LPF. The particular encoder and decoder used in PCM was described in Fig.~\ref{fig:PCM_system}. Therefore, since the optimization in \eqref{eq:dist_def} is over all encoders and decoders, for any signal for which pointwise sampling is well-defined we have $D_S(f_s,R) \leq D_{\PCM}(f_s,R)$. \par
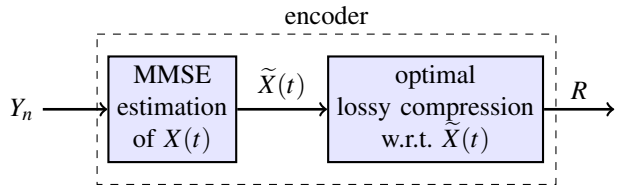
\begin{figure}
\begin{center}
\begin{tikzpicture}[node distance = 2cm]
\node (source) at (0,0) {$Y_n$};
\node[int1, right of = source, node distance = 2cm, align = center] (box1) {MMSE \\ estimation \\
of $X(t)$};
\node[int1, right of = box1, align = center, node distance = 3.5cm] (box2) {optimal \\ lossy compression \\
w.r.t. $\widetilde{X}(t)$};
%\node[int1, right of = box2, node distance = 2cm, align = center] (box3) {decoder};
\draw[dashed] (box1)+(-1,1) -- node[above] {encoder} +(5.1,1) -- +(5.1,-1) -- +(-1,-1) -- +(-1,1);
%\node[right of = box2, node distance = 2.5cm] (dest) {$\widehat{X}(t) = \widetilde{X}(t)$};
\node[right of = box2, node distance = 2.5cm] (dest) {};
\draw[->,line width = 1pt] (source) -- (box1);
\draw[->,line width = 1pt] (box1) -- node[above] {$\widetilde{X}(t)$} (box2);
\draw[->,line width = 1pt] (box2) -- node[above] {$R$} (dest);
%\draw[->,line width = 1pt] (box3) -- (dest);
\end{tikzpicture}
\caption{The optimal encoder in the ADX setting first estimates the analog source from its samples $\Yv$ and then encodes this estimate in an optimal manner. \label{fig:two_steps}}
\end{center}
\end{figure}

Characterizing $D_S(f_s,R)$ gives rise to a source coding problem in which the encoder has no direct access to the source signal it is required to describe. Source coding problems of this type are referred to as \emph{remote} or \emph{indirect} source coding problems \cite{berger1971rate}. More details on this class of problems is provided in the box {\bf Indirect Source Coding}. Under the MSE criterion \eqref{eq:dist_def}, the optimal encoding scheme of most indirect source coding problems is obtained by a simple two step procedure \cite{1054469,1056251,berger1971rate}:
\begin{itemize}
\item[(i)] Estimate $X(t)$ from its samples $\Yv$ subject to the MSE criterion \eqref{eq:dist_def}. Namely, compute the conditional expectation
\[
\widetilde{X}(t) = \mathbb E\left[X(t) | \Yv \right],
\]
where $\Yv$ is the output of the sampler with input $X(t)$, $t\in [-T/2,T/2]$.
\item[(ii)] Encode the estimated signal as in a standard (direct) source coding problem at rate $R$. That is, encode $\widetilde{X}(t)$ as the source signal to the system in Fig.~\ref{fig:shannons_DRF_encoding}.  
\end{itemize}
These two steps are illustrated in Fig.~\ref{fig:two_steps}. We note that although the encoding in step (ii) is with respect to an analog signal and hence prone to the same sampling limitation in processing analog signals mentioned in the previous subsection, the input to step (i) is a discrete-time process. Therefore, the composition of steps (i) and (ii) is a valid coding scheme for the encoder in the ADX setting, since it takes as its input a discrete-time sequence of samples and outputs a binary word.
\par
As explained in the box {\bf Indirect Source Coding}, the above two-step encoding procedure leads to the following decomposition:
\begin{equation} \label{eq:idrf_general_decomp}
D_S(f_s,R) = \mmse_S(f_s) + D_{\widetilde{X}} (R),
\end{equation}
where $\mmse_S(f_s)$ is the asymptotic non-causal MMSE in estimating $X(t)$ from the output $\Yv$ of the sampler $S$, and $D_{\widetilde{X}} (R)$ is Shannon's DRF of the estimated process $\widetilde{X}(t)$. \par
The decomposition in \eqref{eq:idrf_general_decomp} has a few important consequences. First, it reduces the characterization of $D_S(f_s,R)$ to the evaluation of the MMSE in sampling plus the evaluation of Shannon's DRF of another signal, defined as the non-causal instantaneous MMSE estimator of $X(t)$ given its samples. In particular, these two quantities are independent of the time horizon $T$, and the MMSE term $\mmse_S(f_s)$ is independent of the bitrate $R$. 
In addition, this decomposition implies that for any sampler $S$, the minimal distortion is always bounded from below by the MMSE in this estimation, as illustrated in Fig.~\ref{fig:contribution}. Moreover, it follows from \eqref{eq:idrf_general_decomp} that whenever the sampling operation is such that $X(t)$ can be recovered with zero MSE from its samples, then $D_S(f_s,R)$ reduces to Shannon's DRF of the source signal $X(t)$. For example, this last situation occurs when $X(t)$ is bandlimited and the sampling is uniform at any sampling rate exceeding the Nyquist rate of $X(t)$, as seen in Section~\ref{sec:PCM}. \par
This last property implies that oversampling cannot increase $D_S(f_s,R)$, as opposed to the distortion in PCM explored in Section~\ref{sec:PCM} that increases when the sampling rate goes above the Nyquist rate of the input signal. 
This fact highlights an important distinction between the optimal encoder we consider in the definition of $D_S(f_s,R)$ and the encoder in PCM. While the scalar quantizer in PCM encodes each sample instantaneously and independently, the optimal encoder can observe an unlimited number of samples by increasing the time horizon $T$ before deciding on a single index out of $2^{\lfloor TR \rfloor }$. This index is chosen to best describe the realization of $X(t)$ based on the samples stacked in its buffer up until time $T$. Oversampling $X(t)$ provides the encoder with redundant information to make this choice which cannot results in a worse choice and hence cannot results in worse performance. 
\\

%%%% TEXT BOX BEGIN
\begin{textbox} 
\begin{center}
{\bf Indirect Source Coding} \\
\end{center}
The characterization of the optimal encoding scheme and the resulting minimal distortion in Fig.~\ref{fig:combined_sampling_source_coding} can be seen as a special case of a family of source coding problems in which the encoder does not observe the  source process $X$ directly. Instead, it observes another process $Y$, statistically correlated with $X$, where the relation between the two processes is given by a conditional probability distribution $\mathrm{P_{Y|X}}$, as in the following system: 
\begin{center}
\begin{tikzpicture}[node distance = 2cm]
\node (source) at (0,0) {$X$};
\node[int1, right of = source, node distance = 1.5cm, align = center] (box1) {$\mathrm{P_{Y|X}}$};
\node[int1, right of = box1, align = center, node distance = 2cm] (box2) {$\mathrm{encoder}$};
\node[int1, right of = box2, node distance = 2.5cm] (box3) {$\mathrm{decoder}$} ;
\node[right of = box3, node distance = 1.5cm] (dest) {$\hat{X}$};
\draw[->,line width = 1pt] (source) -- (box1);
\draw[->,line width = 1pt] (box1) -- node[above] {$Y$} (box2);
\draw[->,line width = 1pt] (box2) -- node[above] {$R$} (box3);
\draw[->,line width = 1pt] (box3) -- (dest);
\end{tikzpicture}
\end{center}
This setting describes a compression problem in which the encoder is required to describe the source $X$ using a code of rate $R$ bits per source symbol, but with only partial information on $X$ as provided by the signal $Y$.
%\cite{pradhan2002distributed,demaine2003correlation}
In information theory, this problem is referred to as the \emph{indirect}, \emph{remote}, or \emph{noisy} source coding problem, first introduced in \cite{1057738}. The optimal tradeoff between code-rate and distortion in this setting is denoted as the \emph{indirect distortion-rate function} (iDRF). For example, when the source is an i.i.d. Gaussian process $X = X_1,X_2,\ldots$ and the observable process at the encoder is $Y_n = X_n + W_n$, where $W_n$ is an i.i.d. Gaussian noise sequence independent of $X$, the iDRF is given by
\begin{equation} \label{eq:idrf_iid}
D_{X|Y}(R) = \mmse(X|Y) + \mathrm{Var}\left(\mathbb E[X|Y] \right)2^{-2R},
\end{equation}
where $\mmse(X|Y)$ is the MMSE in estimating $X_n$ from $Y_n$ and $\mathrm{Var}\left(\mathbb E[X|Y] \right)$ is the variance of this estimator. Comparing \eqref{eq:idrf_iid} with Shannon's DRF of $X$ in \eqref{eq:drf_iid_Gausian}, we see that the first term in \eqref{eq:idrf_iid} is the MMSE in estimating the source from its observations, and the second term is Shannon's DRF of the MSE estimator. % $\mathbb E[X_n|Y_n] = \frac{\sigma^2}{1+\sigma^2} Y_n$. 
The decomposition of the iDRF into an MMSE term plus the DRF of the estimator is a general property of the indirect source coding setting for any ergodic source pair $(X,Y)$ under quadratic distortion \cite{1054469}. In the ADX setting of Fig.~\ref{fig:combined_sampling_source_coding}, this decomposition takes on the form of \eqref{eq:idrf_general_decomp}. 
\end{textbox}
%%%% TEXT BOX ENDS

Next, we study the behavior of $D_S(f_s,R)$ under various classes of samplers. We begin with samplers that can be described by the concatenation of a linear time-invariant filter and a uniform point-wise evaluation of the filtered signal, as illustrated in Fig.~\ref{fig:ADX_simple} \cite{eldar2015sampling}. We then gradually generalize the sampling mechanism to address more general forms of linear continuous sampling as described in the box {\bf Generalized Sampling of Random Signals}. 

% TEXTBOX BEGINS
\begin{textbox}
\begin{center}
{\bf Generalized Sampling of Random Signals} \\
\end{center}
Let $\mathcal X$ be a class of signals defined over the entire real line. We define \emph{linear continuous sampling} of $\mathcal X$ at sampling rate $f_s$ by $\lfloor T f_s \rfloor$ linear continuous functionals of $\mathcal X$. Namely, denoting the bilinear operation between $\mathcal X$ and its continuous dual $\mathcal X^*$ by an integral, the $n$th sample is given by
\begin{equation} \label{eq:sampling_deterministic}
y_n = \int_{-\infty}^{\infty} x(t) g_n(t) dt,
\end{equation}
where $g_n \in \mathcal X^{*}$. In order to incorporate sampling techniques that arise in practice, the class of signals $\mathcal X$ is chosen such that pointwise evaluation is continuous, i.e., the Dirac distribution $\delta(t)$ belongs to $\mathcal X^{*}$. \\

When the source $X(t)$ is a random signal, the set of functionals are often associated with the statistics of the signal. In order to define the counterpart of \eqref{eq:sampling_deterministic} when $X(t)$ is a stationary process with known statistics, we use the Fourier transform relation between the covariance of $X(t)$ and its PSD: 
\begin{equation} \label{eq:trig_iso}
\mathbb E\left[ X(t) X(s) \right] = \mathbb E\left[ X(t-s) X(0) \right] = \int_{-\infty}^\infty e^{2\pi i(t-s)f} S_X(f)df.
\end{equation}
This equation defines an isomorphism between the Hilbert space generated by the closed linear span of the random source signal $X(t) = \left\{ X(t),\,t\in \mathbb R \right\}$ with norm $\|X(t)\|^2 = \mathbb E[X^2(t)]$ and the Hilbert space $\Ltwo(S_X)$ of complex valued functions generated by the closed linear span (CLS) of the exponentials $\mathcal E = \left\{e^{2\pi i f t},\, t \in \mathbb R\right\}$ with an $\Ltwo$ norm weighted by $S_X(f)$ \cite{dym1978gaussian}. This isomorphism allows us to define sampling of the random signal $X(t)$ by describing its operation on the exponentials $\mathcal E$. Specifically, for any linear continuous functional $h$ on the CLS of $\mathcal E$, denote
\begin{equation} \label{eq:trig_sampling}
\phi_h(f) = \int_{-\infty}^{\infty} e^{2\pi i f t} h(t)dt.
\end{equation}
As long as $\phi_h$ is in $\Ltwo(S_X)$, the sample of $X(t)$ by the functional $h$ is defined by the inverse map of $\phi_h$ under the aforementioned isomorphism. For example, pointwise evaluation of $X(t)$ at time $n/f_s$ is obtained when $h$ is the Dirac distribution at $t = n/f_s$, and is well-defined as long as the $\mathrm L_1$ norm of $S_X(f)$ is finite. The last condition requires that $X(t)$ is bounded in energy, which is one of the few assumptions in our ADX setting. \\

The SI uniform sampler of Fig.~\ref{fig:ADX_simple} corresponds to sampling with functionals $h(t-n/f_s)$, $n\in \mathbb Z$, where $h$ is an arbitrary linear continuous functional on the CLS of $\mathcal E$. Similarly, uniform multi-branch sampling is obtained by sampling with respect to $h_1(t - nL/f_s),\ldots,h_L(t-nL/f_s)$, where $h_1,\ldots,h_L$ are $L$ such functionals. 
\end{textbox}
% TEXTBOX ENDS

% \begin{figure}
% \begin{center}
% \begin{tikzpicture}[node distance=2cm,auto,>=latex]
% \node at (0,0) (source) {$X(t)$};
% \node[coordinate] (source_up) [above of = source,node distance = 1cm]{};

% \node[coordinate] (first_jnc) [right of = source, node distance=1.5cm] {};
  
% \draw[-, line width=1pt] (source)--(first_jnc);   

% \node[int1]  (pre_sampling2) [right of = first_jnc, node distance=0.8cm]{$H(f)$};  

% \draw[->, line width=1pt] (first_jnc)--(pre_sampling2);   
  	  
% \node [coordinate, right of = pre_sampling2,node distance = 1.7cm] (smp_in2) {};
%   \node [coordinate, right of = smp_in2,node distance = 0.7cm] (smp_out2){};
% 	\node [coordinate,above of = smp_out2,node distance = 0.4cm] (tip2) {};
% \fill  (smp_out2) circle [radius=2pt];
% \fill  (smp_in2) circle [radius=2pt];
% \fill  (tip2) circle [radius=2pt];
% \node[left,left of = tip2, node distance = 0.5 cm] (ltop2) {$f_s$};

% \node [right of = smp_out2, node distance=3cm]  (out) {$Y_n$};

% \draw[->,densely dotted,line width = 1pt,thin] (ltop2) to [out=0,in=70] (smp_out2.north);
%  \draw[line width=1pt]  (smp_in2) -- (tip2);
%  \draw[-,line width=1pt]   (pre_sampling2)-- 
%  %node[above] {\small $Z(\cdot)$}
%  (smp_in2);

% \draw[line width=1pt]  (smp_in2) -- (tip2);

% \draw[->,line width = 1pt] (smp_out2) -- (out); 

% \draw[line width=1pt, dashed] (0.9,0.8) rectangle (6,-0.7) ;
    
% \end{tikzpicture}
% \caption{Shift invariant uniform sampler. \label{fig:sampler_scalar} }
% \end{center}
% \end{figure}

\subsection{Shift Invariant Sampling}
\begin{figure}
\begin{center}
\begin{tikzpicture}[node distance=2cm,auto,>=latex]
  \node at (0,0) (source) {$X(t)$} ;
  \node[int1, right of = source, node distance =2cm] (anti) {$H(f)$};   
 \node [coordinate,right of = anti, node distance =3.5cm] at (0,0) (smp_in) {};
  \node [coordinate, right of = smp_in,node distance = 0.7cm] (smp_out){};
	\node [coordinate,above of = smp_out,node distance = 0.4cm] (tip) {};
\fill  (smp_out) circle [radius=2pt];
\fill  (smp_in) circle [radius=2pt];
\fill  (tip) circle [radius=2pt];
\node[left,left of = tip, node distance = 0.6 cm] (ltop) {$f_s$};
\draw[->,dashed,line width = 1pt] (ltop) to [out=0,in=90] (smp_out.north);
\draw[line width=1pt]  (smp_in) -- (tip);
\draw[dashed] (smp_in)+(-2.5,0.8) -- node[above] {\scriptsize sampler} +(0.9,0.8) --  +(0.9,-0.5)-- +(-2.5,-0.5) -- +(-2.5,0.8);

\node[int1,right of=smp_out, node distance = 2cm, align = center] (enc) {encoder};

\node [right of = enc, node distance = 2cm] (right_edge) {};
\node [below of = right_edge, node distance = 2cm] (right_b_edge) {};
\node [right] (dest) [below of=source, node distance = 2cm]{$\hat{X}(t)$};

\node[int1,below of = enc, node distance = 2cm] (dec) {decoder};

\draw[-,line width=2pt] (smp_out) -- node[above] {$ Y_n$} (enc);

\draw[-,line width=2pt] (enc) -- node[above, xshift = 0.9cm]{$R [\frac{\mathrm{bits}}{\mathrm{sec}}]$}(right_edge);

\draw[-,line width = 2]  (right_edge.west) -|  (right_b_edge.east);
\draw[->,line width = 2]  (right_b_edge.east) -- (dec);
\draw[->,line width = 2]  (dec) -- (dest);
\draw[->,line width=2pt] (source) -- (anti);
\draw[->,line width=2pt] (anti) -- (smp_in);
\end{tikzpicture}
\end{center}
\caption{\label{fig:ADX_simple} ADX with a shift invariant uniform sampler.}
\end{figure}
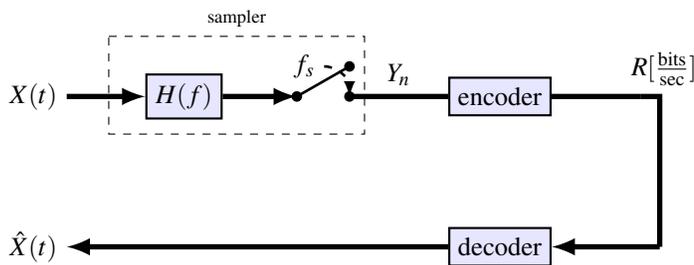

The system of Fig.~\ref{fig:ADX_simple} described the combined sampling and source coding system under a specific class of samplers. Each sampler in this class consists of a linear time-invariant filter applied to the analog source followed by pointwise evaluation of the filter's output every $T_s = f_s^{-1}$ time units. Therefore, this sampler is characterized only by its sampling rate $f_s$ and the frequency response $H(f)$ of the pre-sampling operation. Samplers of this form are called \emph{shift invariant} (SI) since their operation is equivalent to taking $\lfloor T f_s \rfloor$ inner products with respect to the functions $h(t-nT_s)$ \cite{eldar2015sampling}, for $n \in \mathbb Z$. When this sampler is used in the combined sampling and coding system of Fig.~\ref{fig:combined_sampling_source_coding}, the resulting system model is described in Fig.~\ref{fig:ADX_simple}. 
In this system, at each time $T$ the encoder observes the length $\lfloor T f_s \rfloor$ vector of samples of the filtered source at instances $\ldots,-T_s,0,T_s,\ldots$ inside the interval $[-T/2,T/2]$. The decoder receives the length $\lfloor TR \rfloor$ binary sequence produced by the encoder from this vector. We denote the MMSE in recovering the source from this binary sequence as $T$ goes to infinity by $D_{\SI}(f_s,R)$. \par
%
%%%TEXT BOX BEGINS
\begin{textbox}
\begin{center}
{\bf MMSE under Sub-Nyquist Sampling}
\end{center}
Consider the non-causal estimation of the process $X(t)$ from the discrete-time process $Y_n$ at the output of the SI sampler of Fig.~\ref{fig:ADX_simple}. Since all signals are Gaussian, the optimal estimator and the resulting MMSE can be found using linear estimation techniques that generalize the Wiener filter \cite{815501, 4663942}. In our case the optimal estimator $\widetilde{X}(t) = \mathbb E\left[X(t) |Y \right]$ is given by
\begin{equation}
\label{eq:estimator}
\widetilde{X}(t) = \sum_{n \in \mathbb Z} Y_n w(t-nT_s),\quad t \in \mathbb R,
\end{equation}
where the Fourier transform of $w(t)$ equals:
\[
W(f) = \frac{S_X(f)\left|H(f)\right|^2}{\sum_{k\in \mathbb Z} S_X(f-kT_s)\left|H(f-kT_s)\right|^2 }.
\]
The resulting MMSE is given by
\begin{equation}
\mmse_\SI(f_s) = \sum_{n\in \mathbb Z} \int_{-\frac{f_s}{2}}^\frac{f_s}{2} \left[ S_X(f-nf_s) - \widetilde{S}_{X|Y}(f) \right] df, \label{eq:mmse_single}
\end{equation}
where 
\begin{equation} \label{eq:s_tilde}
\widetilde{S}_{X|Y}(f) \triangleq \frac{\sum_{n\in \mathbb Z} S_X^2(f-f_s n ) \left|H(f-f_sn)\right|^2 }{\sum_{n\in \mathbb Z} S_X(f-f_sn)\left|H(f-f_sn)\right|^2 }.
\end{equation}
We interpret the fraction above to be zero whenever both numerator and denominator are zero. \\

When $f_s$ is above the Nyquist rate of $X(t)$, the support of $S_X(f)$ is contained within the interval $(-f_s/2, f_s/2)$. It can be seen from \eqref{eq:estimator} that in this case, provided that $H(f)$ is non-zero over the support of $S_X(f)$, we have that $\widetilde{X}(t) = X(t)$, $\widetilde{S}_{X|Y}(f)$ coincides with $S_X(f)$, and therefore $\mmse_\SI(f_s)=0$. Hence, as the time horizon goes to infinity, it is possible to reconstruct $X(t)$ from its samples with zero MSE. On the other hand, when $f_s$ is below the Nyquist rate, the expression \eqref{eq:mmse_single} shows how the MMSE in this estimation is affected by \emph{aliasing}, i.e., interference of different frequency components of the signal due to sampling. %In this case, the function $\widetilde{S}_{X|Y}(f)$ represents the amount of energy in that can be recovered on the original signal due to aliasing. 
\end{textbox}
%%%TEXT BOX ENDS
%
From the general decomposition \eqref{eq:idrf_general_decomp}, it follows that the minimal distortion for a SI sampler is obtained as the sum of the MMSE in estimating $X(t)$ from its filtered and uniform samples at rate $f_s$, plus Shannon's DRF of the non-causal estimator from these samples. 
%, and the MMSE in  non-causal estimation of $X(t)$ from the output of the sampler by $\mmse(X|Y)$. 
As explained in the box {\bf MMSE under Sub-Nyquist Sampling}, this MMSE vanishes whenever $f_s$ exceeds the Nyquist rate of $X(t)$ provided that the pre-sampling filter $H(f)$ does not block any part of the signal's spectrum $S_X(f)$. In this situation, the estimator $\widehat{X}(t)$ coincides with the original signal $X(t)$ in the $\Ltwo$ sense, and the decoder essentially encodes $X(t)$ directly as in the previous section. Therefore, for bandlimited signals, we conclude that $D_{\SI}(f_s,R)$ equals Shannon's DRF of $X(t)$ when the sampling rate is above the Nyquist rate. Moreover, when $X(t)$ is not bandlimited, a similar equality holds as the sampling rate goes to infinity \cite{neuhoff2013information}. \par
When the sampling rate is below the Nyquist rate, the expression for the optimal estimator and the resulting MMSE are obtained by standard linear estimation techniques as explained in the box {\bf MMSE under Sub-Nyquist Sampling}. In this case, the estimator $\widetilde{X}(t)$ has the form of a stationary process modulated by a deterministic pulse, and is therefore a \emph{block-stationary} or a \emph{cyclostationary} process \cite{Gardner:2006}. It is shown in \cite{KipnisCyclo} that Shannon's DRF for this class of processes can be described by a generalization of the orthogonal transformation and rate allocation that leads to the water-filling expression \eqref{eq:skp}, in a way analogous to the description in the box {\bf The Water-filling Scheme}.
%: the code-rate is allocated along a set of frequency components obtained after an orthogonalization over the \emph{polyphase decomposition} of $\widetilde{X}(t)$ \cite{vetterli2014foundations}.
By evaluating the resulting expression for the DRF of the cyclostationary process $\widetilde{X}(t)$ and using the decomposition \eqref{eq:idrf_general_decomp}, we obtain the 
%the expression for Shannon's DRF of the estimator leads to the 
following closed-form formula for $D_{\SI}(f_s,R)$, initially derived in \cite{Kipnis2014}:
\begin{subequations}
\label{eq:idrf_single}
\begin{align}
\label{eq:idrf_single_D}
D_{\SI}(f_s,R_\theta) & = \mmse_\SI(f_s) + \intfstofs \min\left\{\widetilde{S}_{X|Y}(f),\theta \right\} df \\
R_\theta & = \frac{1}{2} \intfstofs \log_2^+ \left[\widetilde{S}_{X|Y}(f)/\theta \right]df, \label{eq:idrf_single_R}
\end{align}
\end{subequations}
where $\mmse(X|Y)$ and $\widetilde{S}_{X|Y}(f)$ are given by \eqref{eq:mmse_single} and \eqref{eq:s_tilde}, respectively. The parametric expression \eqref{eq:idrf_single} combines the MMSE \eqref{eq:mmse_single} which depends only on $f_s$ and $H(f)$, with the reverse water-filling expression \eqref{eq:skp}, which also depends on the bitrate $R$. The function $\widetilde{S}_{X|Y}(f)$ arises in the MMSE estimation of $X(t)$ from its samples. As explained in \cite{KipnisCyclo}, this function is the average over the PSD of each polyphase component of the cyclostationary process $\widetilde{X}(t)$. 
To summarize, \eqref{eq:idrf_single} provides the MMSE distortion in encoding a Gaussian stationary signal at rate $R$ from its uniform samples taken at rate $f_s$. Moreover, according to Fig.~\ref{fig:two_steps}, the coding scheme that attains this minimal distortion can be described by the composition of the non-causal MMSE estimate of $X(t)$ as in \eqref{eq:estimator}, followed by an optimal encoding of the estimated process to attain its Shannon's DRF. \par
% >> HERE Maybe first show modification, then explain its importance. Remove Fig. 13
It is possible to extend the system model of Fig.~\ref{fig:combined_sampling_source_coding} to include a noisy input signal before the sampler. In this extended model, the excess distortion is a result of lossy compression, sampling, and independent noise. Therefore, the problem of estimating the source signal from the digital output of the encoder combines a linear filtering problem, an interpolation problem and a lossy compression problem. The only adjustment to the description of the minimal distortion under this extension is to replace the function $\widetilde{S}_{X|Y}(f)$ in \eqref{eq:idrf_single} and \eqref{eq:mmse_single} with \cite{Kipnis2014}:
\begin{equation} \label{eq:s_tilde_new}
\widetilde{S}_{X|Y}(f) = \frac{\sum_{n\in \mathbb Z} S_X^2(f-f_s n ) \left|H(f-f_sn)\right|^2 }{\sum_{n\in \mathbb Z} \left( S_X(f-f_sn) + S_\eta(f-f_s n) \right)\left|H(f-f_sn)\right|^2 }.
\end{equation}
Equations 
 \eqref{eq:mmse_single}, \eqref{eq:idrf_single}, and \eqref{eq:s_tilde_new} describe the MMSE in non-causal filtering, the MSE due to uniform sampling, and the distortion under optimal lossy compression. Namely, these equations determine the combined effect of three of the most fundamental operations in signal processing: quantization, sampling and interference by noise. Most importantly, these equations provide a unified representation for the distortion in these three fundamental operations, allowing us to explore the interaction among them. In this article we consider a less general case; we explore the interaction between sampling and lossy compression and assume that the noise is zero, hence the simplified form \eqref{eq:s_tilde} for $\widetilde{S}_{X|Y}(f)$ is used. \\

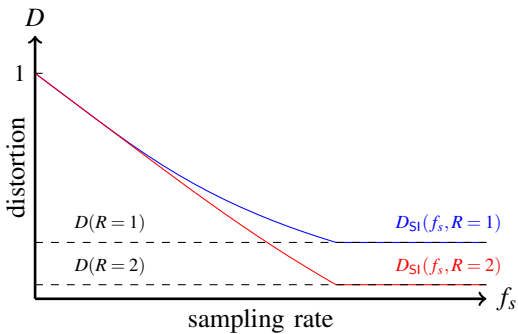
\begin{figure}
\begin{center}

\begin{tikzpicture}

\draw[->, line width = 1pt] (0,0) -- node[below] {sampling rate} (6,0) node[right] {$f_s$};
\draw[->, line width = 1pt] (0,0) -- node[above, rotate = 90] {distortion} (0,3.5) node[above] {$D$};

\draw[blue]  plot[domain=0.001:4, samples=20]  (\x, {3*(1-\x/4+\x/4*2^(-8/\x)) } ) node[above, xshift = 1.5cm] {\scriptsize $D_{\SI}(f_s,R=1)$ } --  plot[domain=4:6, samples=20]  (\x, {3*2^(-8/4) } )	;
\draw[dashed] plot[domain = 0:6, samples = 20] (\x,{3*2^(-8/4)}) node[above, xshift = -5cm] {\scriptsize $D(R=1)$} ; 

\draw[red]  plot[domain=0.001:4, samples=20]  (\x, {3*(1-\x/4+\x/4*2^(-16/\x)) } ) node[above, xshift = 1.5cm] {\scriptsize $D_{\SI}(f_s,R=2)$ } --  plot[domain=4:6, samples=20]  (\x, {3*2^(-16/4) } )	;
\draw[dashed] plot[domain = 0:6, samples = 20] (\x,{3*2^(-16/4)}) node[above, xshift = -5cm] {\scriptsize $D(R=2)$} ; 

%\draw[->] (0,1.1) -- (2.7,1.1) node[right] {\tiny $f$};
%\draw[->] (1.5,1.1) -- (1.5,2.6) node[above] {\tiny $S_{\Pi}(f)$};
%\draw[line width=1pt] (0,1) rectangle (3,3);
%\draw[line width=1pt, color=blue] node[below] {\tiny $-W$} (0.5,1.1) -- (0.5,2.1) -- (2.5,2.1) -- (2.5,1.1) node[below] {\tiny $W$};
%
%\draw plot[domain = 0:4, samples = 20] (\x,{3*(1-\x/4)}) node[below, xshift = 0cm, rotate = -35, yshift = 0] {\scriptsize $\mmse_\SI(f_s)$}; 

\draw (0,3) node[left] {\small $1$} --  (0.1,3) ;
\end{tikzpicture}
\caption{\label{fig:idrf_rect} Distortion as a function of sampling rate for the source with PSD $S_{\Pi}(f)$ of \eqref{eq:psd_rect} and source coding rates $R=1$ and $R=2$ bits per time unit. }
\end{center}
\end{figure}
As a simple example for using formula \eqref{eq:idrf_single}, we consider $X(t)$ to be a stationary Gaussian signal with a flat bandlimited PSD, i.e.,
\begin{equation} \label{eq:psd_rect}
S_{\Pi}(f) = \begin{cases} \frac{1}{2W} & |f|<W, \\
0& \mathrm{otherwise}.
\end{cases}
\end{equation}
It can be shown that as long as the pre-sampling filter pases all frequencies $f \in (-W,W)$, the relation between the distortion in \eqref{eq:idrf_single_D} and the bitrate in \eqref{eq:idrf_single_R} is given by
\begin{equation} \label{eq:idrf_rect}
D_{\SI}(f_s,R) = \begin{cases} 
\mmse_{\SI}(f_s) + \frac{f_s}{2W} 2^{-\frac{2R}{f_s}},& \frac{f_s}{2W} < 1 \\
2^{-\frac{R}{W}},&  \frac{f_s}{2W} \geq 1,
\end{cases}
\end{equation}
where $\mmse_{\SI}(f_s) = 1- \frac{f_s}{2W}$. Expression \eqref{eq:idrf_rect} is shown in Fig.~\ref{fig:idrf_rect} for two fixed values of the bitrate $R$. It has a very intuitive structure: for frequencies below the signal's Nyquist rate $2W$, the distortion as a function of the rate increases by a constant factor due to the error as a result of non-optimal sampling. This factor completely vanishes once the sampling rate exceeds the Nyquist frequency, in which case $D_{\SI}(f_s,R)$ coincides with Shannon's DRF of $X(t)$. \\

In the example above with PSD $S_\Pi(f)$, the filter $H(f)$ has no effect on the distortion as long as its passband contains the support of $S_\Pi(f)$. However, as we shall see below, when the spectrum is non-flat over its support there is a precise way to choose the passband of the pre-sampling filter in order to minimize the function $D_{\SI}(f_s,R)$.

\begin{figure*}
\begin{center}
\begin{tikzpicture}
\draw [fill=yellow!50, line width=1pt] (0,0) rectangle  (0.5,0.5) node[right, xshift=0cm, yshift = -0.2cm] {preserved spectrum};

\draw [fill=red!50, line width=1pt, pattern=north west lines, pattern color=red] (4.5,0) rectangle  (5,0.5) node[right, xshift = 0cm, yshift = -0.2cm] {sampling distortion};

\draw [fill=blue!50, line width=1pt] (9,0) rectangle  (9.5,0.5) node[right, xshift = 0cm, yshift = -0.2cm] {lossy compression distortion};
\end{tikzpicture}
\end{center}

\begin{subfigure}[h]{0.32\textwidth}
\begin{center}
\begin{tikzpicture}[scale=1.4]

\draw[fill=blue!50] (-0.8,0) -- plot[domain=-0.8:0.8, samples=15]  (\x, {(-(\x)*(\x)+2.25)/1.3 }) -- (0.8,0) ;

\draw[line width=1pt, pattern=north west lines, pattern color=red] (-1.5,0) -- plot[domain=-1.5:-0.8, samples=15]  (\x, {(-(\x)*(\x)+2.25)/1.3 }) -- (-0.8,0);

\draw[line width=1pt, fill=red, pattern=north west lines, pattern color=red] (0.8,0) -- plot[domain=0.8:1.5, samples=15]  (\x, {(-(\x)*(\x)+2.25)/1.3 }) -- (1.5,0) ;

\draw[fill=yellow!50] (-0.8,0.5) -- plot[domain=-0.8:0.8, samples=10]  (\x, {(-(\x)*(\x)+2.25)/1.3 }) -- (0.8,0.5) ;

\draw[<->] (-0.79,0.25) -- node[above, xshift = 0.4cm, yshift = -0.2cm, fill=blue!50] {\scriptsize $f_s$} (0.79,0.25);

\draw[line width=1pt] (-1.5,0) -- plot[domain=-1.5:1.5, samples=10]  (\x, {(-(\x)*(\x)+2.25)/1.3 }) -- (1.5,0) ;

\draw [dashed,line width=1pt] (-1.5,0.5) --  (1.5,0.5) node[right] {\small $\theta_a$};

  \draw[dashed,line width=0.5pt] (-1.1,-0.05) node[below] {{\scriptsize -}\small $\frac{f_{R}}{2}$} -- (-1.1,0.72);
    \draw[dashed,line width=0.5pt]  (1.1,-0.05) node[below] {\small $\frac{f_{R}}{2}$} -- (1.1,0.72);

  \draw[->,line width=1pt] (-1.8,0) -- (1.8,0) node[right, xshift=-0.1cm] {\small $f$} ;
  \draw[->,line width=1pt] (0,0) -- (0,2) node[above] {\small $S_X(f)$};   
    \node at (0,-0.5) {(a)};
\end{tikzpicture}
\end{center}
%\caption{$D(f_s,R)>D(R)$ for $f_s<f_{DR}$}
\end{subfigure}
\begin{subfigure}[h]{0.32\textwidth}
\begin{center}
\begin{tikzpicture}[scale=1.4]

\draw[fill=blue!50] (-1.25,0) -- plot[domain=-1.25:1.25, samples=15]  (\x, {(-(\x)*(\x)+2.25)/1.3 }) -- (1.25,0) ;

\draw[line width=1pt, pattern=north west lines, pattern color=red] (-1.5,0) -- plot[domain=-1.5:-1.25, samples=15]  (\x, {(-(\x)*(\x)+2.25)/1.3 }) -- (-1.25,0);

\draw[line width=1pt, fill=red, pattern=north west lines, pattern color=red] (1.25,0) -- plot[domain=1.25:1.5, samples=15]  (\x, {(-(\x)*(\x)+2.25)/1.3 }) -- (1.5,0) ;

\draw[fill=yellow!50] (-1.1,0.72) -- plot[domain=-1.1:1.1, samples=10]  (\x, {(-(\x)*(\x)+2.25)/1.3 }) -- (1.1,0.72) ;

\draw[<->] (-1.25,0.25) -- node[above, xshift = 0.4cm, yshift = -0.2cm, fill=blue!50] {\scriptsize $f_s$} (1.25,0.25);

\draw[line width=1pt] (-1.5,0) -- plot[domain=-1.5:1.5, samples=10]  (\x, {(-(\x)*(\x)+2.25)/1.3 }) -- (1.5,0) ;

\draw [dashed,line width=1pt] (-1.5,0.72) --  (1.5,0.72) node[right] {\small $\theta$};

  \draw[dashed,line width=0.5pt] (-1.1,-0.05) node[below] {{\scriptsize -}\small $\frac{f_{R}}{2}$} -- (-1.1,0.72);
    \draw[dashed,line width=0.5pt]  (1.1,-0.05) node[below] {\small $\frac{f_{R}}{2}$} -- (1.1,0.72);

  \draw[->,line width=1pt] (-1.8,0) -- (1.8,0) node[right, xshift=-0.1cm] {\small $f$} ;
  \draw[->,line width=1pt] (0,0) -- (0,2) node[above] {\small $S_X(f)$};   
    \node at (0,-0.5) {(b)};
\end{tikzpicture}
%\caption{$D(f_s,R)$ for $f_s > f_{DR}$}
\end{center}
\end{subfigure}
\begin{subfigure}[h]{0.32\textwidth}
\begin{center}
\begin{tikzpicture}[scale=1.4]

\draw[fill=blue!50] (-1.5,0) -- plot[domain=-1.5:1.5, samples=15]  (\x, {(-(\x)*(\x)+2.25)/1.3 }) -- (1.5,0) ;

\draw[fill=yellow!50] (-1.1,0.7) -- plot[domain=-1.1:1.1, samples=15]  (\x, {(-(\x)*(\x)+2.25)/1.3 }) -- (1.1,0.7) ;

\draw[line width=1pt] (-1.5,0) -- plot[domain=-1.5:1.5, samples=15]  (\x, {(-(\x)*(\x)+2.25)/1.3 }) -- (1.5,0) ;

\draw [dashed,line width=1pt] (-1.5,0.72) --  (1.5,0.72) node[right] {\small $\theta$};

  \draw[dashed,line width=0.5pt] (-1.1,-0.05) node[below] {{\scriptsize -}\small $\frac{f_{R}}{2}$} -- (-1.1,0.72);
    \draw[dashed,line width=0.5pt]  (1.1,-0.05) node[below] {\small $\frac{f_{R}}{2}$} -- (1.1,0.72);

  \draw[->,line width=1pt] (-1.8,0) -- (1.8,0) node[right, xshift=-0.1cm] {\small $f$} ;
  \draw[->,line width=1pt] (0,0) -- (0,2) node[above] {\small $S_X(f)$};   

%  \draw[dashed,line width=1pt] (-1,-0.05) node[below] {\small $\frac{\tiny{-}f_{DR}}{2}$} -- (-1,0.6);

 \node at (0,-0.5) {(c)};
\end{tikzpicture}
%\caption{Fix $R$ and find $D_X(R)$ from \eqref{eq:skp}.}
\end{center}
\end{subfigure}

\vspace{-5pt}
\caption{\label{fig:proof_sketch} 
Water-filling interpretation of \eqref{eq:idrf_single} with $H(f)$ a low-pass filter of cutoff frequency $f_s/2$. The distortion is the sum of the sampling and the lossy compression distortions. All figures correspond to the same bitrate $R$ and different sampling rates: (a) $f_s < f_{R}$, (b) $f_s \geq f_{R}$ and (c) $f_s > f_{\nyq}$. The DRF of $X(t)$ is attained for all $f_s$ greater than $f_R < f_{\nyq}$. %Note: $\theta_a < \theta$ since equals in all cases. 
}
\end{figure*}
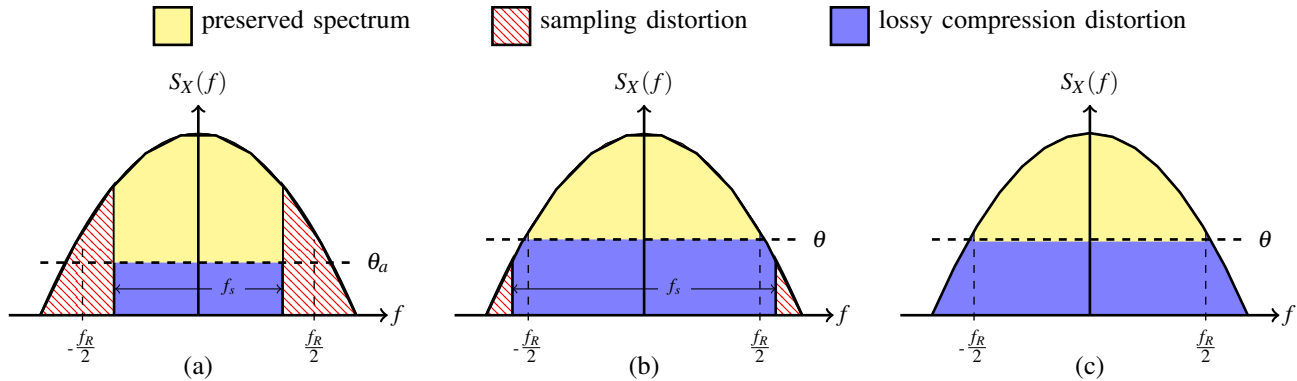

\subsection{Optimal sampling rate under bitrate constraint}
\begin{figure}
\begin{center}
\begin{tikzpicture}
\node[coordinate] (origin) at (0,0) {};
\node[above right = -0.25cm and -0.23cm of origin]  {\includegraphics[scale=0.5]{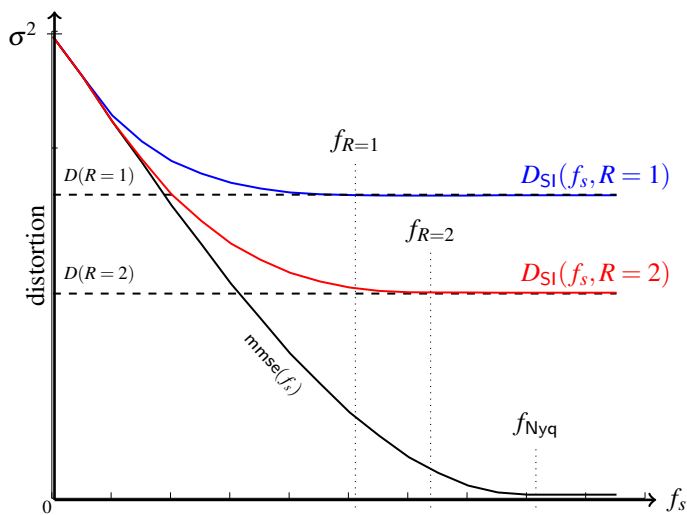}}; 
\draw[->, line width = 1pt]  (origin)-- (8,0) node[right] {$f_s$};
\draw[->, line width = 1pt] (origin) -- node[above, rotate = 90] {distortion} (0,6.5);

\draw[dotted] (4,-0.1) -- (4,4.5) node[above] {$f_{R=1}$};
\draw[dotted] (5,-0.1) -- (5,3.3) node[above] {$f_{R=2}$};

\draw[dotted] (6.4,-0.1) -- (6.4,0.7) node[above] {$f_{\nyq}$};
\node at (7.2,4.3) {\color{blue} $D_{\SI}(f_s,R=1)$};
\node at (7.2,3) {\color{red} $D_{\SI}(f_s,R=2)$};
\node[rotate = -48] at (2.9,1.8) {\scriptsize $\mmse(f_s)$};
\node at(0.6,4.3) {\scriptsize $D(R=1)$};
\node at(0.6,3) {\scriptsize $D(R=2)$};
\node at (-0.1,-0.1) {\scriptsize $0$};
\draw (-0.1,6.2) node[left] {$\sigma^2$} -- (0.1,6.2);
%\draw[help lines,step=5mm,gray!20] (0,0) grid (6,4);
\end{tikzpicture}
\end{center}
\caption{The function $D_{SI}(f_s,R)$ for the PSD of Fig.~\ref{fig:proof_sketch} with a LPF with cutoff frequency $f_s/2$ and two values of the bitrate $R$. This function describes the minimal distortion in recovering a Gaussian signal with this PSD from a bitrate-$R$ encoded version of its uniform samples. This minimal distortion is bounded from below by Shannon's DRF of $X(t)$, where the latter is attained at the sub-Nyquist sampling rate $f_{R}$.
\label{fig:DRF_optimal}}
\end{figure}

We now consider the expression $D_{\SI}(f_s,R)$ of \eqref{eq:idrf_single} for the unimodal PSD shown in Fig.~\ref{fig:proof_sketch}, where the pre-sampling filter $H(f)$ is an ideal LPF with cutoff frequency $f_s/2$. This LPF operates as an anti-aliasing filter, and therefore the part of $D_{\SI}(f_s,R)$ associated with the sampling distortion is only due to those energy bands blocked by the filter. As a result, the function $D_{\SI}(f_s,R)$ can be described by the sum of the red and the blue parts in Fig.~\ref{fig:proof_sketch}(a). Fig.~\ref{fig:proof_sketch}(b) describes the function $D_{\SI}(f_s,R)$ under the same bitrate $R$ and a higher sampling rate, while the cutoff frequency of the low-pass filter is adjusted to this higher sampling rate. As can be seen from the figure, at this higher sampling rate $D_{\SI}(f_s,R)$ equals the DRF of $X(t)$ in Fig.~\ref{fig:proof_sketch}(c), although this sampling rate is still below the Nyquist rate of $X(t)$. In fact, it follows from Fig.~\ref{fig:proof_sketch} that the DRF of $X(t)$ is attained at some critical sampling rate $f_R$ that equals the spectral occupancy of the preserved part in the Pinsker-Kolmogorov water-filling expression \eqref{eq:skp}. The existence of this critical sampling rate can also be seen in Fig.~\ref{fig:DRF_optimal}, which illustrates $D_{\SI}(f_s,R)$ as a function of $f_s$ with $H(f)$ a low-pass filter. 
\par
In the previous subsection we concluded that the DRF of $X(t)$ can be attained by sampling at or above the Nyquist rate, since then the MMSE term in \eqref{eq:skp} vanishes. Now we see that by using the low-pass filter with cut-off frequency $f_s/2$, the equality between $D_{\SI}(f_s,R)$ and the DRF, which is the minimal distortion subject to a bitrate constraint, occurs at a sampling rate smaller than the Nyquist rate. \par
An intriguing way to explain the above phenomena is as an alignment of the degrees of freedom in the signal after the pre-sampling operation with the degrees of freedom that the lossy compression with bitrate $R$ can capture in this sampled signal. For stationary Gaussian signals, the degrees of freedom in the signal representation are those spectral bands where the PSD is non-zero. When the signal energy is not uniformly distributed over these bands (unlike in the example of the PSD \eqref{eq:psd_rect}), the optimal lossy compression scheme calls for discarding those bands with the lowest energy, i.e., the parts of the signal with the lowest uncertainty. 
The pre-sampling operation removes these low-energy signal components such that the resulting signal has the same degrees of freedom as those that can be captured by the lossy compressed signal representation that follows the sampler. Thus, the pre-sampling operation ``aligns'' the degrees of freedom of the pre-sampled signal with those of the post-sampled lossy compression operation. \par
\begin{figure}
\begin{center}
\begin{tikzpicture}[scale=1]
\node at (3.82,2.3) {\includegraphics[scale=0.5, trim = 0cm 0cm 0cm 3cm, clip = true]{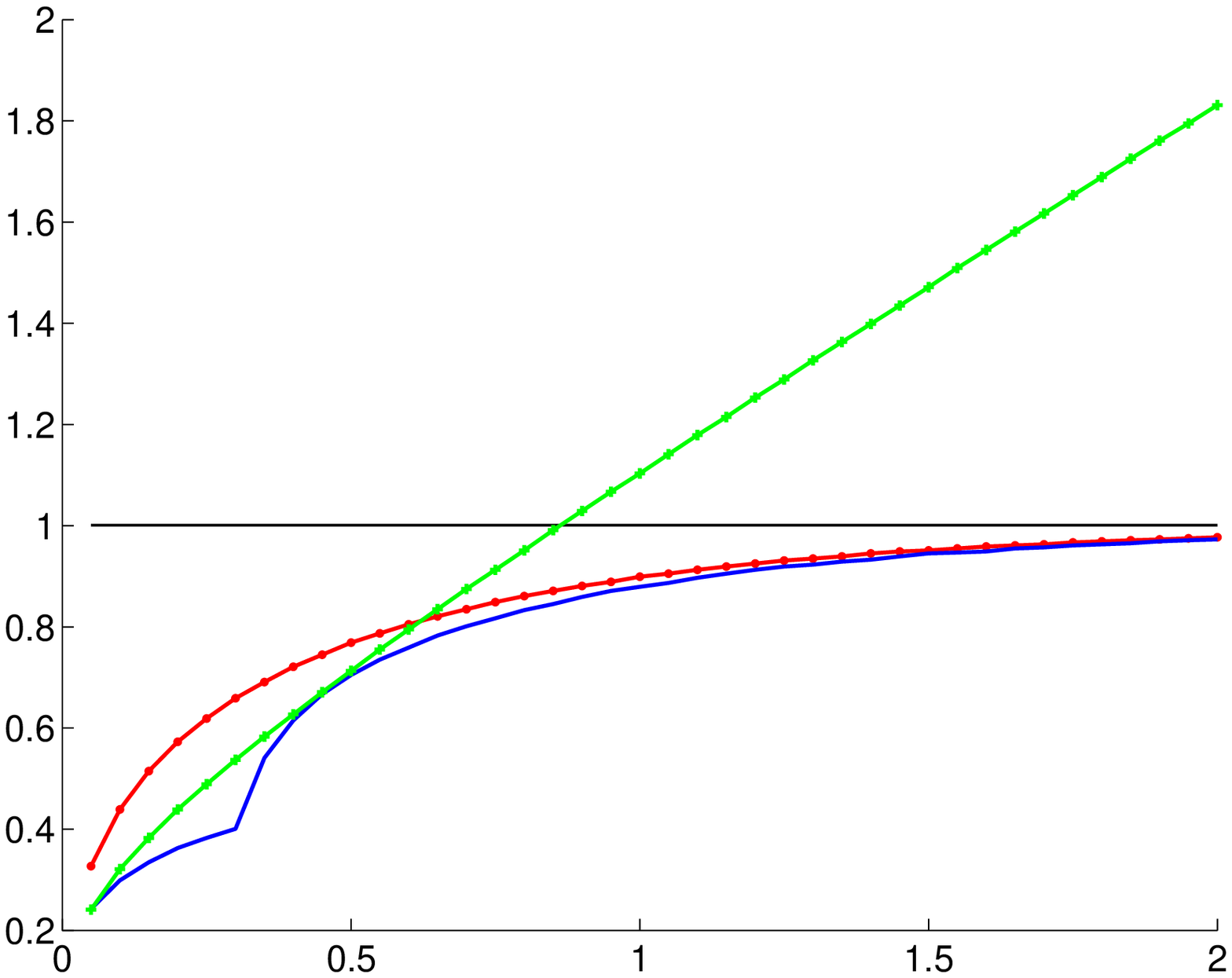} };

\node (rect) at (1,4.7) {\includegraphics[scale=0.1]{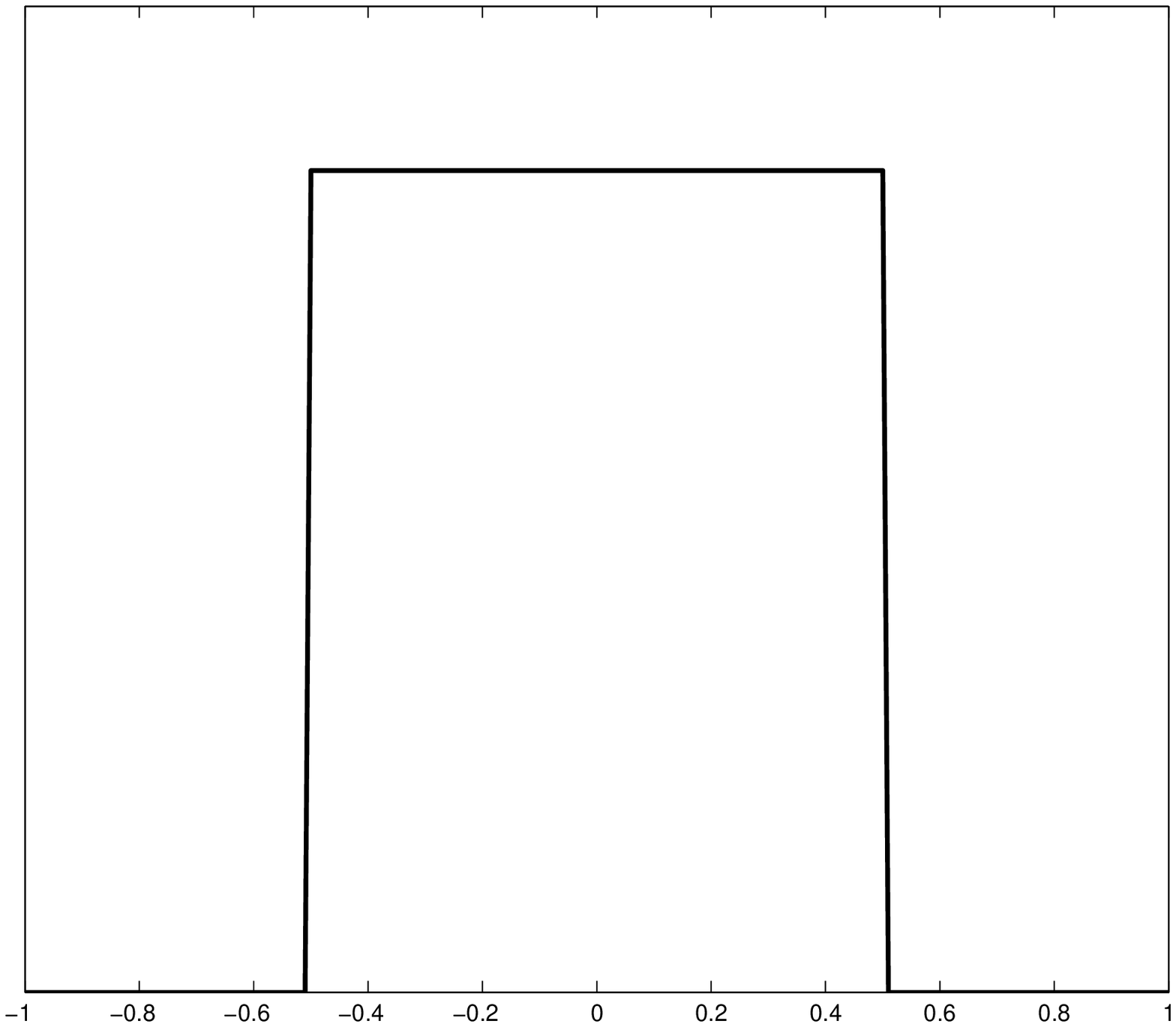} };
\node (triangle) at (3,4.7) {\includegraphics[scale=0.1]{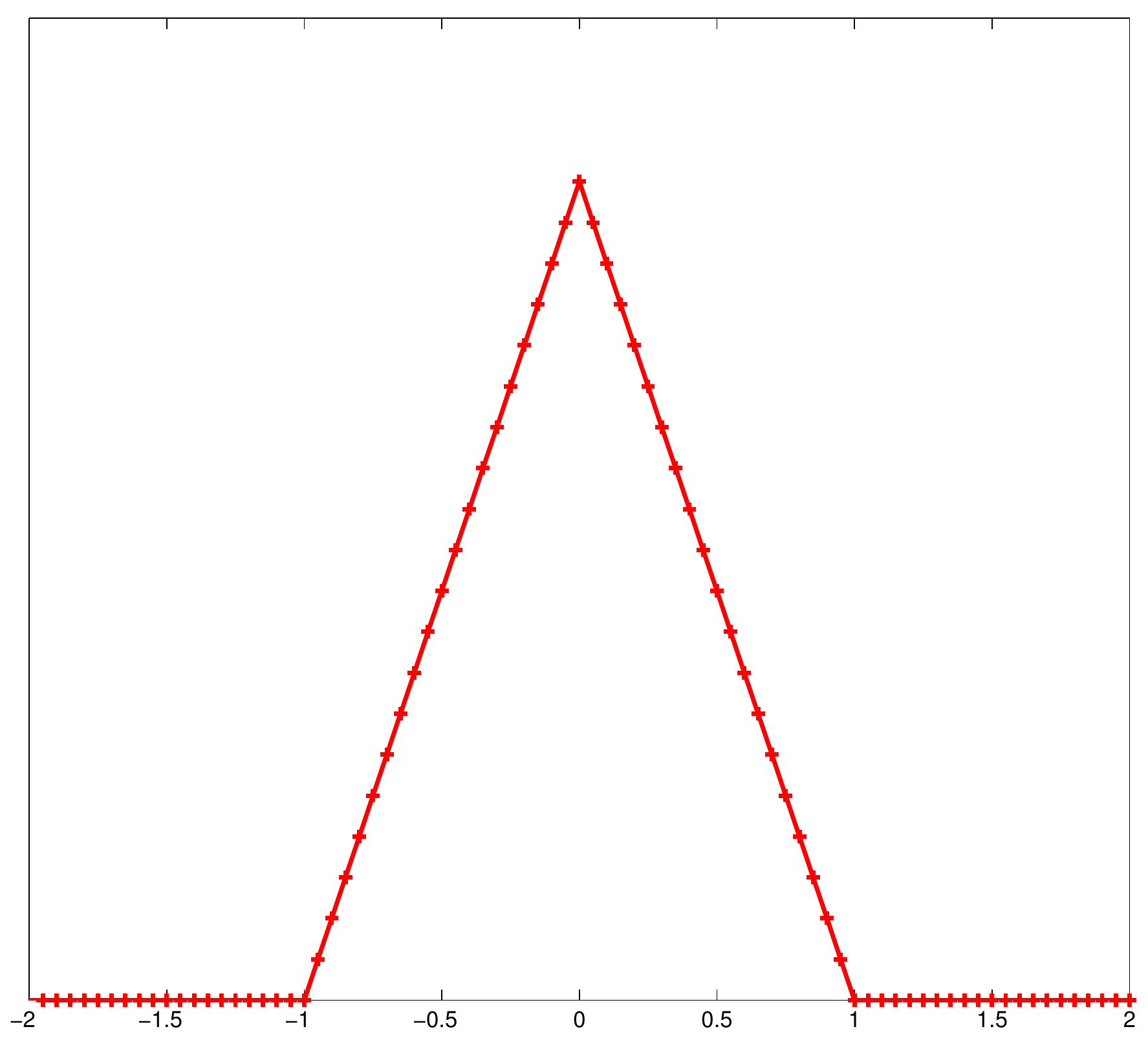} };
\node (wiggly) at (5,4.7) {\includegraphics[scale=0.1]{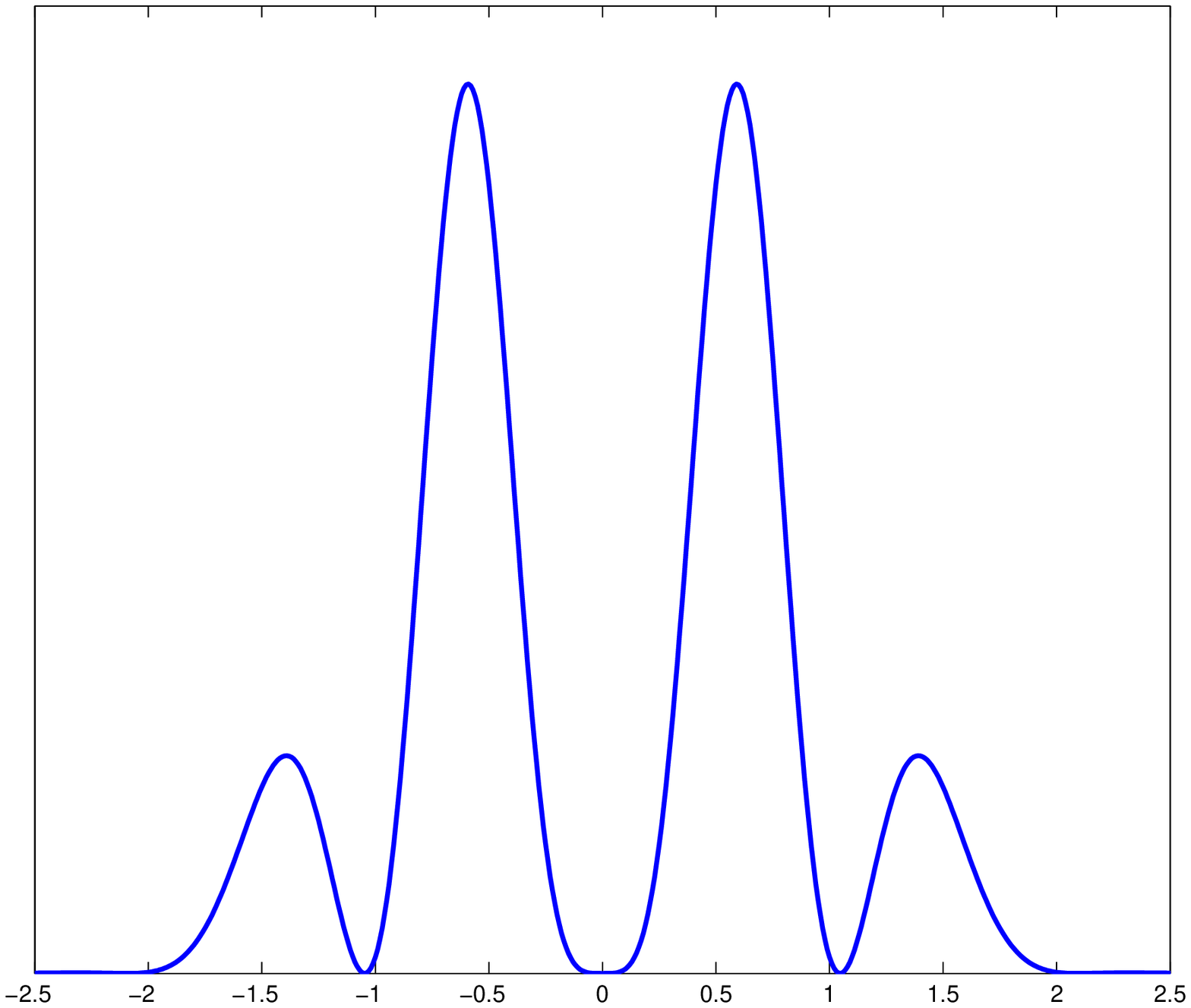} };
\node (gauss) at (7,4.7) {\includegraphics[scale=0.1]{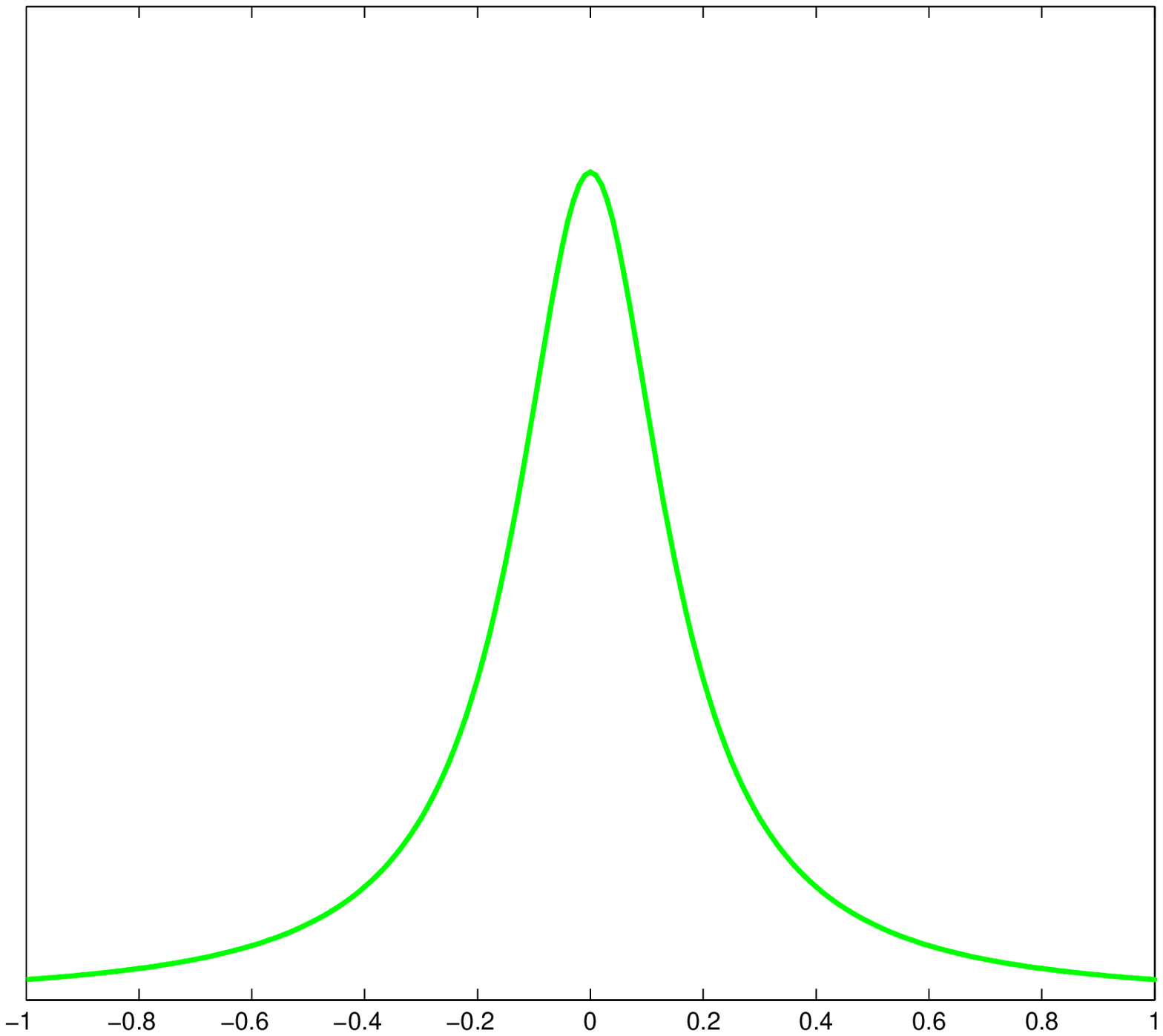} };

\node[above of = rect, node distance = 1cm] {\small $S_\Pi(f)$};
\node[above of = triangle, node distance = 1cm] { \color{red} \small $S_\Lambda(f)$};
\node[above of = gauss, node distance = 1cm] {\color{green} \small $S_\Omega(f)$};
\node[above of = wiggly, node distance = 1cm] {\color{blue} \small $S_\omega(f)$};
\draw[->, line width=1pt] (0,0) -- node[below,yshift=-0.2cm] {\small $R$ \small [bit/sec]} (7.9,0);
\draw[->, line width=1pt] (0,0) -- node[above,rotate=90, yshift = 0.4cm] {\small $f_{R}$ \small [Hz]} (0,4.8);
\end{tikzpicture}
\vspace{-5pt}
\caption{\label{fig:critical_f} The critical sampling rate $f_{R}$ as a function of the bitrate $R$ for the PSDs given in the small frames. For the bandlimited PSDs
$S_\Pi(f)$, $S_\Lambda(f)$ and $S_\omega(f)$, the critical sampling rate is always below the Nyquist rate. The critical sampling rate is finite for any $R$ even for the non-bandlimited PSD $S_\Omega(f)$.}
\end{center}
\end{figure}

The degree to which the new critical rate $f_R$ is smaller than the Nyquist rate depends on the energy distribution of $X(t)$ along its spectral occupancy. The more uniform it is, the more degrees of freedom are required to represent the lossy compressed signal and therefore $f_R$ is closer to the Nyquist rate. Figure~\ref{fig:critical_f} illustrates the dependency of $f_R$ on $R$ for various PSD functions. Note that whenever the energy distribution is not uniform and the signal is bandlimited, the critical rate $f_R$ converges to the Nyquist rate as $R$ goes to infinity and to zero as $R$ goes to zero. \\

In the discussion above we considered only signals with unimodal PSD (for example, the PSD in Fig.~\ref{fig:waterfilling} is not unimodal). The main challenge in extending the above conclusions to signals with non-unimodal PSD is the design of a sub-Nyquist sampling system that samples the signal components containing the most information about the signal (i.e. the signal's high-energy bands) in order to obtain the optimal lossy compressed signal representation when these samples are encoded at the fixed bitrate $R$. Before describing this extension, we consider the general structure of a pre-sampling transformation that minimizes the distortion in the ADX setting.

%Aside from the case of a unimodal PSD, it is in general impossible to attain this lower bound by the single-branch sampling system of Fig.~\ref{fig:sampler_simple}. This fact motivate us to consider further on more general sampling techniques such as multi-branch sampling and nonuniform sampling. 

%An immediate conclusion from \eqref{eq:idrf_single} is that when the distortion due to sampling is high, i.e., $\widetilde{S}_{X|Y}(f)$ is small compared to ${S}_X(f)$, any increase in bitrate will not lead to a significant change in distortion.

\subsection{Optimal pre-sampling transformation}
We now consider the pre-sampling filter $H(f)$ that minimizes the function $D_{SI}(f_s,R)$ subject to a fixed bitrate $R$ and sampling rate $f_s$. By examining expressions \eqref{eq:idrf_single} and \eqref{eq:mmse_single} we conclude that this minimization is equivalent to maximization of $\widetilde{S}_{X|Y}(f)$ for any $f$ in the interval $(-f_s/2,f_s/2)$. This fact is not surprising, since we have seen in \eqref{eq:mmse_single} that $\widetilde{S}_{X|Y}(f)$ represents the part of the source available to the encoder. Due to the fact that the function $\widetilde{S}_{X|Y}(f)$ is independent of $R$, the optimal filter $H(f)$ that minimizes $D_{\SI}(f_s,R)$ is only a function of the sampling rate, and it is therefore identical to the pre-sampling filter that minimizes $\mmse(X|Y)$, i.e. the MMSE without the bitrate constraint. Note that since $\widetilde{S}_{X|Y}(f)$ is indifferent to scaling in $H(f)$, the only effect of the pre-sampling filter on the distortion is through its \emph{passband}, i.e. the support of $H(f)$. It is explained in the box {\bf Optimal pre-sampling transformation} that the passband of the pre-sampling filter that minimizes $\mmse(X|Y)$ can be completely characterized by the following two properties:
\begin{enumerate}
\item[(i)] {\bf Aliasing-free} - the passband is such that the filter eliminates aliasing in sampling at frequency $f_s$, i.e., all integer shifts of the support of the filtered signal by $f_s$ are disjoint. 
\item[(ii)] {\bf Energy maximization} - the passband is chosen to maximize the energy of $X(t)$ at the output of the filter, subject to the aliasing-free property (i).
\end{enumerate}

\begin{figure}
\begin{center}
\begin{tikzpicture}
\node at (3.2,2.45) {\includegraphics[scale=0.4, trim = 0cm 0cm 0cm 0cm, clip = true]{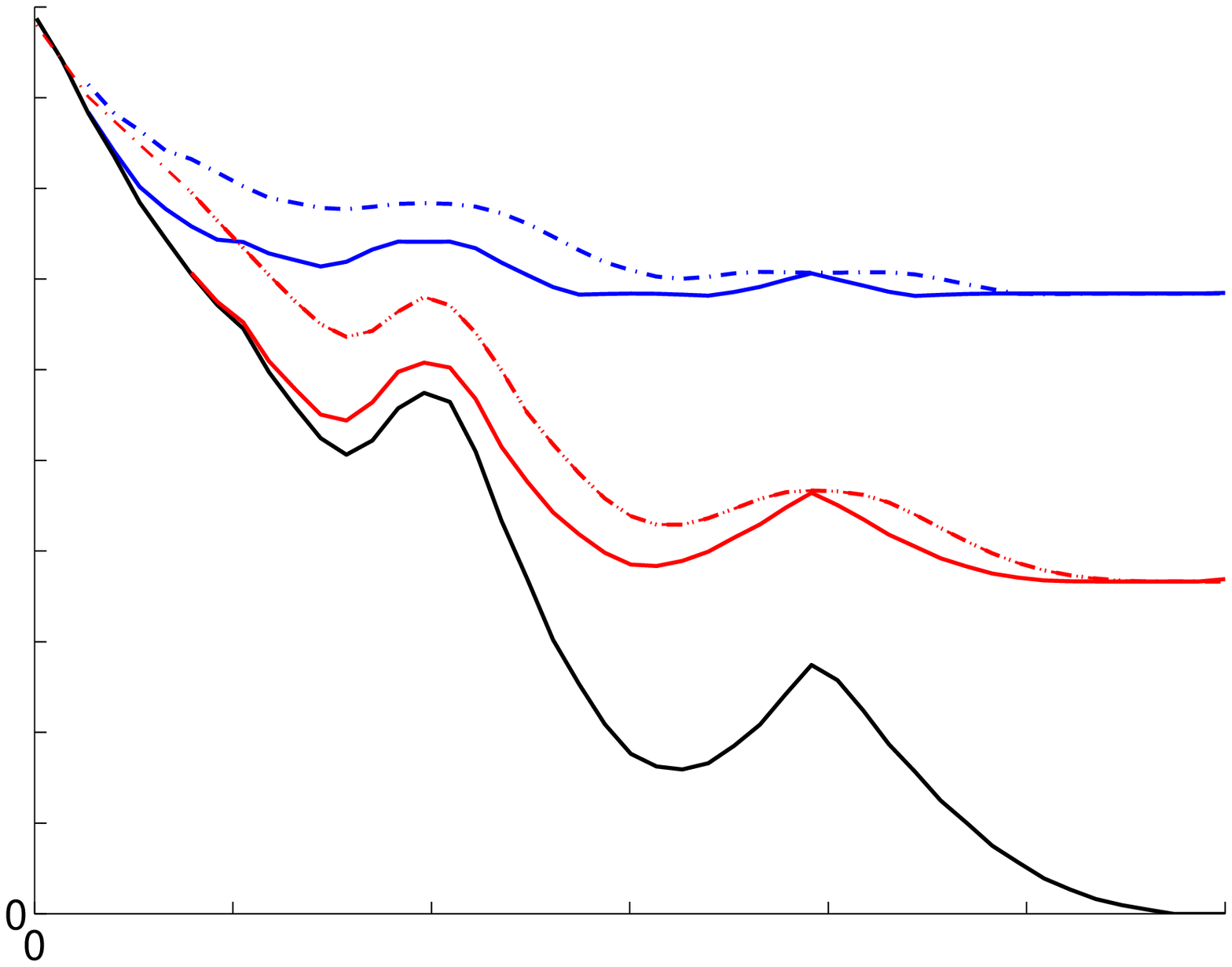}};

\draw[->, line width = 1pt] (0,0) -- node[below] {sampling rate} (6.7,0) node[right] {$f_s$};
\draw[->, line width = 1pt] (0,0) -- node[above, rotate = 90] {distortion} (0,5.2) node[above] {};

\draw[dashed] (6.2,-0.1) node[below] {$f_{\nyq}$}-- (6.2,3.8);

\draw  (0,4.8) node[left] {\scriptsize $\sigma^2$} -- (0.1,4.8);

\draw[dashed] (0,3.45) node[below, xshift = 0.6cm] {\scriptsize $D(R=1)$}  -- (5.5,3.45) node[above, xshift = -1cm, yshift = 0.2cm] {\scriptsize \color{blue} $D_{\SI}(f_s,R=1)$};

\draw[dashed]  (0,1.85) node[below, xshift = 0.6cm] {\scriptsize $D(R=3)$}  -- (5.5,1.85) node[above, xshift = -1cm,yshift = 0.4cm] {\color{red} \scriptsize $D_{\SI}(f_s,R=3)$};
\node[rotate = -40] at (5.1,0.8) {\scriptsize $\mmse_{\SI}(f_s)$};
\end{tikzpicture}
\begin{tikzpicture}
\node[int] at (6.7,2.5) {\includegraphics[scale=0.26, trim = 4cm 0cm 1cm 0cm, clip = true]{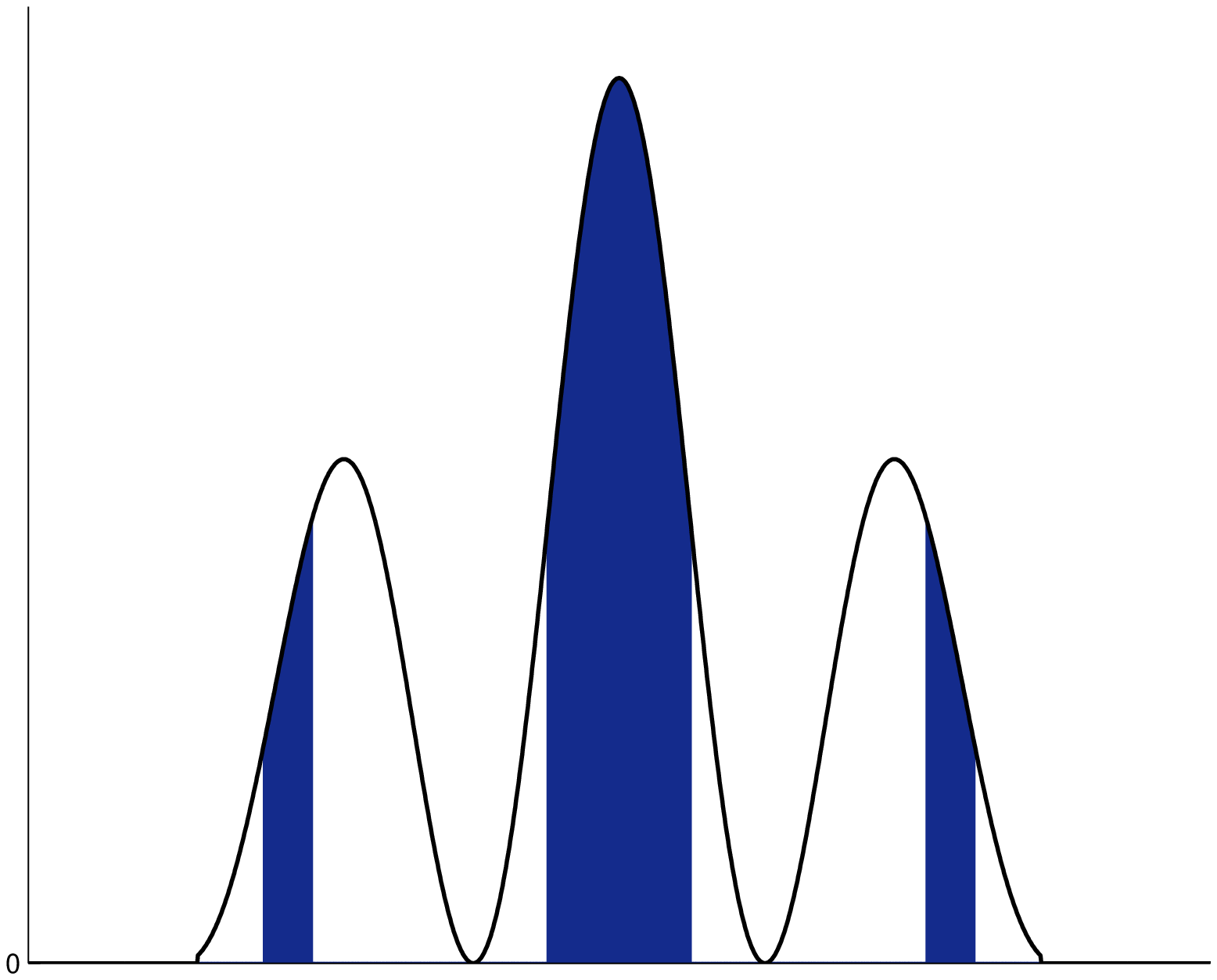}};
\draw[<->, color = red] (6.9,2) -- node[above] {\color{red} $f_s$} (7.65,2) ;
\draw[<->, color = red] (6.15,2) -- (6.85,2) ;
\draw[<->, color = red] (5.4,2) -- (6.1,2) ;
\end{tikzpicture}
\caption{Left: minimal distortion $D_{\SI}(f_s,R)$ using an optimal pre-sampling filter as a function of the sampling rate for two values of the bitrate $R$. The faint lines represent the distortion with an all-pass pre-sampling filter that allows aliasing. Right: support of the optimal pre-sampling filter over the source PSD for a particular sub-Nyquist sampling rate $f_s$. The difference between any two bands in the support is not an integer multiple of $f_s$. 
\label{fig:opsf_PSD}  }
\end{center}
\end{figure}

In the case of a unimodal PSD, a low-pass filter with cut-off frequency $f_s/2$ satisfies both the aliasing free and the energy maximization properties, and is therefore the optimal pre-sampling filter that minimizes $D_{\SI}(f_s,R)$. For this reason Fig.~\ref{fig:proof_sketch} describes the minimal value of $D_{\SI}(f_s,R)$ for the PSD considered there. In general, however, the set that maximizes the passband energy is not aliasing-free. As an example, consider the PSD illustrated in Fig.~\ref{fig:opsf_PSD} (right): the colored area represents the support of the optimal pre-sampling filter. This support is aliasing-free since the difference between any two bands in the support are not an integer multiple of $f_s$. The example in Fig.~\ref{fig:opsf_PSD} (left) also shows that although $D_{\SI}(f_s,R)$ is guaranteed to coincide with $D(R)$ for $f_s > f_{\nyq}$, the convergence to this value may not be monotonic in $f_s$. That is, some sub-Nyquist sampling rates may introduce more aliasing than sampling rates that are lower than them. This phenomena does not occur in sampling signals with unimodal PSD.\\

\begin{textbox} 
\label{box:optimal_pre_sampling_filter} 
\begin{center} {\bf Optimal Pre-sampling Transformation} \end{center}
Properties (i) and (ii) of the optimal pre-sampling filter imply that in order to minimize the MSE and hence the overall distortion, it is preferred to eliminate all information on lower energy sub-bands in the case where they interfere with higher energy bands. In order to provide an intuitive explanation for this phenomena, we consider two independent Gaussian random variables $X_1$ and $X_2$ with zero mean and variances $\sigma_1^2$ and $\sigma_2^2$, respectively. These random variables can be seen as two different spectral lines in the spectrum of $X(t)$ that interfere with each other due to aliasing in uniform sampling. Assume that we are given the linear combination $U = h_1X_1 + h_2 X_2$, and are interested in the joint estimation of $X_1$ and $X_2$ subject to a MSE criterion. Namely, we want to minimize
\[
\mmse(X_1,X_2 | U ) \triangleq \mathbb E \left(X_1- \widehat{X}_1 \right)^2 + \mathbb E \left(X_2- \widehat{X}_2 \right)^2. 
\]
The optimal estimator of each variable as well as the corresponding estimation error can be easily found since the optimal estimator is linear. We further ask how to choose the coefficients $h_1$ and $h_2$ in the linear combination such that the MSE is minimized. A simple optimization over the expression for $\mmse(X_1,X_2 | U )$ shows that $h_1 \neq 0$, $h_2 = 0$ is the answer whenever $\sigma_1^2 > \sigma_2^2$, and $h_1 = 0$, $h_2 \neq 0$ whenever $\sigma_1^2 < \sigma_2^2$. That is, the optimal linear combination eliminates all the information on the part of the signal with the lowest variance and passes only the part with the highest uncertainty. Going back to spectral components, the MSE is minimized by a pre-sampling filter $H(f)$ that eliminates all spectral components of low energy whenever they interfere with high energy spectral components due to the aliasing that results from uniform sampling. 
%A similar analysis holds in the case when the observations are contaminated by noise \cite{Kipnis2014}. The only difference is that the part of the spectrum to be passed is now the part with the highest SNR. Also note that the aliasing-free property of the filter is intimately related to the squared error fidelity criteria we use, and may not hold under a different performance metric. 
\end{textbox}

\begin{figure}
\begin{center}
\begin{tikzpicture}[scale=1]
%\node at (0,0) { \includegraphics[scale=0.4]{classic_waterfilling.eps}};

 \fill[fill=red!50, pattern=north west lines, pattern color=red] (-4,0) -- (-3.5,0.22)--plot[domain=-3.5:3.5, samples=100] (\x, {(0.7+3*cos(24*3.14*\x)^2)*exp(-\x*\x/10)}) -- (3.5,0.22)--(4,0);
 
 \fill[fill=blue!50] (-2.8,0) -- plot[domain=-2.8:-1.7, samples=100] (\x, {(0.7+3*cos(24*3.14*\x)^2)*exp(-\x*\x/10)}) -- (-1.7,0)-- cycle;
   
 \fill[fill=blue!50] (-0.8,0) -- plot[domain=-0.8:0.8, samples=100] (\x, {(0.7+3*cos(24*3.14*\x)^2)*exp(-\x*\x/10)}) -- (0.8,0)--
   cycle;

 \fill[fill=blue!50] (1.7,0) -- plot[domain=1.7:2.8, samples=100] (\x, {(0.7+3*cos(24*3.14*\x)^2)*exp(-\x*\x/10)}) -- (2.8,0)--
   cycle;

 \fill[fill=yellow!80] (-2.8,1.3) -- plot[domain=-2.8:-1.7, samples=100] (\x, {(0.7+3*cos(24*3.14*\x)^2)*exp(-\x*\x/10)}) -- (-1.7,1.3)--
   cycle;

 \fill[fill=yellow!80] (-0.8,1.3) -- plot[domain=-0.8:0.8, samples=100] (\x, {(0.7+3*cos(24*3.14*\x)^2)*exp(-\x*\x/10)}) -- (0.8,1.3)--
   cycle;

 \fill[fill=yellow!80] (1.7,1.3) -- plot[domain=1.7:2.8, samples=100] (\x, {(0.7+3*cos(24*3.14*\x)^2)*exp(-\x*\x/10)}) -- (2.8,1.3)--
   cycle;

 \draw (-4,0) -- (-3.5,0.22)--plot[domain=-3.5:3.5, samples=100] (\x, {(0.7+3*cos(24*3.14*\x)^2)*exp(-\x*\x/10)}) -- (3.5,0.22)--(4,0);

\foreach \x/\xtext in {-3,-2,-1,0,1,2,3}
 \draw[shift={(\x,0)}] (0pt,2pt) -- (0pt,-2pt) node[below] {$\xtext$};
  \foreach \y/\ytext in {1/,2/,3/}
    \draw[shift={(0,\y)}] (2pt,0pt) -- (-2pt,0pt) node[left] {$\ytext$};

\draw [fill=yellow!50, line width=1pt] (1,3.5) rectangle  (1.5,4) node[right, xshift=0cm, yshift = -0.2cm] {preserved spectrum};

\draw [fill=red!50, line width=1pt, pattern=north west lines, pattern color=red] (1,2.5) rectangle  (1.5,3) node[right, xshift = 0cm, yshift = -0.2cm] {sampling distortion};

\draw [fill=blue!50, line width=1pt] (1,3) rectangle  (1.5,3.5) node[right, xshift = 0cm, yshift = -0.2cm] {lossy compression distortion};

%\node[rectangle,draw,fill=red!50,inner sep=2pt,align=left, pattern=north west lines, pattern color=red]  at (-3,2.6) {$mmse(f_s)$};
%\node[rectangle,draw,fill=yellow!80,inner sep=1pt,align=left] at (-2.15,3.8) {\small $R=\frac{1}{2}\int_{F^\star}\log^+\left[{S_X}(f)/\theta \right]df $};
%\node[rectangle,draw,fill=blue!50,inner sep=1pt,align=left]  at (-2.4,3.2) {$ \int_{F^\star} \min\left\{{S_X}(f),\theta \right\} df$};
\node at (0.15,1.45) {$\theta$};
\draw[dashed, line width=1pt] (-4,1.3) -- (4,1.3);
\draw[->,line width=1pt]  (-4,0)--(4,0) node[right] {$f$};
\node at (-2.9,1.7) [rotate=60] {\small $S_X(f)$};
\draw[->,line width=1pt]  (0,0)--(0,4);
%\node [rotate=76] at  (-0.9,0.3) (Sx_or) {$S_X(f)$};
\end{tikzpicture}
\end{center}
\vspace{-10pt}
\caption{\label{fig:idrf_lower_bound} Water-filling interpretation of the fundamental minimal distortion $D(f_s,R)$ in ADX. The overall distortion is the sum of the sampling distortion and the lossy compression distortion. The set $F^\star(f_s)$ defining $D(f_s,R)$ is the support of the preserved spectrum.} 
\end{figure}
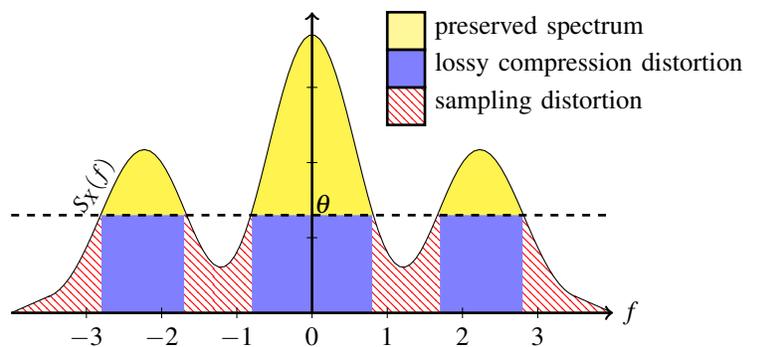

%\subsection{Fundamental distortion lower bound under sampling}
The dependency of the passband of $H(f)$ on the sampling frequency $f_s$ comes from the aliasing-free property. In particular, this property restricts the Lebesgue measure of the passband of any aliasing free filter to be smaller than $f_s$ \cite[Prop. 2]{Kipnis2014}. It follows from this last fact that a lower bound on the function $D_{\SI}(f_s,R)$ is obtained by taking the part of the spectrum of highest energy and overall Lebesgue measure not exceeding $f_s$. Namely, denote by $F^\star(f_s)$ the part of the spectrum that maximizes $\int_F S_X(f)df $ over all sets $F$ of Lebesgue measure not exceeding $f_s$. The following expression bounds the function $D_{\SI}(f_s,R)$ from below:
\begin{subequations} \label{eq:idrf_lower_bound}
\begin{align}
D(f_s,R) & = \mmse(f_s) + \int_{F^\star(f_s)} \min\left\{ S_X(f), \theta \right\} \\
R (\theta) & = \frac{1}{2} \int_{F^\star(f_s) } \log_2^+ \left[ S_X(f) /\theta \right] df,
\end{align}
\end{subequations}
where
\begin{equation}
\label{eq:mmse_lower_bound}
\mmse(f_s) =\int_{-\infty}^\infty S_X(f)df  - \int_{F^\star(f_s)}  S_X(f) df. 
\end{equation}
A graphical water-filling interpretation of the above expression is given in Fig.~\ref{fig:idrf_lower_bound}. In the next section we describe how to attain this lower bound by extending SI samplers to an array of such samplers. 

\subsection{Multi-Branch Sampling}
\begin{figure}
\begin{center}
\begin{tikzpicture}[node distance=2cm,auto,>=latex]
\node at (0,0) (source) {$X(t)$};
\node[coordinate] (source_up) [above of = source,node distance = 1cm]{};
\node[coordinate] (first_jnc) [right of = source, node distance=1.2cm] {};
\fill  (first_jnc) circle [radius=2pt];
\draw[->, line width=1pt] (source)--(first_jnc);   
\node[int1]  (pre_sampling2) [right of = first_jnc, node distance=1cm]{$H_2(f)$};  
\draw[->, line width=1pt] (first_jnc)--(pre_sampling2);   
\node [coordinate, right of = pre_sampling2,node distance = 1.7cm] (smp_in2) {};
\node [coordinate, right of = smp_in2,node distance = 0.7cm] (smp_out2){};
\node [coordinate,above of = smp_out2,node distance = 0.4cm] (tip2) {};
\fill  (smp_out2) circle [radius=2pt];
\fill  (smp_in2) circle [radius=2pt];
\fill  (tip2) circle [radius=2pt];
\node[left,left of = tip2, node distance = 0.5 cm] (ltop2) {$f_s/L$};
\node [coordinate] (enc_left) [right of = source, node distance = 6cm] {};
\node [right of = enc_left]  (out) {$\Yv_n$};
\draw[->,densely dotted,line width = 1pt,thin] (ltop2) to [out=0,in=70] (smp_out2.north);
 \draw[line width=1pt]  (smp_in2) -- (tip2);
 \draw[-,line width=1pt]   (pre_sampling2)-- node[above] {\small }(smp_in2);
 \draw[->,line width=1pt] (smp_out2) -- node[above] {$Y_2[n]$} (enc_left);
\draw[line width=1pt]  (smp_in2) -- (tip2);
\node[coordinate] (enc_bot) [below of = enc_left, node distance = 1.3cm] {};
\node[coordinate] (enc_top) [above of = enc_left, node distance = 0.8cm] {};
\fill (enc_bot) circle [radius=3pt];
\fill (enc_top) circle [radius=3pt];
\fill (enc_left) circle [radius=3pt];
\draw[line width = 3pt] (enc_top) -- (enc_bot); 
\draw[->,line width = 3pt] (enc_left) -- (out); 
\node [below of=enc_bot, node distance = 1cm] (right_down){};
\node[int1]  (pre_sampling3) [below of = pre_sampling2, node distance=1.3cm]{$H_L(f)$};  
\draw[->,line width=1pt] (first_jnc)|-(pre_sampling3);   
\draw[-,dotted, line width = 1pt] (pre_sampling2) -- (pre_sampling3) ;  	  
\node [coordinate, right of = pre_sampling3,node distance = 1.7cm] (smp_in3) {};
  \node [coordinate, right of = smp_in3,node distance = 0.7cm] (smp_out3){};
	\node [coordinate,above of = smp_out3,node distance = 0.4cm] (tip3) {};
\fill  (smp_out3) circle [radius=2pt];
\fill  (smp_in3) circle [radius=2pt];
\fill  (tip3) circle [radius=2pt];
\node[left,left of = tip3, node distance = 0.5 cm] (ltop3) {\small $f_s/L$};
\draw[->,dashed,densely dotted,line width = 1pt,thin] (ltop3) to [out=0,in=70] (smp_out3.north);
 \draw[line width=1pt]  (smp_in3) -- (tip3);
 \draw[-,line width=1pt]   (pre_sampling3)--node[above] {\small } (smp_in3);
 \draw[->,line width=1pt] (smp_out3) -- node[above] {\small $ Y_L[n]$} (enc_bot);
\draw[line width=1pt]  (smp_in3) -- (tip3);
\node[int1]  (pre_sampling1) [above of = pre_sampling2, node distance=0.8cm]{$H_1(f)$};  
\draw[->, line width=1pt] (first_jnc)|-(pre_sampling1);   
\node [coordinate, right of = pre_sampling1,node distance = 1.7cm] (smp_in1) {};
  \node [coordinate, right of = smp_in1,node distance = 0.7cm] (smp_out1){};
	\node [coordinate,above of = smp_out1,node distance = 0.4cm] (tip1) {};
\fill  (smp_out1) circle [radius=2pt];
\fill  (smp_in1) circle [radius=2pt];
\fill  (tip1) circle [radius=2pt];
\node[left,left of = tip1, node distance = 0.5 cm] (ltop1) {\small $f_s/L$};
\draw[->,dashed,densely dotted,line width = 1pt,thin] (ltop1) to [out=0,in=70] (smp_out1.north);
 \draw[line width=1pt]  (smp_in1) -- (tip1);
 \draw[-,line width=1pt]   (pre_sampling1)--node[above] {\small }(smp_in1);
 \draw[->,line width=1pt] (smp_out1) -- node[above] {$Y_1[n]$}  (enc_top);
\draw[line width=1pt]  (smp_in1) -- (tip1);
\draw[line width=1pt, dashed] (0.9,1.5) rectangle (7,-2) ;
 \end{tikzpicture}
\end{center}
\caption{\label{fig:sampler_multi} Multi-branch filter-bank uniform sampler.}
\end{figure}
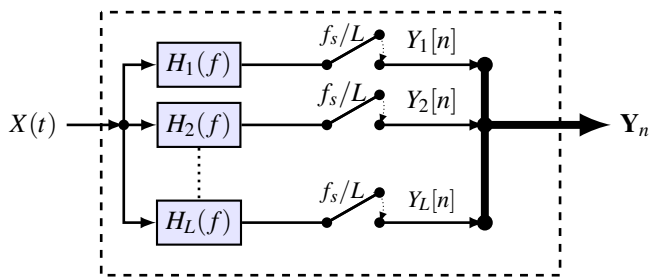

In contrast to the case of a unimodal PSD, it is in general impossible to attain the function $D(f_s,R)$ of \eqref{eq:idrf_lower_bound} using a single SI sampler. Indeed, once we fix a band, no other bands located at integer multiples of the sampling rate are included in the support of the optimal pre-sampling filter due to the aliasing-free property. This limitation implies that the support of the optimal pre-sampling filter does not necessarily consist a set of measure $f_s$ with largest signal energy as in the definition of $D(f_s,R)$. By using more sampling branches, the global aliasing-free property is relaxed to a local aliasing-free property at each sampling branch. Therefore, while each branch has constraints on the position of the bands in the support of its filter in order to avoid aliasing, the increment in sampling branches allows for more freedom in selecting the overall part of the spectrum passed by all filters. As a result, the union of the supports of an optimal set of $L$ filters that are aliasing-free with respect to $f_s/L$, approximates the set of maximal energy of measure $f_s$ better than is possible with a single filter that is aliasing-free with respect to $f_s$. This situation is illustrated in Fig.~\ref{fig:opsf_PSD}. In particular, it can be seen there that components that needed to be eliminated in the single branch case due to aliasing with higher energy components can now be retained as these two components can be preserved on separate branches without interference with each other after sampling. 
In other words, multi-branch sampling reduces part of the constraint on retaining desired signal components that arises as a result of the aliasing-free requirement in a single SI filter, leading to higher energy frequency components in the resulting signal representation before encoding, and therefore lower distortion after encoding. \par
%
%As an example, Fig.~\ref{fig:opsf_PSD} illustrates a case where for some $f_s$ the passband of the optimal pre-sampling filter blocks part of the frequency components in the side-lobes of $S_X(f)$, since these interfere with frequency bands in the central lobe. It follows that the amount of energy passed by an aliasing-preventing pre-sampling operation can be increased if more sampling branches were allowed, leading to higher energy active bands in the resulting signal representation before encoding and therefore lower distortion post encoding. 
This intuition motivates replacing the SI sampler in Fig.~\ref{fig:ADX_simple} with an array of such samplers, as illustrated in Fig.~\ref{fig:sampler_multi}. Within each branch, the pre-sampling filter may pass only a narrow part of the signal's spectrum and apply passband sampling \cite{vaughan1991theory}. This multi-branch uniform sampler covers a wide class of sampling systems used in practice, including single-branch SI sampling, nonuniform periodic sampling and multi-coset sampling \cite{eldar2015sampling, 52200}. 
\par
The analysis of the system is greatly simplified if all sampling branches have the same sampling rate. Thus, we assume that the sampling rate at each branch equals $f_s/L$, so that the overall \emph{effective} sampling rate is $f_s$. Similarly to the case of a single SI sampler, the optimal selection of the pre-sampling filters across all branches leads to a collections of filters with the aliasing free property at each branch, such that the net energy passed by these filters is maximal \cite{Kipnis2014}. Since the measure of the passband of each aliasing-free filter for sampling at rate $f_s/L$ is at most $f_s/L$, the overall part of the spectrum passed by the $L$ filters is of size at most $f_s$. This property implies that the lower bound $D(f_s,R)$ of \eqref{eq:idrf_lower_bound} is kept under this form of sampling. \par
\begin{figure}
\begin{center}
\begin{tikzpicture}
\node at (0,0) {\includegraphics[scale=0.27, trim = 0cm 0cm 0cm 0cm, clip = true]{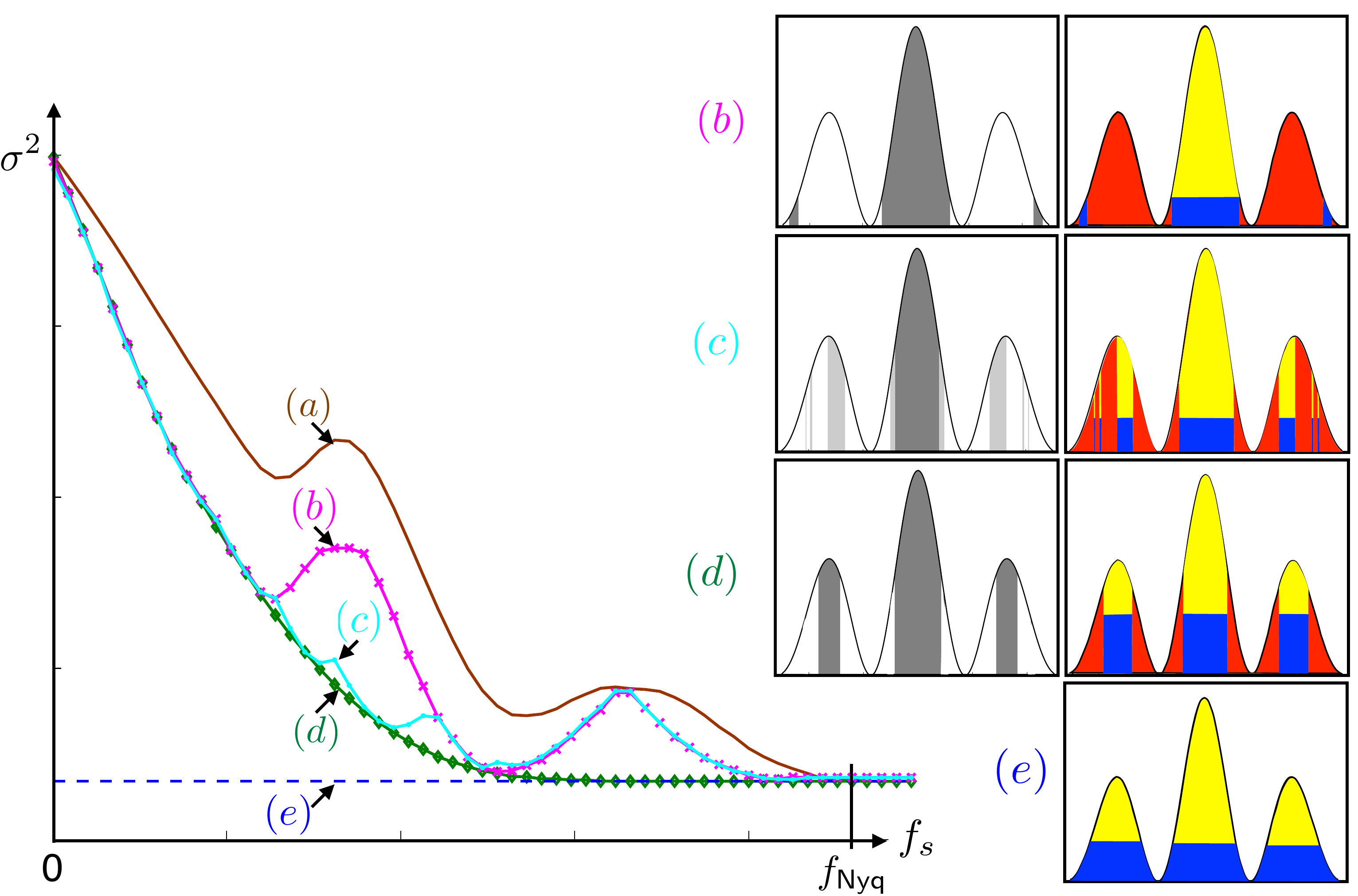}};
\node[align = center, scale = 0.8] at (1.7,3.6) {support of \\ optimal filters \\
(cases b,c,d)};
\node[align = center, scale = 0.8] at (3.7,3.6) {water-filling \\ 
representation \\ (cases b,c,d,e)
};
\end{tikzpicture}
\caption{Minimal distortion versus the sampling rate $f_s$ for a fixed value of $R$. 
The case of no sampling pre-filter is given in case (a), and the case of one, two, and five sampling branches with optimal branch pre-filtering are considered in cases (b)-(d), respectively. For each of these cases and a fixed $f_s$, the union of support for the optimal filters, which equals $f_s$, is shown in the grayscale image above on the left and how these bands are identified through water-filling and the sampling distortion that results is shown on the right. Case (d) of five SI sampling branches preserves the part of the spectrum of measure $f_s$ with highest energy, and therefore achieves $D(f_s,R)$. (e) Shannon's DRF with its water-filling representation. \label{fig:opsf_DRF}}
\end{center}
\end{figure}

The next question is whether this lower bound is attainable, provided that we are allowed to increase the number of sampling branches $L$ and the pre-sampling filters $H_1(f),\ldots,H_L(f)$. A positive answer to this question was given in \cite{Kipnis2014}, where it was shown that for any PSD, the distortion level $D(f_s,R)$ can be attained using some finite number $L^\star$ of sampling branches and a particular set of filters, each of which is anti-aliasing for sampling at rate $f_s/L^\star$. The reduction of the distortion in ADX using the optimal filter-bank sampler as the number of branches increases is illustrated in Fig.~\ref{fig:opsf_DRF}. Also shown in this figure are the supports of the optimal pre-sampling filters at a specific sampling rate $f_s$. 
\par 
We conclude that the function $D(f_s,R)$ describes an achievable lower bound for the distortion in the ADX setting with a multi-branch uniform sampler. In the next subsection we extend this result to nonuniform and generalized linear sampling procedures.
%It is interesting to note that with any $L>1$ sampling branches each of sampling rate $f_s/L$, the distortion under optimal pre-filter design must not exceed that of a single sampling branch of sampling rate $f_s$. The reason is that the $L$ sampling branches can effectively realize the single branch in the sense that the vector process $\Yv$ resulting from the multi-branch sampler contains the same samples as the process $Y$ resulting from the single branch sampler. The same argument implies that using $L_1$ sampling branches always outperform any $L_2$ sampling branches whenever $L_1$ is an integer multiple of $L_2$. It is possible, however, that for some PSD, sampling rate and $L1 < L_2 < L^*$ co-primes, we have that the distortion using $L_1$ sampling branches is smaller than using $L_2$ branches.

% \begin{figure} 
% \begin{center}
% \begin{tikzpicture}
% \node[int1] at (6.7,2.5) {\includegraphics[scale=0.26, trim = 2cm 0cm 1cm 2cm, clip = true]{OPSF_painted_spectrum_P2}};
% \end{tikzpicture}
% \caption{Support of optimal pre-sampling filters over the PSD of the analog source for the multi-branch sampling system with two sampling branches ($L=2$). The support of each filter is an aliasing-free set for sampling at rate $f_s/L$.}
% \end{center}
% \end{figure}

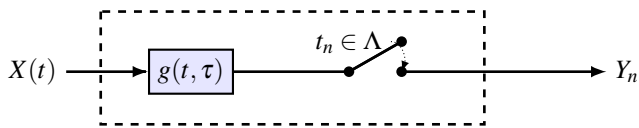
\begin{figure}[b]
\begin{center}
\begin{tikzpicture}[node distance=2cm,auto,>=latex]
\node at (0,0) (source) {$X(t)$};
\node[coordinate] (source_up) [above of = source,node distance = 1cm]{};

\node[coordinate] (first_jnc) [right of = source, node distance=1.5cm] {};
  
\draw[-, line width=1pt] (source)--(first_jnc);   

\node[int1]  (pre_sampling2) [right of = first_jnc, node distance=0.6cm]{$g(t,\tau)$};  

\draw[->, line width=1pt] (first_jnc)--(pre_sampling2);   
  	  
\node [coordinate, right of = pre_sampling2,node distance = 2.1cm] (smp_in2) {};
  \node [coordinate, right of = smp_in2,node distance = 0.7cm] (smp_out2){};
	\node [coordinate,above of = smp_out2,node distance = 0.4cm] (tip2) {};
\fill  (smp_out2) circle [radius=2pt];
\fill  (smp_in2) circle [radius=2pt];
\fill  (tip2) circle [radius=2pt];
\node[left,left of = tip2, node distance = 0.7 cm] (ltop2) {$t_n \in \Lambda $};

\node [right of = smp_out2, node distance=3cm]  (out) {$Y_n$};

\draw[->,densely dotted,line width = 1pt,thin] (ltop2) to [out=0,in=70] (smp_out2.north);
 \draw[line width=1pt]  (smp_in2) -- (tip2);
 \draw[-,line width=1pt]   (pre_sampling2)-- node[above, xshift =-0.2cm] {}(smp_in2);

\draw[line width=1pt]  (smp_in2) -- (tip2);

\draw[->,line width = 1pt] (smp_out2) -- (out); 

\draw[line width=1pt, dashed] (0.9,0.8) rectangle (6,-0.7) ;
\end{tikzpicture}
\end{center}
\caption{\label{fig:sampler_nonuniform} Nonuniform sampler with time-varying pre-processing.}
\end{figure}

\subsection{Nonuniform and Generalized Sampling}
We now extend the ADX setting to include a nonuniform sampling system with time-varying pre-processing. We show that under some mild assumptions on the sampling set, it is impossible to achieve distortion lower than $D(f_s,R)$, where here $f_s$ equals the \emph{density} of the sampling set. The definition of this density and more detailed background on nonuniform sampling can be found in the box {\bf Nonuniform Sampling}. This extension includes all cases of linear continuous sampling, as given in the box {\bf Generalized Sampling of Random Signals}. 
\\

A nonuniform time-varying sampler is illustrated in  Fig.~\ref{fig:sampler_nonuniform}. It is characterized by a discrete and ordered sampling set of sampling times $\Lambda = \left\{ \ldots,t_{-1},t_0,\ldots, t_n,\ldots \right\} \subset \mathbb R$ and a time-varying impulse response $g(t,\tau)$. The sampling set is assumed to be uniformly discrete, in the sense that there exists a universal constant $\epsilon>0$ such that each two elements of $\Lambda$ are at least $\epsilon$ apart. The $n$th output of the sampler is the convolution of $g(t_n,t)$ with $X(t)$, where $t_n \in \Lambda$. For every finite time lag $[-T/2,T/2]$, the vector $\Yv$ is the sampler output at times $[-T/2,T/2] \cap \Lambda$. Our goal is to map this vector to one of $2^{\lfloor TR \rfloor}$ elements, and, by observing this element, recover $X(t)$ over this time interval under MSE distortion. We note that although the sampler in Fig.~\ref{fig:sampler_nonuniform} has only a single sampling branch, it can be shown that the multi-branch sampling system of Fig.~\ref{fig:sampler_multi} may be realized by this filter using a particular choice of the time-varying operation \cite{YuxinNonUniform}. 
\\

%Our general setting calls for a few regularity assumptions on the sampling set $\Lambda$. First, we assume that $\Lambda$ is uniformly discrete, in the sense that there exists a universal constant $\epsilon>0$ such that each two elements of $\Lambda$ are at least $\epsilon$ apart. In addition, we assume that $\Lambda$ is such that the discrete-time process $Y_n$ at the output of the nonuniform sampler is \emph{asymptotic mean stationary} (AMS) \cite{gray2011entropy}. An AMS process can be decomposed as the sum of ergodic processes, and AMS is therefore a weaker property than ergodicity. The AMS assumption is required in order to provide a coding theorem that connects the problem of encoding and recovering these processes to the information expression involving the optimization over probability distributions subject to a mutual information rate constraint \cite{gray2011entropy}. Although our current setting is an indirect source coding problem where the goal is to recover $X(t)$ and not the AMS process at the output of the sampler, the existence of a general coding theorem for AMS processes leads to a similar characterization of the distortion in the indirect problem \cite{KipnisBitrate}. To avoid pathological scenarios, we also evaluate the distortion in this case in the limit as the time horizon $T$ goes to infinity, rather than considering $T$ as part of the optimization problem. \\

As in the case of uniform sampling, it is instructive to begin our discussion with the lower bound on the minimal distortion obtained by the MMSE in estimating $X(t)$ from its nonuniform sampled version $Y_n$. A classical result in functional analysis and signal processing due to Landau asserts that a signal can be perfectly recovered from its nonuniform samples if and only if the density of $\Lambda$ exceeds its \emph{spectral occupancy} \cite{Landau1967}. See the box "nonuniform Sampling" for an overview of this result. In our setting, the spectral occupancy takes the form of the support of the PSD. Therefore, the function $D(f_s,R)$ of \eqref{eq:idrf_lower_bound} agrees with Landau's characterization since it implies that as $R$ goes to infinity, zero MSE is attained if and only if the sampling rate exceeds the spectral occupancy. \par
The ADX with the nonuniform sampler extends the above result since it considers the case of a limited finite bitrate, and linear pre-processing of the samples. For this setting, it is shown in \cite{KipnisBitrate} that the lower bound on the distortion $D(f_s,R)$ still holds, provided $f_s$ is replaced by the density of $\Lambda$. That is, for any time-varying system $g(t,\tau)$ and any sampling set $\Lambda$ for which a density exists, the minimal distortion in the ADX setting with a time-varying nonuniform sampler is lower bounded by $D(f_s,R)$ where $f_s$ equals the density of $\Lambda$. \par
It follows that minimal distortion in the ADX setting under the class of linear pointwise samplers at rate $f_s$ is fully characterized by the function $D(f_s,R)$. In general, and according to Landau's condition for stable sampling, an equality between $D(f_s,R)$ and Shannon's DRF of the analog source is expected for sampling rates higher than the spectral occupancy of $X(t)$. We have seen, however, that this equality usually occurs already as the sampling rate $f_s$ exceeds the support of the preserved part of the spectrum in the Pinsker-Kolmogorov water-filling expression \eqref{eq:skp}. In other words, the sampling structure that attains $D(f_s,R)$ utilizes the special structure associated with optimal lossy compression of analog signals given by the Pinsker-Kolmogorov result: it ``aligns'' the degrees of freedom of the pre-sampled signal with those of the post-sampled lossy compressed signal, so that the part of the signal removed prior to the sampling stage matches the part of the signal removed under optimal lossy compression of the signal subject to the bitrate constraint. \\

As a final remark, we note that any linear continuous sampler as defined in the box {\bf Generalized Sampling of Random Signals} can be expressed as the time-varying nonuniform sampler of Fig.~\ref{fig:sampler_nonuniform}. Indeed, the kernel of the time-varying operation $g(t,\tau)$ defines a set of linear continuous functionals $g_n(t) = g(t_n,t)$, $t_n \in \Lambda$. %Implicit in our setting it the assumption that for each $n\in \mathbb Z$, the function
%\[ \int_{-\infty}^{\infty} g_n(t) e^{2\pi i f t} dt, \]
%is in $\Ltwo (S_X)$ (as a function of $f$).

\begin{textbox}
\begin{center}
{\bf Nonuniform Sampling} 
\end{center}
Consider a sampling set $\Lambda$ for which there exists an $\epsilon >0$ such that $|t_k-t_n|>\epsilon$ for every $t_n\neq t_k \in \Lambda$. The \emph{lower Beurling density} of $\Lambda$ is defined as the minimal number of elements of $\Lambda$ contained in a single interval of length $r$ divided by $r$, in the limit as $r$ goes to infinity. %When the upper Beurling density $d^+(\Lambda)$ equals the lower Beurling density $d^-(\Lambda)$, we say that $\Lambda$ has Beurling density $d(\Lambda)=d^+(\Lambda)$. \\
For example, the (lower) Beurling density of a uniform sampling set $\Lambda = f_s \mathbb Z$ is $f_s$. \\

%The Fourier transform relation between the covariance of a stationary process $X(t)$ and its PSD 
%\begin{equation} \label{eq:trig_iso}
%\mathbb E\left[ X(t) X(s) \right] = \mathbb E\left[ X(t-s) X(0) \right] = \int_{-\infty}^\infty e^{2\pi i(t-s)f} S_X(f)df
%\end{equation}
%defines an isomorphism between the Hilbert space generated by the closed linear span of $\left\{ X(t),\,t\in \mathbb R \right\}$ with norm $\|X(t)\|^2 = \mathbb E[X^2(t)]$ and the Hilbert space $\Ltwo(S_X)$ of complex valued functions generated by the closed linear span of the exponentials $\left\{e^{2\pi i f t},\, t \in \mathbb R\right\}$ with an $\Ltwo$ norm weighted by $S_X(f)$ \cite{dym1978gaussian}. 
%This isomorphism 
The isomorphism described by \eqref{eq:trig_iso} establishes an equivalence between the problem of estimating a Gaussian stationary process from its samples at times $\Lambda$ under the MSE criterion, and the problem of orthogonal projection onto the space spanned by $\mathcal E(\Lambda) \triangleq \left\{e^{2\pi i f t_n},\, t_n \in \Lambda \right\}$. The conditions for this MSE to vanish are related to the fact that every element of $\Ltwo (S_X)$ can be approximated by a linear combination of exponentials in $\mathcal E(\Lambda)$ \cite{10.2307/2028209,1447892, landau1964sparse}. 
%closeness of these exponential in $\Ltwo (S_X)$ \cite{10.2307/2028209}, or equivalently, to the fact that $\Lambda$ defines a \emph{set of uniquness} for the Paley-Wiener space $\Pw(\supp S_X)$ of complex valued functions whose Fourier transform is supported on $\supp S_X$ \cite{1447892}. 
This property, however, turns out to be too weak for practical sampling systems, since it does not guarantee stability: the approximation may not be robust to small perturbations in the time instances which inevitably are present in practice \cite{1447892,FEICHTINGER1992530,1082745}. As a result, only \emph{stable sampling} schemes \cite{young2001introduction} should be considered in applications. %a notion which is equivalent to the fact that $\mathcal E(\Lambda)$ comprises a Reisz basis for $\Ltwo(\supp S_X)$. 
A necessary and sufficient condition for stable sampling was given by Landau \cite{Landau1967}, who showed that it can be obtained if and only if the lower Beurling density of $\Lambda$ exceeds the spectral occupancy of $X(t)$. 
\end{textbox}

\subsection{Summary of Analog-to-Digital Compression}
We have shown that the optimal tradeoff among distortion, bitrate and sampling rate under the class of linear samplers with pointwise operations is fully described by the function $D(f_s,R)$ of \eqref{eq:idrf_lower_bound}. Moreover, the procedure for attaining an optimal point in this tradeoff is summarized in the following steps:
\begin{enumerate}
\item[(i)] Given the bitrate constraint $R$, use Pinsker-Kolmogorov water-filling \eqref{eq:skp} over the PSD $S_{X}(f)$. The critical sampling rate $f_R$ is the support of the frequency components associated with the preserved part of the  spectrum in this expression.
\item[(ii)] Use a multi-branch uniform sampler with a sufficient number of sampling branches optimized such that the combined passband of all samplers is the support of the preserved part of the spectrum \cite[Sec. IV]{Kipnis2014}. 
\item[(iii)] Recover the part of the signal associated with the preserved part of the spectrum from all branches as in standard MSE interpolation \cite{678466}.
\item[(iv)] Fix a large time lag $T$ and use a vector quantizer with $\lfloor TR \rfloor $ bits to encode the estimate in (iii) over this lag. 
\end{enumerate}
%The above steps are illustrated in Fig.~\ref{fig:ADX_procedure}.
%\begin{figure}
%\begin{center}
%\caption{ ADX methodology for sampling signals at the minimal sampling rate subject to a bitrate constraint \label{fig:ADX_procedure}}
%\end{center}
%\end{figure}
The procedure above calls for a few comments and extensions. First, we note that although our description determines the minimal distortion and sampling rate as a function of the bitrate, this dependency can be inverted. Namely, given a target distortion $D$, the Pinsker-Kolmogorov expression \eqref{eq:skp} leads to a minimal bitrate $R$ and a corresponding sampling rate required to attain this target. Second, procedure (i)-(iv) can be easily adjusted to consider a different distortion criterion according to a spectral importance masking, as described in Section~\ref{sec:theoretical_ADC}. In addition, steps (iii) and (iv) may be replaced by different techniques to attain the optimal lossy compression performance \cite{berger1971rate}. For example, the output of each sampling branch can be encoded independently of the other outputs using a separate bitstream. The bitrate of each bitstream is determined by the water-filling principle of \eqref{eq:skp_R}, with the PSD replaced by the PSD of the filtered signal at each sampling branch. Finally, we note that the multi-branch uniform sampler can be replaced by a nonuniform sampler with a single branch and possibly time-varying operation \cite{YuxinNonUniform}, or fewer uniform sampling branches of different sampling rates. That is, although uniform multi-branch sampling attains the minimal distortion $D(f_s,R)$, it may not achieve it using the most compact system implementation. In addition to these extensions, we note that the characterization of the minimal distortion in ADX has also been derived for the Wiener process and for sparse source signals \cite{KipnisWiener, KipnisReeves2017}.\\

We conclude by exploring various applications of the optimal sampling and encoding scheme in the ADX setting. 

\section{Applications \label{sec:applications}}
The most straightforward application of sampling according to the optimal ADX scheme is the possibility to reduce the sampling rates in systems operating under bitrate restrictions. Examples are listed in the box {\bf System Constraints on Bitrate}. These systems process information that originated in an analog signal under a bitrate constraint. Therefore, in these cases the rate at which the analog input is sampled can be reduced to be as low as the critical sampling rate $f_R$, without increasing the overall distortion. How low this $f_R$ is compared to the Nyquist rate or the spectral occupancy of the signal depends on our assumptions on the source statistics through its PSD. Examples for the dependency between the two are illustrated in Fig.~\ref{fig:critical_f}. Evidently, reducing the sampling rate allows the saving of other system parameters, such as power and thermal noise resulting from lower clock cycles. Alternatively, this reduction provides a way to sample wideband signals that cannot be sampled at their Nyquist rate without introducing additional distortion due to sampling, on top of the distortion due to a bitrate constraint. \\

In the rest of this section we explore additional theoretical and practical implications of our ADX scheme.

\subsection{Sampling Infinite Bandwidth Signals}
While a common assumption in signal processing is that for all practical purposes the bandwidth of the source signal is bounded, there are many important cases where this assumption does not hold. 
%In particular, any signal that cannot be sampled without loss of information can be considered non-bandlimited. 
These cases include Markov processes, autoregressive processes and the Wiener process or other semi-martingales. An important contribution of the ADX paradigm is in describing the optimal tradeoff among distortion, sampling rate and bitrate, even if the source signal is bandlimited. This tradeoff is best explained by an example. \par
Consider a Gaussian stationary process $X_\Omega(t)$ with PSD
\begin{equation} \label{eq:psd_gaussmarkov}
S_{\Omega}(f) = \frac{1/f_0}{(\pi f/f_0)^2+1},\quad f_0>0.
\end{equation}
The signal $X_\Omega(t)$ is also a Markov process, and it is in fact the unique Gaussian stationary process that is also Markovian (a.k.a the \emph{Ornstein-Uhlenbeck} process). The PSD $S_{\Omega}(f)$ is illustrated in Fig.~\ref{fig:critical_f}, along with the relation between the bitrate $R$ and the minimal sampling frequency $f_R$ required to achieve Shannon's DRF of $X_\Omega(t)$. This relation is obtained by evaluating $D(f_s,R)$ for the PSD $S_{\Omega}(f)$. In fact, the exact equation describing the green curve in Fig.~\ref{fig:critical_f} can be evaluated in closed form, from which it follows that \cite{KipnisBitrate} 
\begin{equation}
\label{eq:fDR_R_GaussMarkov}
%f_R \approx R \ln2   + f_0.
R = \frac{1}{\ln 2} \left( f_{R} -f_0 \frac{\arctan\left(\pi f_{R}/f_0\right)}{\pi/2} \right).
\end{equation}
Notice that although the Nyquist frequency of the source in this example is infinite, for any finite $R$ there exists a critical sampling frequency $f_R$, satisfying \eqref{eq:fDR_R_GaussMarkov}, such that Shannon's DRF of $X_\Omega(t)$ can be attained by sampling at or above $f_R$. \par
The asymptotic behavior of \eqref{eq:fDR_R_GaussMarkov} as $R$ goes to infinity is given by $R \sim \frac{f_{R}}{\ln 2}$. Thus, for $R$ sufficiently large, the optimal sampling rate is linearly proportional to $R$ and, in particular, in the limit of zero distortion when $R$ grows to infinity.
%
%This asymptotic behavior is required to determine the ratio between the bitrate and the sampling rate $R/f_s$ in order to encode a realization of the process in digital with vanishing distortion, as both the bitrate and the sampling rate goes to infinity. 
%
The ratio $R/f_s$ is the average number of bits per sample used in the resulting digital representation. It follows from \eqref{eq:fDR_R_GaussMarkov} that, asymptotically, the ``right'' number of bits per sample converges to $1/\ln 2 \approx 1.45$. If the number of bits per sample is below this value, then the distortion in ADX is dominated by Shannon's DRF of $X_\Omega(t)$, as there are not enough bits to represent the information acquired by the sampler. If the number of bits per sample is greater than this value, then the distortion in ADX is dominated by the sampling distortion, as there are not enough samples for describing the signal up to a distortion equals to its Shannon's DRF. \par
As a numerical example, assume that we encode $X_\Omega(t)$ using two bits per sample, i.e. $f_s = 2R$. As $R\rightarrow \infty$, the ratio between the minimal distortion $D(f_s,R)$ and Shannon's DRF of the signal converges to approximately $1.08$, whereas the ratio between $D(f_s,R)$ and $\mmse(f_s)$ converges to approximately $1.48$. In other words, it is possible to attain the optimal encoding performance within an approximate $8\%$ gap by providing one sample per each two bits per unit time used in this encoding. On the other hand, it is possible to attain the optimal sampling performance within an approximate $48\%$ gap by providing two bits per each sample taken.
%With four bits per sample the above ratios converges to $\approx 1.62$ and $\approx 1.12$, respectively.  
%\par
%The asymptotic ratio between the bitrate and the the sampling rate required to attain vanishing distortion distinguish between two classes of signals: signals for which the ratio between bits to samples go to infinity, and signals for which this ratio is bounded. For example, the first class includes all bandlimited signals since for these signals $f_R$ is bounded by $f_{\nyq}$, but also signals whose PSD vanishes too quickly. The second class includes the Gauss-Markov process with PSD \eqref{eq:psd_gaussmarkov} and the Wiener process \cite{KipnisWiener}. 
%This is also the case for non-bandlimited signals whose PSDs vanish too rapidly. 
%We note that when the bandwidth is finite, the number of bits per sample $R/f_R$ required in order to achieve Shannon's DRF goes to infinity as $R$ goes to infinity, since $f_R$ is bounded. This is also the case for non-bandlimited sources whose PSDs vanish too rapidly. 

\subsection{Theoretic limits on estimation from sampled and quantized information}
The limitation on bitrate in the scenarios mentioned in the box {\bf System Constraints on Bitrate} are the result of engineering limitations. However, sampling and quantization may also be inherit in the system model and the estimation problem. As an example, consider the estimation of an analog signal describing the behavior of the price of a financial asset. Although we assume that the price follows some continuous-time behavior, the value of the asset is only ``observed'' whenever a transaction is reported. This limitation on the observation can be described by a sampling constraint. If the transactions occur at non-uniform time lags then this sampling is non-uniform. Moreover, it is often assumed that the instantaneous change of the price is given by a deterministic signal representing the \emph{drift} plus an additive infinite bandwidth and stationary noise \cite{karatzas1998methods}. Therefore, the signal in question is of infinite bandwidth and sampling occurs below the Nyquist rate. In addition to the sampling constraint, it may be the case that the values of the transactions are hidden from us. The only information we receive is through a sequence of actions taken by the agent controlling this asset. Assuming that the set of possible actions is finite, this last limitation corresponds to a quantization constraint. Therefore, the MMSE in estimating the continuous-time price based on the sequence of actions is described by the minimal distortion in the ADX. \par
While in this case we have no control on the actual way the samples are encoded (into actions), the minimal distortion in the ADX setting provides a lower bound on the distortion in estimating the continuous-time price. This distortion can be expressed by an additional noise in a model that takes decisions based on the estimated price. 

\subsection{Removing Redundancy at the Sensing Stage}
At the end of Section~\ref{sec:PCM} we concluded that under an optimal encoder, oversampling does not affect the fundamental distortion limit since the introduced redundancy is removed in the encoding. However, oversampling may still be undesirable to the overall system performance since it results in redundant data that must be removed by additional processing. In fact, since analog signal processing is not constrained by memory or bitrate, when information originating in an analog signal is converted to digital it may bloat the system's memory with a large amount of redundant data. The processing of this data requires additional resources that are proportional to its size, and may severely restrict the system's ability to extract useful information. Indeed, lack of computational resources for the  extraction of useful information from large datasets is one of the most pressing issues of the digital age \cite{hilbert2011world}. One way to address this ``big data'' challenge is by collecting only relevant information from the analog world, i.e., attaining a non-redundant digital representation of the analog signal. For example, oversampling in PCM of Section~\ref{sec:PCM} leads to a redundant digital representation of the quantized samples, since these become more correlated with one another as the sampling rate increases. Indeed, properties of PCM imply that the optimal sampling rate that minimizes the distortion also maximizes the entropy rate of its digital output. 
\par
The counterpart of the redundancy phenomena of PCM in the more general setting of ADX is the representation attained by optimal sampling at the critical rate. This optimal sampling can be seen as a mechanism to remove redundancy at the sampling stage; it guarantees that the signal post-sampling does not contain any parts that would be removed under an optimal lossy compression. 
%Indeed, choosing the set of maximal energy leads to a sampled signal whose distribution is as close as possible to a white noise, over all signals obtained by sampling the original one using linear uniform sampling. 
\par
\begin{figure}[b]
\begin{center}
\includegraphics[scale = 0.3]{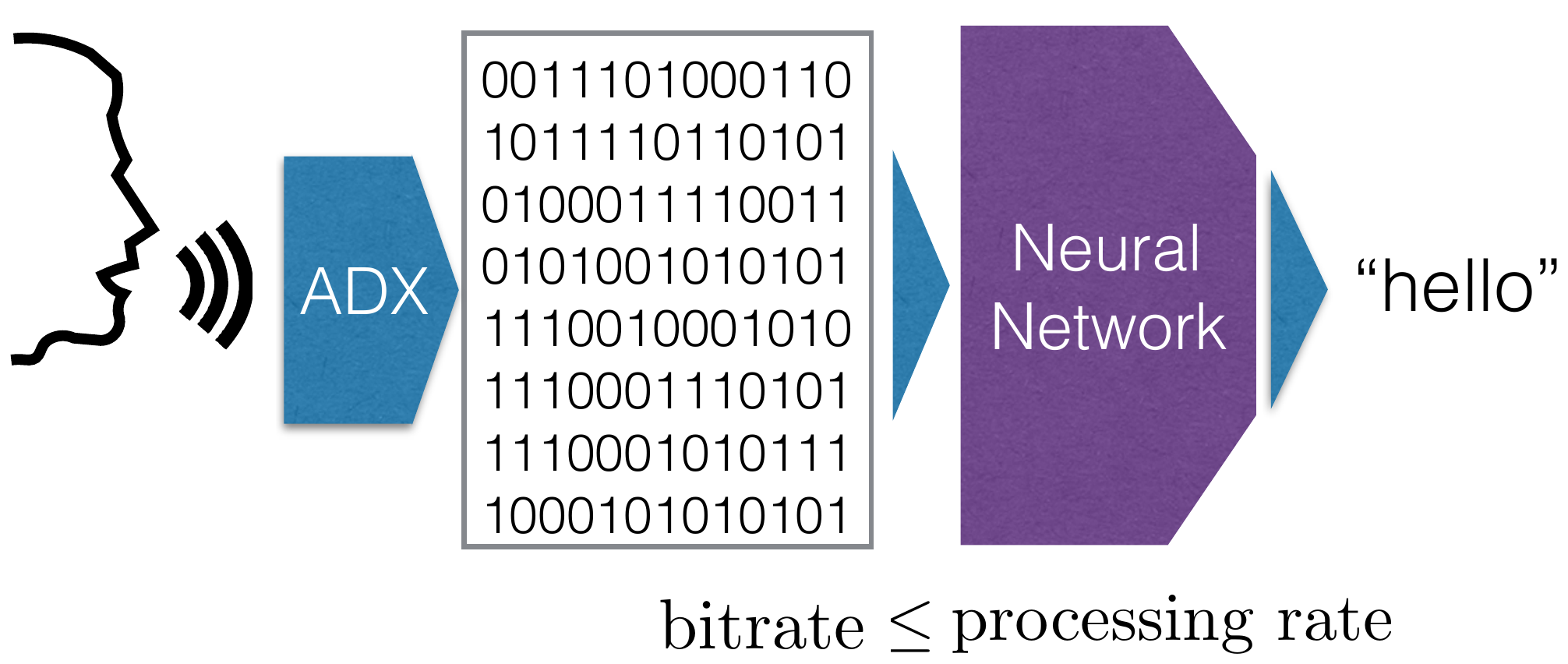}
\caption{ \label{fig:neural_net}
The bitrate of a digital representation of the sound of the word ``hello'' should not exceed the processing rate of the neural net. Sampling and lossy compression according to ADX preserves the most relevant part of the analog signal with respect to the distortion criterion and subject to the bitrate constraint. }
\end{center}
\end{figure}

As an example for a system that benefits from operating according to the principle above, we envision a real-time voice to text transcriber based on an artificial neural network \cite{graves2013speech}. 
%consider an artificial intelligent system that extract a singing voice from within musical a mixture \cite{Simpson2015}.
Such a system consists of an artificial neural network that maps a sequence of bits to words, where this sequence is obtained by an ADX unit as illustrated in Fig.~\ref{fig:neural_net}. Since the rate of information per unit time that can be processed by the neural net is limited, an optimal design of the ADX would provide bits into the neural network consistent with this rate. The challenge is therefore to sample and encode the audio signal at the rate of the neural network processing so as to provide the most relevant information subject to that rate constraint for the network to perform its classification task. If we assume that the most relevant information is described by a spectral psychoacoustic distortion function, then the optimal ADX scheme with signal PSD weighted by this distortion function provides the most relevant information for classification subject to the processing constraint. \\

\section{Conclusions}
Processing, communication and/or digital storage of an analog signal is achieved by first representing it as a bit sequence. The restriction on the bitrate of this sequence is the result of constraints on power, memory, communication and computation. In addition, hardware and modeling constraints in processing analog information imply that the digital representation is obtained by first sampling the analog waveform, and then quantizing or encoding its samples. That is, the transformation from analog signals to bits involves the composition of sampling and quantization or, more generally, lossy compression operations. \par
In this article we explored the minimal sampling rate required to attain the fundamental distortion limit subject to a strict constraint on the bitrate of the system. We concluded that when the energy of the signal is not uniformly distributed over its spectral occupancy, the optimal signal representation can be attained by sampling at a rate lower than the Nyquist rate, which depends on the actual bitrate constraint. This reduction in the optimal sampling rate under finite bit-precision is made possible by designing the sampling mechanism to sample only those parts of the signals that are not discarded due to optimal lossy compression. \par
The characterization of the fundamental distortion limit and the sampling rate required to attain it has several important implications. Most importantly, it provides an extension of the classical sampling theory of Whittaker-Kotelnikov-Shannon-Landau, as it describes the minimal sampling rate required for attaining the minimal distortion in sampling an analog signal. It also leads to a theory of representing signals of infinite bandwidth with vanishing distortion. In particular it provides the average number of bits per sample, i.e. the ratio of the bitrate (bits per unit time) and the sampling rate (samples per unit time) so that, as the number of bits and samples per unit time go to infinity, the distortion under optimal sampling and encoding decreases to zero.%
Our results also indicate that sampling at the Nyquist rate is not necessary when working under a bitrate constraint for signals of either finite or infinite bandwidth. Such a constraint may be due to power, cost or memory limitations of hardware. Moreover, sampling a signal at its critical sampling rate associated with a given bitrate constraint results in the most compact digital representation of the analog signal, and thus provides a mechanism to remove redundant information
at the sensing stage. \\

%%%%%%%%%%%%%%%%%%%%%%%%%%%%%%%%%%%%%%%%%%%%%%%%%%%%%%%%%%%%%%%%%%%%%%%%%%%%%%%%
\bibliographystyle{IEEEtran}
\bibliography{IEEEfull,sampling}

\end{document}